\documentclass[iop,appendixfloats]{emulateapj}

\usepackage{graphics,graphicx}
\usepackage{subfigure}
\usepackage{color}
\usepackage{natbib}

\begin{document}

\title{Uncovering Obscured Active Galactic Nuclei in Homogeneously Selected Samples of Seyfert 2 Galaxies}

\author{Stephanie M. LaMassa$^{1}$, T. M. Heckman$^{1}$, A. Ptak$^{2}$, L. Martins$^{3}$, V. Wild$^{4}$, P. Sonnentrucker$^1$, A. Hornschemeier$^{2}$}

\affil{$^1$The Johns Hopkins University, Baltimore, MD, USA
$^2$ NASA/GSFC
$^3$ NAT - Universidade Cruzeiro do Sul, S\~{a}o Paulo, Brazil 
$^{4}$ Institut d'Astrophysique de Paris, 75014 Paris, France
}

\begin{abstract}

We have analyzed archival \textit{Chandra} and \textit{XMM-Newton} data for two nearly complete homogeneously selected samples of type 2 Seyfert galaxies (Sy2s). These samples were selected based on intrinsic Active Galactic Nuclei (AGN) flux proxies: a mid-infrared (MIR) sample from the original IRAS 12$\mu$m survey and an optical ([OIII]$\lambda$ 5007 \AA\ flux limited) sample from the Sloan Digital Sky Survey (SDSS), providing a total of 45 Sy2s. As the MIR and [OIII] fluxes are largely unaffected by AGN obscuration, these samples can present an unbiased estimate of the Compton-thick (column density N$_H > 10^{24}$ cm$^{-2}$) subpopulation. We find that the majority of this combined sample is likely heavily obscured, as evidenced by the 2-10 keV X-ray attenuation (normalized by intrinsic flux diagnostics) and the large Fe K$\alpha$ equivalent widths (several hundred eV to over 1 keV). A wide range of these obscuration diagnostics is present, showing a continuum of column densities, rather than a clear segregation into Compton-thick and Compton-thin sub-populations. We find that in several instances, the fitted column densities severely under-represent the attenuation implied by these obscuration diagnostics, indicating that simple X-ray models may not always recover the intrinsic absorption. We compared AGN and host galaxy properties, such as intrinsic luminosity, central black hole mass, accretion rate, and star formation rate with obscuration diagnostics. No convincing evidence exists to link obscured sources with unique host galaxy populations from their less absorbed counterparts. Finally, we estimate that a majority of these Seyfert 2s will be detectable in the 10-40 keV range by the future NuSTAR mission, which would confirm whether these heavily absorbed sources are indeed Compton-thick.

\end{abstract}

\section{Introduction}

A subset of galaxies are active, indicating that the central supermassive black hole (SMBH) is accreting material. Within such active galactic nuclei (AGN), this accretion disk is in turn surrounded by an obscuring medium of dust and gas, thought to have a toroidal geometry (e.g. Antonucci 1993, Urry \& Padovani 1995). In Type 1 AGN, the system is oriented such that the line of sight intercepts the opening of this ``torus,'' exposing the accretion disk and broad load region (BLR). Conversely, this central region is blocked in Type 2 AGN, as the system is inclined such that the line of sight is through the obscuring medium. In these obscured AGNs, the narrow line region (NLR) is visible as well as scattered and reflected emission from the central engine.  

Previous studies have shown that obscured AGN constitute at least half of the local population (Risaliti 1999, Bassani 1999, Guainazzi et al. 2005). Obscuration can result from the putative torus or even the host galaxy where dust from nuclear star formation processes (e.g. Ballantyne 2008), extranuclear dust (Rigby et al. 2006) or dilution of AGN emission by the galaxy (Trump et al. 2009) can attenuate optical signatures of AGN activity. However, in this study we focus on ``toroidal'' obscured sources (where the absorption is intrinsic to the AGN, on the scale of a few parsecs, enshrouding the BLR) since a substantial fraction of heavily obscured, Compton-thick (column density N$_H \geq$ 1.5$\times 10^{24}$ cm$^{-2}$) AGN are often invoked to explain the unresolved portions of the X-ray background at 30 keV (e.g. Gilli et al. 2007). An accurate census of these AGN is necessary to constrain X-ray background synthesis models (e.g. Treister et al. 2009) and studying their properties are crucial in understanding the full AGN population. Unfortunately, such obscured AGN are missed in 2-10 keV X-ray surveys as absorption and Compton-scattering severely attenuate this X-ray emission.

Selecting AGN samples based on intrinsic AGN flux proxies (F$_{intrinsic}$), which are ideally unaffected by the amount of toroidal obscuration present, is therefore necessary to uncover Compton-thick AGN. Such diagnostics include emission lines which are primarily ionized by accretion disk photons and are formed in the NLR, making them not subject to torus obscuration, and include the optical [OIII] 5007\AA\ line (e.g. Heckman et al. 2005) and the infrared [OIV] 25.89 $\mu$m line (e.g. Mel\'endez et al. 2008, Diamond-Stanic et al. 2009, Rigby et al 2009). The obscuring medium absorbs continuum photons and re-radiates them in the mid-infrared (MIR). This emission constitutes approximately 20\% of the bolometric luminosity in most type 1 and type 2 AGN \citep{12m}, making it another isotropic indicator of intrinsic AGN flux. Follow-up X-ray observations can then reveal which sources are potentially Compton-thick.

Several studies have used infrared emission to locate such obscured AGN (Daddi et al. 2007, Fiore et al. 2009). These studies have selected sources with infrared emission in excess of that attributable to star formation, indicating the presence of an AGN. Using either \textit{Chandra} and/or \textit{XMM-Newton} spectra (Fiore et al. 2009) or stacked X-ray spectra for non-detections (Daddi et al. 2007, Fiore et al. 2009), the column densities of these sources were estimated to constrain the Compton-thick fraction. However, these studies focus on high redshift sources ($z$ = 0.7-2.5), rely on assumptions of the source spectrum shape (to convert count rate to flux for detections or hardness ratio to flux for stacked spectra) and in the case of stacked sources, only characterize the aggregate population rather than individual sources. These analyses are useful in estimating a potential Compton-thick fraction at early times in the universe, but local sources generally have the advantage of X-ray spectra with higher signal-to-noise, where each source can be individually analyzed and the spectra better characterized to more robustly constrain obscuration diagnostics.

We have undertaken such an analysis on two homogeneous samples of local ($z <$ 0.15) Seyfert 2 galaxies (Sy2s) selected based on intrinsic flux proxies: a MIR sample from the original \textit{IRAS} 12$\mu$m survey \citep{12m} and an [OIII]-selected sample culled from the Sloan Digital Sky Survey \citep{me1}. The \textit{Chandra} and/or \textit{XMM-Newton} spectra of these sources were fit in a homogeneous and systematic manner. However, column densities derived from spectral fitting in the 2-10 keV band are highly model dependent and thus may not always reflect the intrinsic toroidal absorption. Other proxies are therefore necessary to identify potentially Compton-thick sources. As the 2-10 keV X-ray emission is suppressed in absorbed sources, the ratio of this emission to intrinsic flux indicators can probe the amount of obscuration. In Compton-thick sources, the ratio of the 2-10 keV X-ray flux (F$_{2-10keV}$) to F$_{intrinsic}$ is about an order of magnitude or lower than what is observed in unobscured sources (e.g. Bassani et al. 1999, Cappi et al. 2006, Panessa et al. 2006, Mel\'endez et al. 2008). X-ray spectral signatures, most notably the equivalent width (EW) of the neutral Fe K$\alpha$ line at 6.4 keV, can also aid in uncovering heavily obscured sources. As the EW is measured against a suppressed continuum, it rises with increasing column density, reaching values of several hundred eV to over 1 keV in Compton-thick sources (e.g. Levenson et al. 2002). Similar to LaMassa et al. 2009, we use F$_{2-10keV}$/F$_{intrinsic}$ and the Fe K$\alpha$ EW as obscuration diagnostics in this work.

This paper is organized as follows: we describe the sample selection in Section 2 and the spectral fitting procedures in Section 3. We then estimate the potential Compton-thick population using ancillary optical and IR data and discuss the various possible absorber geometries revealed by our obscuration diagnostics. Utilizing  IR data to parametrize host galaxy characteristics, namely star formation processes, and AGN activity, such as intrinsic AGN luminosity, central black hole mass and Eddington ratio, we investigate whether Compton-thick sources trace a unique population. We also comment on the feasibility of detecting these sources in an upcoming hard X-ray (5-80 keV) mission, NuSTAR.

\section{Sample Selection}

Our combined sample consists of 28 MIR selected and 17 [OIII] selected Sy2s. The MIR sample is a subset of the 31 Sy2s from the original \textit{IRAS} 12$\mu$m survey \citep{12m},\footnote{Of the 32 Sy2s in the original sample, one was re-classified as a Sy1 (NGC 1097).} which was drawn from the \textit{IRAS} Point Source Catalog (Version 2).  The Sy2s in the 12$\mu$m sample were selected via a weak color cut (i.e. 12$\mu$m flux less than the 60 or 100 $\mu$m flux) and is complete to a flux-density limit of 0.3 Jy at 12$\mu$m, with latitude $|b| > 25^o$ which avoids Galactic contamination \citep{12m}. Archival \textit{Chandra} and \textit{XMM-Newton} data exist for 28 of these 31 sources.

The [OIII]-selected Sy2s were culled from the Main Galaxy sample (Strauss et al. 2002) in the Sloan Digital Sky Survey (SDSS) Data Release 4 \citep{DR4}. Using the SDSS spectra, Sy2s were identified using the the diagnostic line ratio plot of [OIII]/H$\beta$ and [NII]/H$\alpha$ (BPT diagram, Baldwin et al. 1981) and the Kauffmann et al. (2003) and Kewley et al. (2006) demarcations which distinguish Sy2s, composite systems and star-forming galaxies. The 20 local ($z <$ 0.15) Sy2s with [OIII] 5007$\AA$ flux $> 4 \times 10^{-14}$ erg s$^{-1}$ cm$^{-2}$ that lie within the AGN locus of the BPT diagram were selected to comprise this sample. Of these 20 Sy2s, 2 had archival \textit{XMM-Newton} observations and we were awarded \textit{XMM-Newton} time for another 15 sources. The X-ray analysis for these 17 [OIII] selected Sy2s was presented in LaMassa et al. 2009 and is not replicated here; we utilize the results of that study (2-10 keV X-ray fluxes, Fe K$\alpha$ EWs, etc.) in this work. Though both original samples are complete, since X-ray data only exists for 28/31 and 17/20 sources from the 12$\mu$m and [OIII] samples respectively, our resultant sample for X-ray analysis is nearly complete.

We note that this study has the advantage of analyzing samples of Sy2s selected via two different techniques which can mitigate biases from any individual selection criterion. For instance, dusty host galaxies can attenuate the optical emission lines used to identify AGN (and thus potentially miss galaxy obscured AGN), whereas MIR selection can isolate such sources. Star formation processes in the host galaxy can also enhance MIR, limiting its usefulness as an intrinsic indicator of pure AGN flux. However, such effects are minimal in this study as all but two MIR identified Sy2s live in the AGN locus of the BPT diagram and those two sources inhabit the composite (AGN and star-forming) locus and would not be optically identified as pure star-forming galaxies. We also note that many low-luminosity AGNs and some quasars lack the IR signature of a torus (Ho 2008, Hao et al. 2010, respectively), and thus MIR selection is not a useful tool for investigating such AGNs or their obscured counterparts. We explore the issues of selection effect biases further in Section 4.2.

\section{Data Analysis}

We analyzed the available archival \textit{Chandra} and \textit{XMM-Newton} observations for the 12$\mu$m sources with XAssist \citep{xassist}. This program runs the appropriate Science Analysis Systems (SAS) tasks to filter the data and clean for flaring as well as extract spectra and associated response files for user-defined sources. Table \ref{log} lists the X-ray observations used in this analysis, including the ObsIDs and net exposure times after filtering.

Twenty-five out of 28 sources were detected at the 3$\sigma$ level or greater in the 0.5 - 10 keV band. One (NGC5193) was detected in the soft band (0.5 - 2 keV) and we were thus able to fit this part of the spectrum. We obtained upper 2 - 10 keV flux limits on this source and the two undetected sources (F08572+3915 and NGC 7590), discussed in detail below. 

We used simple absorbed power law models to fit the spectra for the detected sources, which may not accurately represent the complex geometry of these systems. However, our main goal is to apply a systematic and homogeneous analysis of the spectra in a similar manner as \citet{me1} to derive an observed X-ray flux, and where possible, EW of the Fe K$\alpha$ line. More extensive X-ray modeling of several sources have been investigated in detail in the literature (e.g. see Brightman \& Nandra 2010 for more detailed X-ray modeling of the extended 12$\mu$m sample) and we do not intend to replicate previously published work. In Appendix A, we discuss individual sources, compare our derived parameters with those quoted in the literature and comment on the impact more complex models have on such parameters. We find that in 18/23 sources, we recover consistent (within 1$\sigma$) observed X-ray fluxes and Fe K$\alpha$ EW values as more complex models. This work also represents the first analysis for a handful of datasets (i.e. \textit{Chandra} spectrum of IC 5063, \textit{XMM-Newton} 2004 EPIC spectra of NGC 7172 and \textit{XMM-Newton} EPIC spectra of NGC 7674).

\subsection{Fitting Spectra from Multiple Observations}
Multiple observations for each source, as well as the spectra from the three \textit{XMM-Newton} detectors (PN, MOS1 and MOS2), were fit simultaneously with a constant multiplicative factor which was allowed to vary by $\sim$20\% to account for calibration differences among detectors/observations. The remaining model parameters were initially tied together, with the residuals inspected to check for inconsistencies among observations. Differences among \textit{XMM-Newton} observations are interpreted as source variability, and were present in 4/28 sources (NGC 4388, NGC 5506, NGC 7172 and NGC 7582).

Nine Sy2s had both \textit{Chandra} and \textit{XMM-Newton} archival data, with 8/9 having flux and/or spectral discrepancies between observations; only NGC 424 had consistent \textit{Chandra} and \textit{XMM-Newton} data. As \textit{Chandra} has higher spatial resolution than \textit{XMM-Newton}, it better isolates the central AGN. Differences in the spectra between the two observatories could thus be due to source variability, or extended emission from the host galaxy (e.g. X-ray binaries, thermal emission from hot gas, etc.) that \textit{XMM-Newton} can not resolve from the AGN emission. To test if such differences were due to variability or contamination, we extracted the \textit{Chandra} source region to have the same size as the \textit{XMM-Newton} region, $\sim$20.'' If the best fit parameters and flux were consistent between the two datasets with the matched aperture extraction areas, we concluded that extended emission is likely contaminating the \textit{XMM-Newton} observation. If a discrepancy still existed, we interpreted this as source variability between observations.

Five sources showed evidence of contamination from extended emission within the \textit{XMM-Newton} aperture, i.e. matched aperture extraction between the \textit{Chandra} and \textit{XMM-Newton} observations resulted in consistent model parameters and flux: NGC 1386, F05189-2524, NGC 3982, NGC 4501 and Mrk 463. For 3 of these sources (NGC 1386, F05189-2524 and Mrk 463), the best-fit parameters with the default \textit{Chandra} extraction region were consistent with the \textit{XMM-Newton} spectra, with the exception of the constant multiplicative factor which was lower in the \textit{Chandra} observations ($\sim40-70$\% of \textit{XMM-Newton}). We therefore fit the \textit{XMM-Newton} and \textit{Chandra} spectra simultaneously to constrain the \textit{Chandra} parameters. However, we report the flux from the \textit{Chandra} observations only in Table \ref{flux}, as this isolates the central AGN. The spectra from the default \textit{Chandra} extraction areas for the other two sources (NGC 3982 and NGC 4501) did not have consistent model parameters with the \textit{XMM-Newton} spectra, likely due to X-ray binaries in the host galaxy affecting the spectral shape in the \textit{XMM-Newton} data, so we therefore fit the \textit{Chandra} spectra from the default extraction area independently and report these parameters in Table \ref{apec}.

Three Sy2s were variable between the two observatories: NGC 4388, Arp 220 and NGC 7582. Arp 220 was fit simultaneously between the \textit{Chandra} and \textit{XMM-Newton} observations with only the absorption component fit independently for the \textit{Chandra} spectrum. NGC 4388 and NGC 7582 exhibited spectral variation between the \textit{Chandra} and \textit{XMM-Newton} observations and were therefore fit independently. We list the best fit parameters for the default \textit{Chandra} extraction spectra and the \textit{XMM-Newton} spectra separately in Table \ref{apec} for these two sources. 

\subsection{Spectral Models}
We initially fit all spectra with an absorbed power law model. Most spectra (18/26) had an adequate number of detected photons to be grouped by a minimum of 15 counts per bin without loss of spectral information and were thus analyzed with $\chi^2$; the remaining 8 (NGC 3982, NGC 4501, TOLOLO 1238-364, NGC 4968, NGC 5135, NGC 5953, NGC 6890 and NGC 7130) were analyzed with the Cash statistic (C-stat, Cash 1979) and binned by 2-3 counts as XSpec handles slightly binned spectra better than unbinned when using C-stat \citep{Teng}. With the exception of 7 sources (NGC 1667, NGC 3982, NGC 4501, TOLOLO 1238-364, NGC 4968, NGC 5953 and Arp 220), a second power law component was needed to accommodate the data (i.e. \textit{phabs$_1$*(pow$_1$ + phabs$_2$*pow$_2$)}). The two power law indices ($\Gamma$) were tied together and the normalizations and absorption components were fit independently. Such a model represents a partial covering geometry with the first power law denoting the soft scattered and/or reflected AGN continuum and the second component describing the absorbed transmitted emission.

Residuals below 2 keV were present in many of the sources, suggesting emission in excess of the scattered AGN continuum. This excess is likely due to thermal emission from hot gas related to star formation processes, and consistent with \citet{me1}, we used a thermal model (APEC in XSpec) with abundances fixed at solar to fit this emission. According to the f-test, addition of this component improved the fit at greater than the 3$\sigma$ level over the best-fit single or double power law model for 16/25 sources.\footnote{Due to the marginal soft detection of NGC 5953, we did not fit this source with APEC.} In Table \ref{apec}, we present the best-fit parameters from the APEC plus power law models, along with the $\chi^2$ values from the single absorbed power law fit, and where applicable, the double absorbed power law fit. We required a lower limit on the first absorption component ($N_{H,1}$) to be equal to the Galactic absorption. In some cases, the best-fit absorption was equal to the Galactic $N_H$ and we subsequently froze $N_{H,1}$ to the Galactic value for these sources. We were only able to obtain an upper limit on $N_{H,1}$ for three sources (Mrk 463, NGC 6890 and NGC 7130), as the lower error bound pegged at the Galactic absorption; the upper 90\% limit is thus listed in Table \ref{apec}. We also quote the 90\% upper limit on kT for the six cases where the lower error on the temperature pegged at the limit of 0.1 keV (F05189-2524, NGC 3982, the \textit{Chandra} observation of NGC 4388, NGC 4968, NGC 5347 and the \textit {Chandra} observation of NGC 7582). We included Gaussian components to accommodate the Fe K$\alpha$ emission when present (see below) and additional Gaussian components for other emission features in NGC 1068 and NGC 7582 (see Appendix A). In Table \ref{pow}, we list the best-fit parameters for the absorbed single/double power law fit for NGC 5953 and the 9 sources which according to the f-test, are not statistically significantly improved ($\geq$ 3$\sigma$) by adding the APEC component and are therefore better described by the simpler single/double power law model (NGC 424, the \textit{Chandra} observation of NGC 4388,  NGC 4968, NGC 5135, NGC 5347, NGC 6890, IC 5063, the \textit{Chandra} observation of NGC 7582 and NGC 7674).

We list the observed 2-10 keV X-ray flux from these best-fit models in Table \ref{flux}. For the cases where addition of the APEC model improved the fit, we excluded this component when deriving the X-ray flux. The flux was averaged among multiple observations when these observations were consistent. For variable sources, the flux is listed independently for each observation. For Arp 220, only the absorption varied between the \textit{XMM-Newton} and \textit{Chandra} observations, which had a negligible impact on the flux. We therefore averaged the \textit{XMM-Newton} and \textit{Chandra} fluxes for this source. We note that NGC 7582 has a higher observed \textit{Chandra} flux, compared to the \textit{XMM-Newton} fluxes, despite the smaller \textit{Chandra} spectral extraction area; aperture effects could contribute to the lower \textit{Chandra} flux (compared with \textit{XMM-Newton}) for NGC 4388.

\begin{figure}[ht]
\subfigure[Chandra spectrum.]{\includegraphics[scale=0.33,angle=-90]{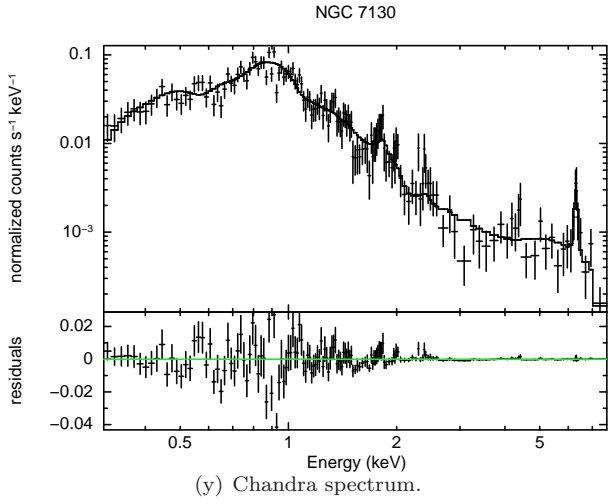}}
\hspace{30pt}
\subfigure[Black - XMM PN spectrum from 2007 observation, red - XMM PN spectrum from 2004 observation, green - XMM PN spectrum from 2002 observation.]{\includegraphics[scale=0.33,angle=-90]{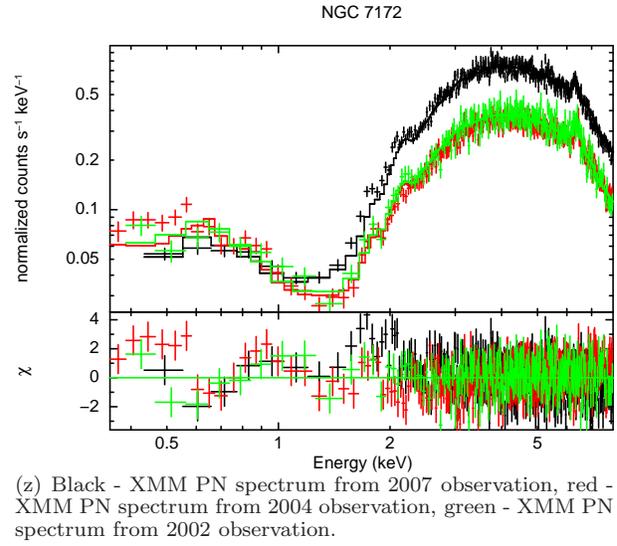}}

\setcounter{subfigure}{0}
\renewcommand{\thesubfigure}{(a\alph{subfigure})}

\subfigure[Black - XMM PN spectrum from 2005 observation, red - XMM PN spectrum from 2001 observation.]{\includegraphics[scale=0.33,angle=-90]{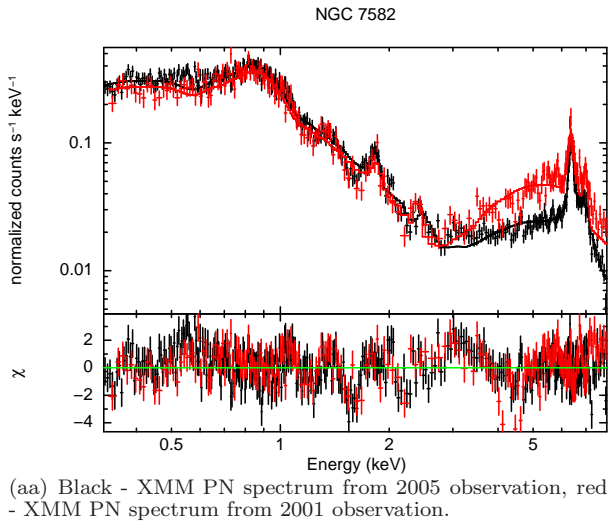}}
\hspace{30pt}
\subfigure[Black - Chandra spectrum from Oct 14, 2006 observation, red - Chandra spectrum from Oct 15, 2006 observation.]{\includegraphics[scale=0.33,angle=-90]{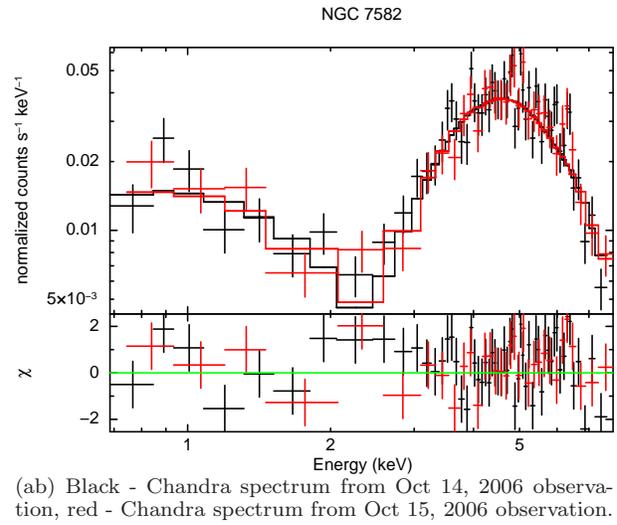}}

\subfigure[XMM PN spectrum.]{\includegraphics[scale=0.33,angle=-90]{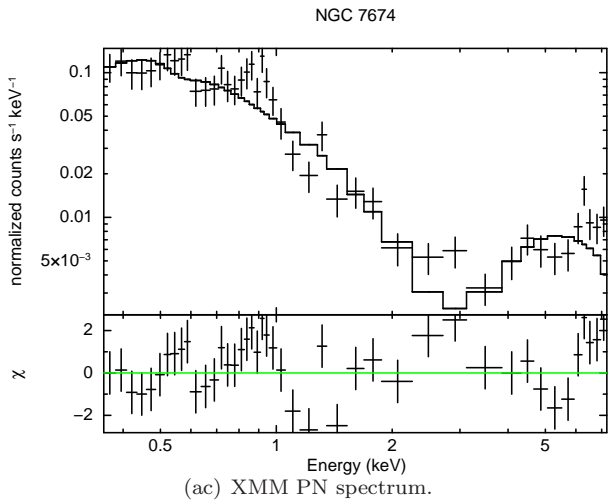}}

\caption{\label{spectra}X-ray spectra with best-fit models.}
\end{figure}

In Figure \ref{spectra}, we plot the spectra with the best-fit models. As many sources have multiple observations, we plot only one spectrum per observation, generally using the PN spectrum for \textit{XMM-Newton} observations unless the MOS spectrum had better signal-to-noise. Though we report the flux of the \textit{Chandra} observations only for NGC 1386, F05189-2524 and Mrk 463, we plot both the \textit{XMM-Newton} and \textit{Chandra} spectra to illustrate how the \textit{XMM-Newton} spectra helped to constrain the fit.

\subsection{Pileup}

Bright X-ray sources can be susceptible to pileup which occurs when a CCD records two or more photons as a single event during the frame integration time. To test if this phenomenon affected our bright sources, we examined the pattern and observed distribution plots from the SAS task \textit{epatplot} for \textit{XMM-Newton} observations and the output of PIMMS\footnote{http://cxc.harvard.edu/toolkit/pimms.jsp} for \textit{Chandra} observations. In a handful of \textit{XMM-Newton} observations (i.e. NGC 1068, NGC 5506 and NGC 7172), one to two of the detectors exhibited evidence of pileup, but at least one of the detectors did not. The ``piled'' spectra were therefore disregarded from the fit without loss of information as we obtained one to two non-piled spectra per observation (see Appendix A for details). As PIMMS uses simple models to test for the presence of pileup (e.g. single absorbed power laws whereas most of our sources needed a second power law component), we fit the \textit{Chandra} spectra with evidence of pileup (i.e. IC 5063 and NGC 7582) in \textit{Sherpa}, using the jdpileup model and best-fit continuum model (with a Gaussian at the Fe K$\alpha$ energy if necessary), to better constrain the pileup percentage. However, we utilized the pileup model in XSpec (with $\alpha$, the ``grade migration'' parameter, as the only free parameter) along with the best-fit models to derive the 2 - 10 keV flux and Fe K$\alpha$ EW, where the pileup component was removed before calculating these quantities. We note that the \textit{Sherpa} and XSpec fits using their respective pileup models give consistent best-fit parameters and observed fluxes.

\subsection{Upper Limits}

Three sources were not detected within the 2 - 10 keV range: F08572+3915, NGC 5953 and NGC 7590. NGC 5953 was detected in the soft band (0.5 - 2 keV) and was therefore fit with an absorbed power law model. It was necessary to freeze the absorption to properly model the photon index. As the soft component generally results from scattered/reflected AGN emission, the absorption attenuating this component results from obscuration along the line of sight rather than intrinsic toroidal absorption. In many cases in this study, such absorption is on the order of Galactic $N_H$ or marginally higher, so we froze $N_H$ to the Galactic value. From this fit in the soft band, we extrapolated an upper limit on the 2 - 10 keV flux.

F08572+3915 and NGC 7590 were not detected over the background in their $\sim$15 ks \textit{Chandra} and $\sim$10 ks \textit{XMM-Newton} observations, respectively. We therefore used a Bayesian approach to estimate an upper limit on the flux based on the total number of counts within the spectral extraction region and an assumed spectral shape for the AGN. We used a region size of $\sim$2'' for F08572+3915 and $\sim$7.5'' for NGC 7590 (though \textit{XMM-Newton} has lower resolution and the extraction region is generally $\sim$20'', we constrained this region to a smaller size to exclude contamination from a nearby ultraluminous X-ray source \citep{ULX}). For NGC 7590, we coadded the MOS spectra together using the ftool \textit{addspec}. We used the total detected and background counts from these spectra to calculate a one-sided 3$\sigma$ (i.e. 99.9\% confidence level) upper limit on the number of source counts. We then obtained an upper limit on the count rate by dividing this source count by the exposure time of the observation. Using an absorbed power law model, which included Galactic absorption, Compton-thick absorption ($N_H = 1.5 \times 10^{24}$ cm$^{-2}$, which is a conservative estimate as neither source was detected in X-rays) at the redshift of the source and a photon index of 1.8, we calculated the 2-10 keV flux that corresponds to the 3$\sigma$ upper limit on the count rate. These upper limits are listed in Table \ref{flux}. We note that applying this method to NGC 5953 results in a higher X-ray flux upper limit than extrapolating the spectral fit of the soft emission to higher energies, $\sim 2\times10^{-13}$ erg s$^{-1}$ cm$^{-2}$ vs  $\sim 5 \times 10^{-14}$ erg s$^{-1}$ cm$^{-2}$. We choose the latter value since this is based on the spectral information we have for this source.

\subsection{Fe K$\alpha$}

We used a Gaussian component to model the neutral Fe K$\alpha$ emission. In many cases, this feature was evident when fitting the 0.5 - 8 keV spectrum and was included in the models mentioned above. For the sources where this line was not visible, we tested for its presence using the ZGAUSS model, freezing the energy at 6.4 keV and the width at 0.01 keV and inputting the galaxy's redshift. From this fit, we can derive either a detection or upper limit on the neutral Fe K$\alpha$ flux and possibly EW. For the sources that had both \textit{XMM-Newton} and \textit{Chandra} observations and had evidence of extended emission in the \textit{XMM-Newton} field of view (i.e. NGC 1386, F05189-2524, NGC 3982, NGC 4501 and Mrk 463), we used only the \textit{Chandra} spectrum to model the Fe K$\alpha$ emission to isolate the AGN contribution.

To better constrain the EW of the neutral Fe K$\alpha$ line, we also fit the local continuum, from 3-4 keV to 8 keV, with a power law or double absorbed power law with an absorption component attenuating the second power law (when the spectral shape required this extra model). We then added a Gaussian or ZGAUSS component to this local continuum fit. The results of the global and local continuum fits to the neutral Fe K$\alpha$line are listed in Table \ref{kalpha}. In some cases (e.g. NGC 424, NGC 1386), the local fit better constrains the underlying continuum and therefore leads to a more reliable value for the EW. We use the EWs from the local fits in the subsequent analysis.

Similar to LaMassa et al. 2009, we tested the significance of the Fe K$\alpha$ EW detections by running simulations based on the power law(s) only component(s) of the local fit. We fit these simulated spectra with a Gaussian (or ZGAUSS) component to estimate the null hypothesis distribution of line normalizations. Then the percentage of times that the observed line normalization exceeded the simulated line normalizations gives the statistical significance of the line.

\section{Discussion}

With the observed X-ray flux and Fe K$\alpha$ EW constrained, we can determine the distribution of the amount of 2-10 keV attenuation associated with the obscuring torus. As both the 12$\mu$m and [OIII] sample were selected on intrinsic AGN properties, such a percentage might represent an unbiased estimate for the global AGN population. Similar to \citet{me1}, we also explore if the fitted column densities agree with the proxies we use for AGN obscuration: if the emission is seen primarily via scattering and/or reflection, do the fitted N$_{H}$ values recover the intrinsic absorption? The obscuration flux diagnostics and Fe K$\alpha$ EWs also provide clues as to the obscuration geometry in these sources. We compare host galaxy and AGN properties with Compton-thick diagnostics to determine if sources with heavy absorption trace a unique populations from their less obscured counterparts. Finally, as higher energy ($>$10 keV) observations are necessary to confirm a source as Compton-thick, we comment on the detectability of these Sy2s by NuSTAR, an upcoming hard X-ray mission.

\subsection{Obscuration Diagnostics}
As fitted column densities are model dependent and could be unreliable, we use other proxies to investigate the amount of toroidal absorption in these systems, including the ratio of the observed X-ray flux to the inherent AGN flux. We consider three diagnostics for intrinsic AGN power (F$_{intrinsic}$): the [OIII]$\lambda$ 5007\AA\ line, the [OIV]25.89 $\mu$m line and the mid-infrared (MIR) continuum. The [OIII] and [OIV] lines are primarily ionized by the central engine, and as they form in the narrow line region, are not subject to torus obscuration. The MIR emission results from the reprocessing of the AGN continuum by the dusty obscuring medium. We use the flux at 13.5$\mu$m, averaged over a 3$\mu$m window, as F$_{MIR}$ since this region is free from strong emission lines and absorption features. These fluxes are published in \citet{me1} and \citet{me2} and are not replicated here. As these proxies are to first-order unaffected by the obscuring medium, whereas the 2-10 keV X-ray flux is attenuated due to absorption and possibly Compton-scattering, the ratio of the X-ray flux to these tracers of intrinsic AGN power can probe the amount of obscuration present and has been used extensively in previous studies (e.g. Bassani et al. 1999, Heckman et al. 2005, Cappi et al. 2006, Panessa et al. 2006, Mel\'endez et al. 2008, LaMassa et al. 2009). We list the values of these obscuration diagnostic flux ratios in Table \ref{obs_table}. There are, however, several limitations to using the [OIII] and MIR fluxes in tracing the intrinsic AGN flux: the [OIII] flux could be heavily affected by dust in the host galaxy and star formation processes can contaminate the MIR flux (see \citet{me2} for a comparison between F$_{MIR}$/F$_{[OIII]}$ between the two samples). In \citet{me2}, we noted that applying the standard R=3.1 reddening correction utilizing the Balmer decrement introduced errors that did not better recover the intrinsic [OIII] emission for the 12$\mu$m sample, likely due to uncertainties in the H$\beta$ measurements from the literature. Due to uncertainties in correcting the [OIII] and MIR fluxes for contamination, we use the observed parameters, with the caveat that these may not accurately probe intrinsic AGN emission for some sources. We discuss the implications of such biases below.

\begin{figure}
\centering
\subfigure[]{\includegraphics[scale=0.30,angle=90]{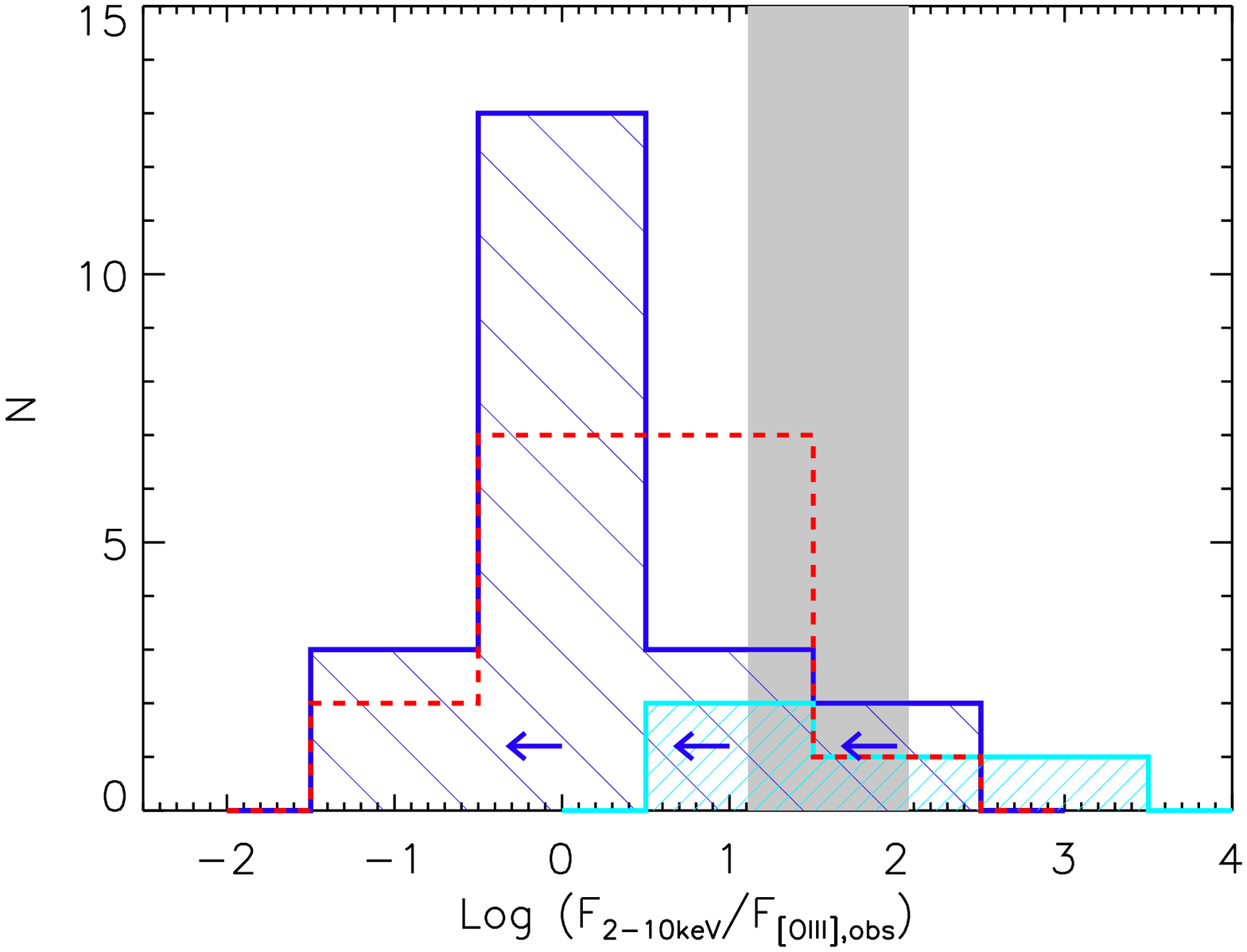}}
\subfigure[]{\includegraphics[scale=0.30,angle=90]{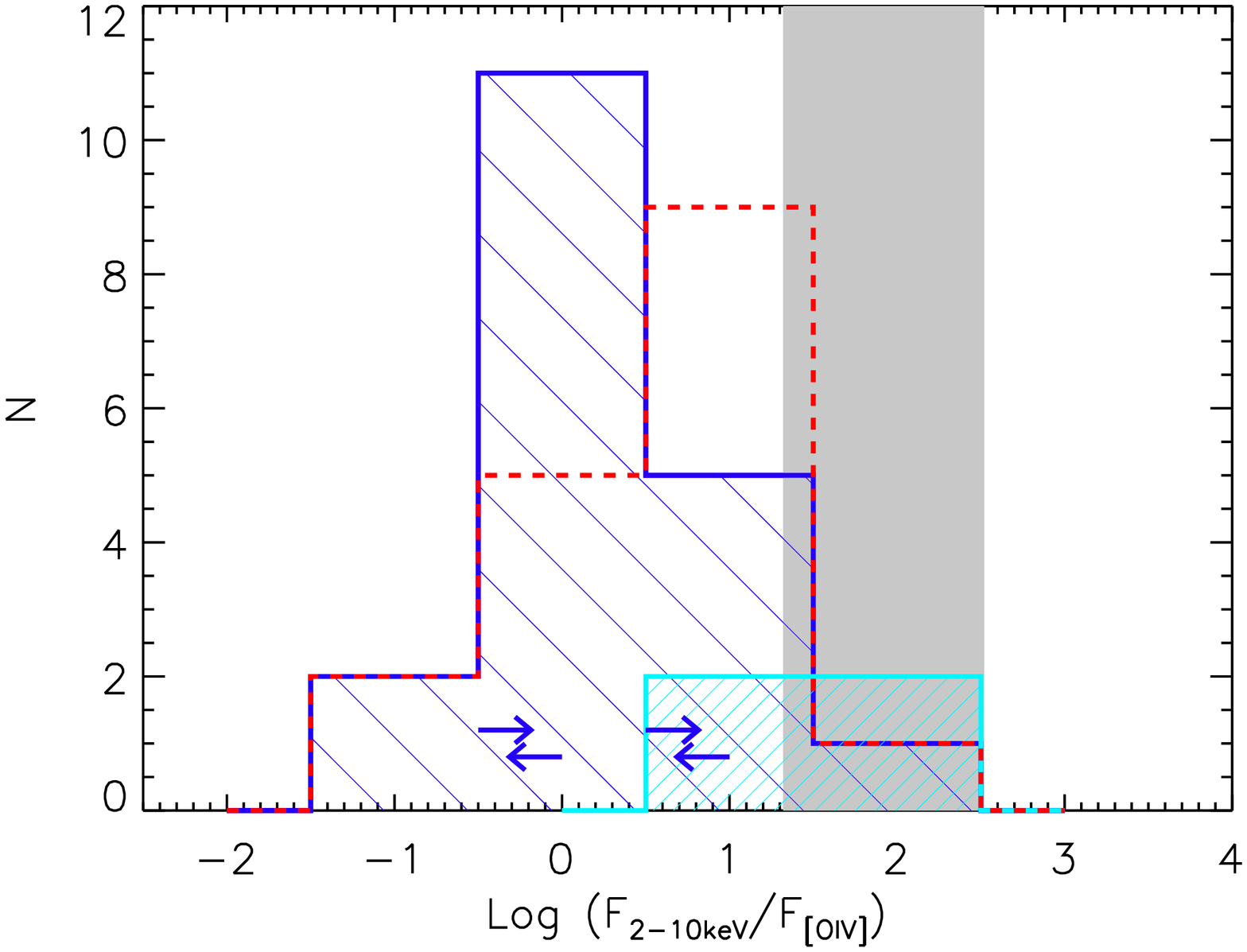}}
\subfigure[]{\includegraphics[scale=0.30,angle=90]{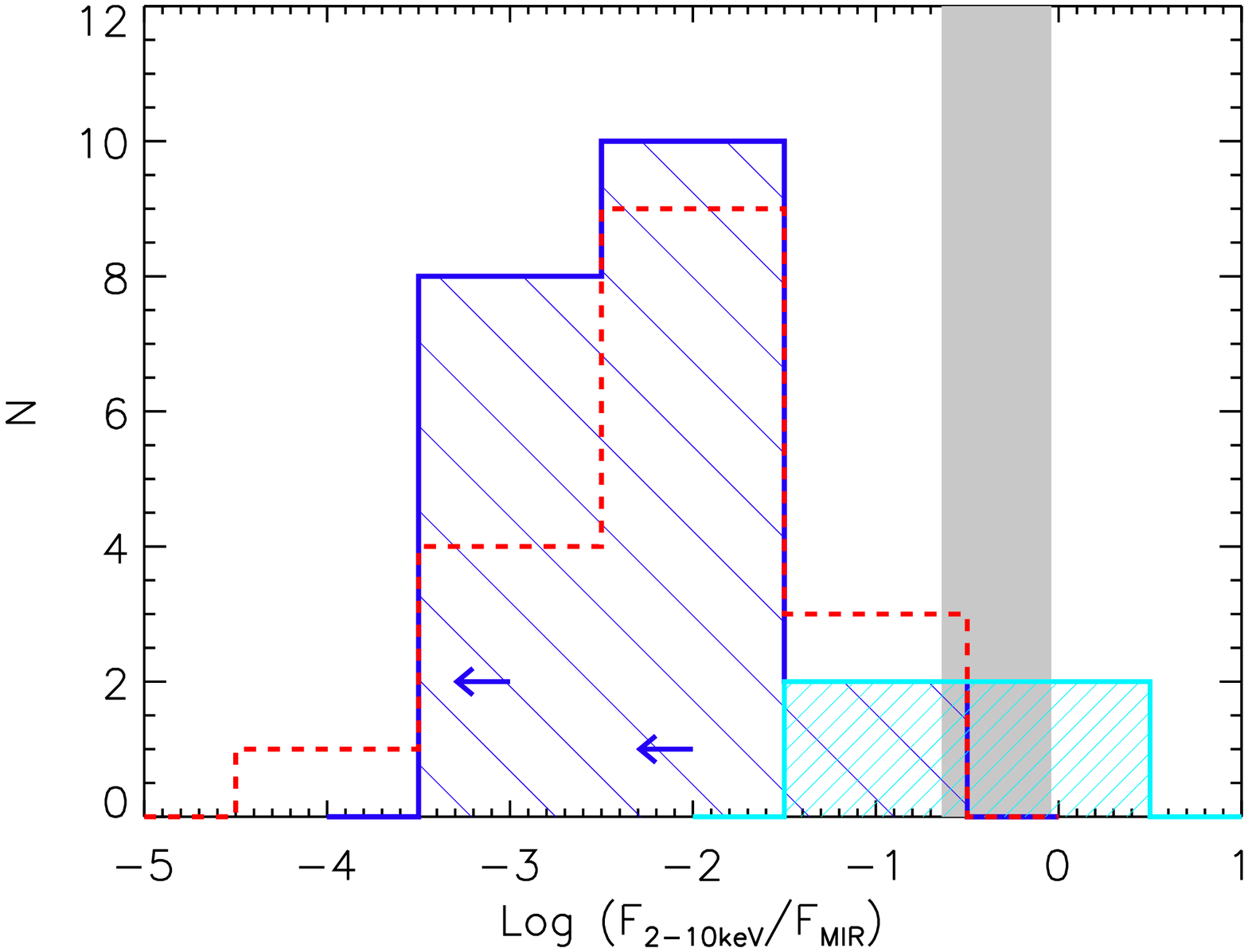}}
\caption{\label{cthick_hist}Histograms showing the distribution of obscuration diagnostic ratios (F$_{2-10keV}$/F$_{isotropic}$). Dark blue histogram represents the non-variable 12$\mu$m sources, cyan reflects the X-ray variable 12$\mu$m sources (X-ray fluxes are averaged for each source), red denotes the [OIII]-sample and the gray shaded region illustrates the average value for Sy1s from a) \citet{H05}, log ($<$F$_{2-10keV}$/F$_{[OIII],obs}>$) =1.59$\pm$0.49 dex, b) \citet{DS}, log ($<$F$_{2-10keV}$/F$_{[OIV]}>$) =1.92$\pm$0.60 dex and c) \citet{Gandhi}, log ($<$F$_{2-10keV}$/F$_{MIR}>$) =-0.34 $\pm$0.30 dex. The left facing arrows represent X-ray upper limits in a), b) and c) and right facing arrows illustrate the [OIV] upper limits in b); these values are not included in the histogram.}
\end{figure}

We plot the distributions of F$_{2-10keV}$/F$_{[OIII],obs}$, F$_{2-10keV}$/F$_{[OIV]}$ and F$_{2-10keV}$/F$_{MIR}$ in Figure \ref{cthick_hist} where the red dashed histogram represents the [OIII]-sample, the dark blue histogram denotes the non-variable 12$\mu$m sources and the cyan histogram reflects the variable 12$\mu$m sources, using the average X-ray flux among the multiple observations for each source. A wide range of values is evident in all three plots.  We compared our values with Sy1 sources, with the average flux ratio and spread delineated by the grey shaded regions in Figure \ref{cthick_hist}. The Sy1 comparison sample are culled from: a) \citet{H05} (heterogeneous [OIII]-bright sample, log ($<$F$_{2-10keV}$/F$_{[OIII],obs}>$) = 1.59$\pm$0.49 dex), b) \citet{DS} (drawn from the revised Shapley-Ames catalog, log ($<$F$_{2-10keV}$/F$_{[OIV]}>$) = 1.92$\pm$0.60 dex ) and c)\citet{Gandhi} (where F$_{MIR}$ is calculated at 12.3$\mu$m with VISIR \citet{VISIR} observations of Sys selected from \citet{Lutz} and those with existing or planned hard (14-195 keV) X-ray observations, log ($<$F$_{2-10keV}$/F$_{MIR}>$) = -0.34$\pm$0.30 dex). We note that \citet{Gandhi} report absorption-corrected X-ray luminosity whereas the other Sy1 comparison samples utilize the observed luminosity. This correction shifts the F$_{2-10keV}$/F$_{MIR}$ Sy1 ratios to higher values, though such a correction could be expected to be minimal for type 1 AGN which are thought to be largely unobscured. Also, not correcting [OIII] flux for reddening and MIR flux for starburst contamination could possibly result in obscuration diagnostic ratios that are larger or smaller respectively, and though this affects several individual galaxies with large amounts of dust and/or greater star formation activity, no such systematic trends for the sample as a whole are evident. Yaqoob and Murphy (2010) have demonstrated that the ratio of F$_{2-10keV}$/F$_{MIR}$ is more sensitive to the X-ray spectral slope and covering factor of the putative torus, rather than column density, indicating that a low ratio does not necessarily imply a Compton-thick source. However, we find all three obscuration diagnostics to agree: the majority of Sy2s have ratios an order of magnitude or lower than their Sy1 counterparts, which may indicate Compton-thick absorption. 

\begin{figure}
\subfigure[]{\includegraphics[scale=0.30,angle=90]{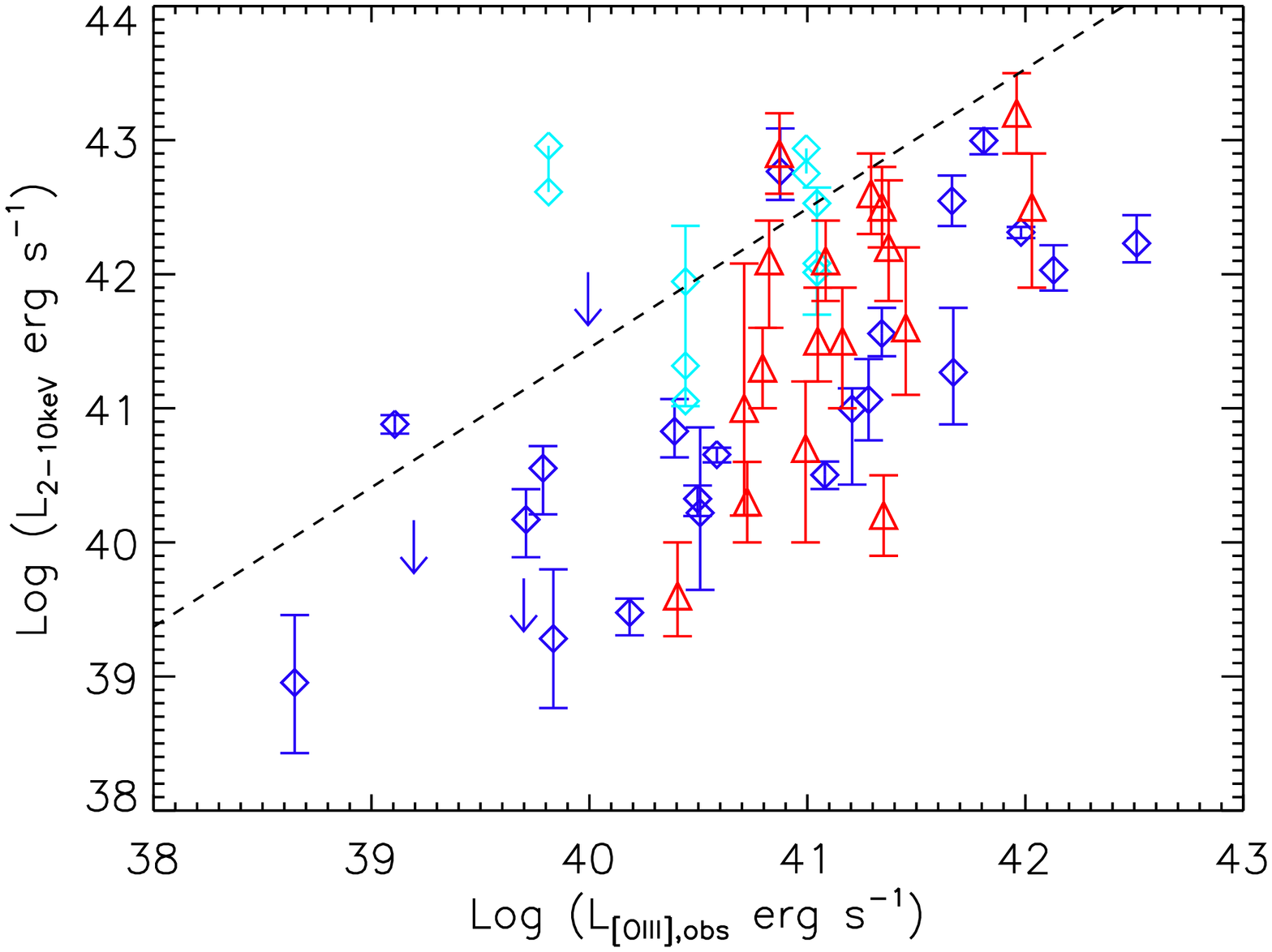}}
\subfigure[]{\includegraphics[scale=0.30,angle=90]{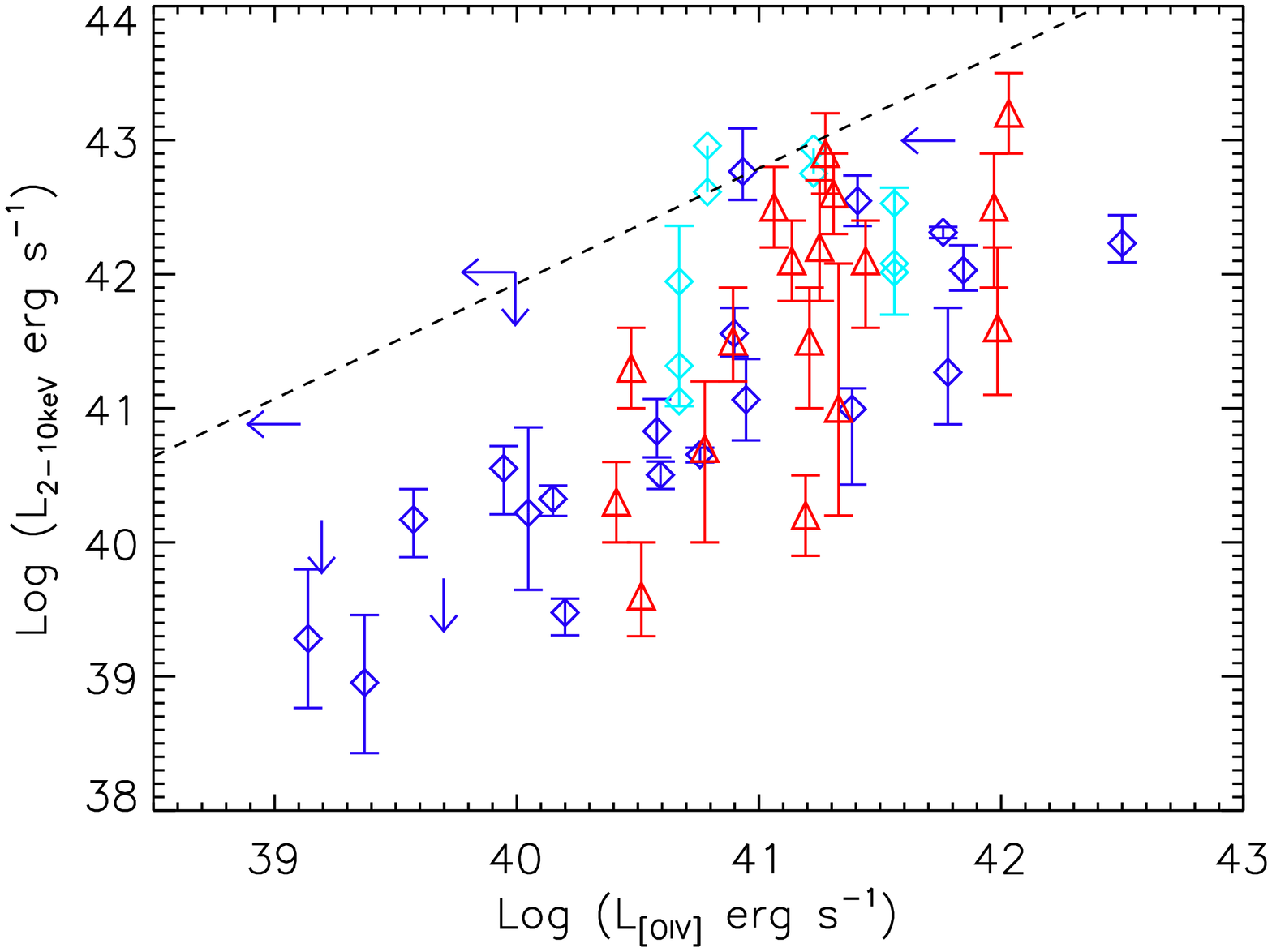}}
\subfigure[]{\includegraphics[scale=0.30,angle=90]{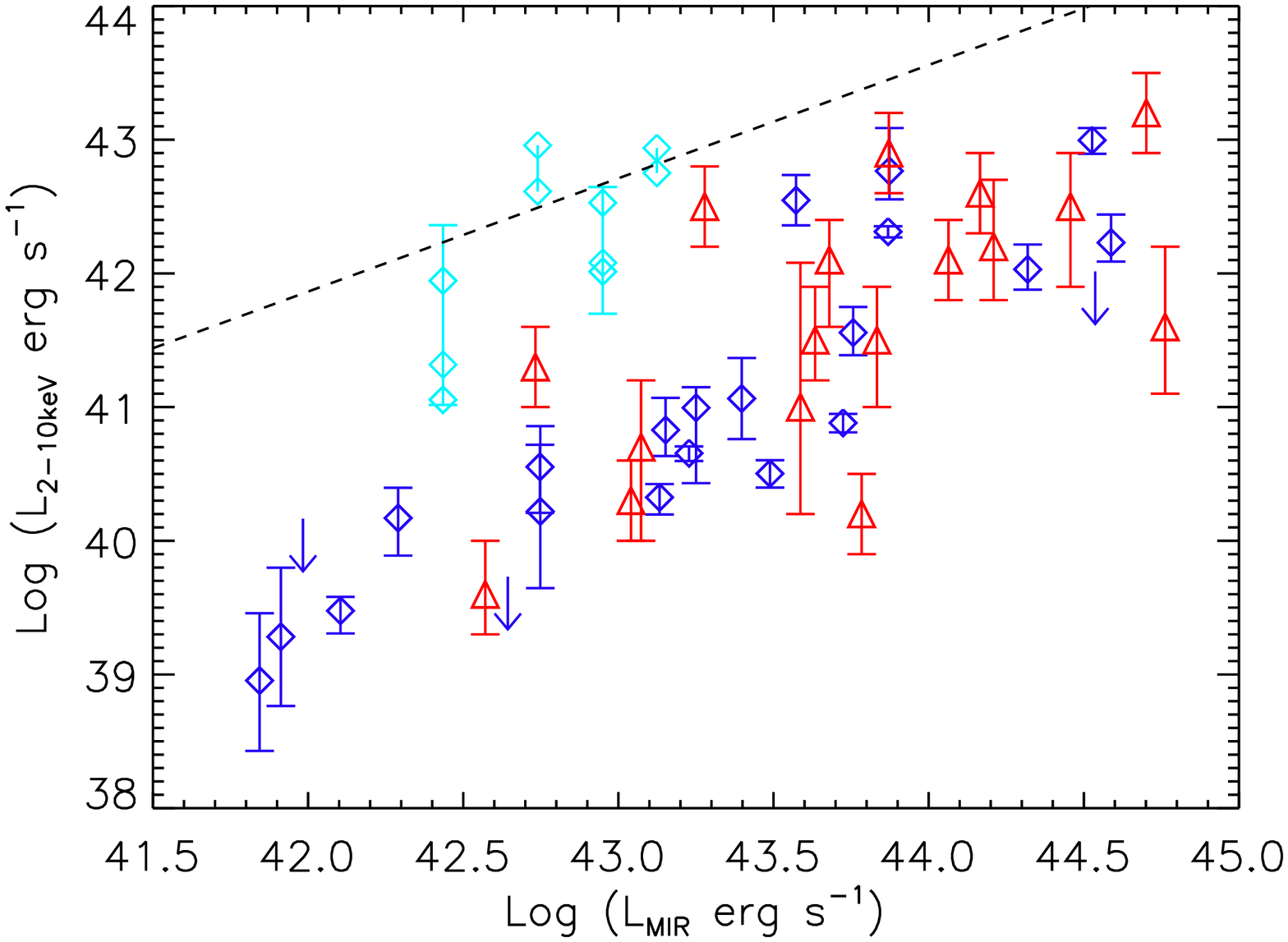}}
\caption{\label{lx_l_iso}X-ray luminosity versus proxies of intrinsic AGN luminosity. Blue diamonds represent the non-variable 12$\mu$m sources, cyan diamonds illustrate the variable 12$\mu$m sources, with the observed flux values from Table \ref{flux} for each source connected by a straight line, and red triangles denote the [OIII]-sources. Error bars for the variable sources represent the upper error on the maximum X-ray flux and lower error on the minimum X-ray flux. Variable sources NGC 5506 and NGC 7172 do not have error bars plotted as they are smaller than the symbol size. The dashed line represents the relationship for Sy1s from  a) \citet{H05}, slope = 1.4 dex with intercept -0.15 dex , b) \citet{DS}, slope = 0.86 dex with intercept = 7.53 dex and c) \citet{Gandhi}, slope = 0.85 dex with intercept 6.27 dex. In all cases, the majority of the Sy2s are below this relationship, illustrating that Sy2s have weaker X-ray emission than their Sy1 counterparts.}
\end{figure}

This trend is further illustrated by Figure \ref{lx_l_iso}, which plots the observed X-ray luminosity as a function of intrinsic AGN luminosity proxies, with the best-fit relationship for Sy1s overplotted. Here, the red triangles represent the [OIII]-sample, the blue diamonds denote the non-variable 12$\mu$m sources and the cyan diamonds reflect the variable 12$\mu$m sources, with the individual X-ray fluxes (see Table \ref{flux}) plotted for each variable source and connected by a solid line. The relationship for the Sy1 sources were calculated by multiple linear regression (i.e. REGRESS routine in IDL) for the \citet{H05} and \citet{DS} samples; for the MIR relationship, we utilized the best-fit parameters from \citet{Gandhi} for their Sy1 subsample. The majority of Sy2s lie well below the relations for Sy1s, demonstrating that these type 2 AGN have weaker observed X-ray emission.

As the X-ray and optical and IR observations were not carried out simultaneously, it is possible that variability in the source could be responsible for the disagreements between the X-ray flux and intrinsic flux proxies. Such a scenario can be realized if the X-ray observations are made after the central source has ``shut-off'' (postulated to explain the discrepancy between the Type 1 optical spectrum yet reprocessing-dominated X-ray spectrum for NGC 4051, see \citet{M03} and references therein), or the converse, where optical observations are made during a sedentary state and X-ray observations catch the source in active state (e.g. \citet{G09}). Though we can not rule out variability as contributing to the discrepancy between the X-ray luminosity and intrinsic AGN luminosity proxies for any individual source, such an effect can not be responsible for the overall trend in this sample. Variability in Sy1 samples contributes to the dispersion in L$_{2-10 keV}$/L$_{isotropic}$ ratios, yet they exhibit systematically higher X-ray luminosity (normalized by intrinsic AGN power) than Sy2s (Figures \ref{cthick_hist} and \ref{lx_l_iso}). This is confirmed by two-sample tests where we employed survival analysis (ASURV Rev 1.2, Isobe and Feigelson 1990; LaValley, Isobe and Feigelson 1992; Feigelson and Nelson 1985 for univariate problems) to account for upper limits in the X-ray flux. The Sy1 and Sy2 obscuration diagnostic ratios differ at a statistically significant level ($\leq$ 1$\times10^{-4}$ probability that they are drawn from the same parent population), which would not be expected if variability was the main driver for the discrepancy between intrinsic AGN flux and observed X-ray flux.

We have demonstrated that the majority of Sy2s in our samples are under-luminous in X-ray emission as compared to Sy1s, but is this trend due to obscuration or inherent X-ray weakness? The EW of the neutral Fe K$\alpha$ line can differentiate between these two possibilities and is thus another obscuration diagnostic. In heavily obscured sources, the AGN continuum is suppressed, whereas the Fe K$\alpha$ line is viewed in reflection, leading to a large Fe K$\alpha$ EW (several hundred eV to several keV, e.g. Ghisellini et al. 1994, Levenson et al. 2002). In Figure \ref{ew_ratios}, Fe K$\alpha$ EW is plotted as a function of obscuration diagnostic ratios. A clear anti-correlation is present which is statistically significant according to survival analysis (Isobe et al. 1986 for bivariate problems): we obtain Spearman's $\rho$ values of -0.648, -0.657 and -0.645 for Fe K$\alpha$ EW vs. F$_{2-10keV}$/F$_{[OIII],obs}$,  F$_{2-10keV}$/F$_{[OIV]}$ and  F$_{2-10keV}$/F$_{MIR}$, respectively. These best-fit correlations are overplotted for each relation in Figure \ref{ew_ratios}. The decrease of observed X-ray flux, normalized by intrinsic AGN flux, with increasing Fe K$\alpha$ EW indicates that obscuration is responsible for attenuating X-ray emission in these Sy2s. These results are consistent with the three-dimensional diagnostic diagram of Bassani et al. 1999 which shows a correlation between Fe K$\alpha$ EW and column density which anti-correlates with F$_{2-10keV}$/F$_{[OIII],corr}$ (where F$_{[OIII],corr}$ is the redenning corrected [OIII] flux).

Not only do a majority of this combined Sy2 sample exhibit trademarks of Compton-thick obscuration (an order of magnitude lower F$_{2-10keV}$/F$_{isotropic}$ ratios than Sy1s and large Fe K$\alpha$ EW values), but a wide range of these diagnostic values are evident. No clear separation exists between Compton-thick and Compton-thin sub-populations. Also, though the diagnostic flux ratios generally point to the same sources as having Compton-thick obscuration, not all three ratios agree for a handful of sources (e.g. F05189-2524, NGC 5347, Arp 220, NGC 4388 and NGC 7582): some ratios indicate a Compton-thin source whereas others suggest Compton-thick. As the various intrinsic AGN indicators exhibit some scatter in inter-comparisons (see e.g. LaMassa et al. 2010), a spread in F$_{2-10keV}$/F$_{isotropic}$ values is expected. For F05189-2524, NGC 5347 and Arp 220, this discrepancy could be due to dust in the host galaxy affecting the [OIII] line, as mentioned above and/or large amounts of dust in the host galaxy boosting the MIR flux. The 2005 \textit{XMM-Newton} observation of NGC 7582 has a F$_{2-10keV}$/F$_{[OIII],obs}$ value marginally higher than the nominal Compton-thick/Compton-thin boundary, so the three flux ratio diagnostics may be considered to agree. However, the biases discussed previously in the observed [OIII] flux and MIR flux can not account for the disagreement of the diagnostic flux ratios in the \textit{Chandra} and July \textit{XMM-Newton} observations of NGC 4388 and the 2001 \textit{XMM-Newton} observation of NGC 7582, where F$_{2-10keV}$/F$_{[OIV]}$ point to the sources being Compton-thick at these stages, but the other ratios suggest a Compton-thin nature. Similarly, an Fe K$\alpha$ EW of 1 keV is often cited as the nominal boundary for a Compton-thick source based on observations (e.g. Bassani et al. 1999, Comastri 2004, Levenson et al. 2006), yet NGC 1068, the archetype for a Compton-thick Sy2 \citep{M97}, has a measured EW of 0.60$^{+0.05}_{-0.05}$ keV (in agreement with Pounds \& Vaughan 2006 but not Matt et al. 2004, see Appendix). Hence, though the diagnostics presented here can help in uncovering the possible Compton-thick nature of a type 2 AGN, nominal boundaries should be considered approximate, especially since a continuum of both diagnostic flux ratios and Fe K$\alpha$ EWs are present.

\subsection{Implications for the Local AGN Population}
As both sub-samples were selected based on intrinsic AGN proxies and the majority is likely Compton-thick, this implies that heavily obscured sources could constitute most of the local AGN population. X-ray surveys in the 2-10 keV range, biased against these Compton-thick type 2 AGNs, would thus miss a significant portion of the population. Indeed, Heckman et al. 2005 find that the luminosity function (which parametrizes the number of sources per luminosity per volume) for X-ray selected AGN is lower than the luminosity function for optically ([OIII]) selected sources. However, recent work (Trouille \& Barger 2010, Georgantopoulos \& Akylas 2010) leads to the opposite conclusion, namely agreement between [OIII] and X-ray luminosity functions. As Georgantopoulos \& Akylas (2010) point out, though the luminosity functions are similar, the selection techniques tend to find different objects, with [OIII]-selection favoring the X-ray weak sources. Hence, the number of sources per volume per luminosity may be comparable, but any one selection technique does not sample the full population. For instance, Yan et al. (2010) found that only 22\% of their 288 optically selected AGNs are detected in the 200 ks \textit{Chandra} Extended Groth Strip survey, and they attribute the non-detection of the majority of the remaining sources to heavy toroidal obscuration. Conversely, X-ray selection can identify AGN that are categorized as star-forming galaxies by optical emission line diagnostics. For instance, Yan et al. (2010) note that about 20\% of the X-ray sources identified as star-forming galaxies from optical emission lines have X-ray emission in excess of that explicable by star-formation, indicating the presence of an AGN. This finding is similar to the results of Trouille \& Barger 2010 who find that at least 20\% of X-ray selected AGN in their sample are identified as star-forming according to optical diagnostics. Perhaps such competing biases work in concert to produce similar [OIII] and X-ray luminosity functions.

\subsection{Investigating Obscuration Geometry}

\subsubsection{Fitted Column Densities}

Here, we explore the relationship between obscuration diagnostics and the column densities (N$_H$) derived from spectral fitting. In Figure \ref{nh_cthick}, we plot the fitted column densities as a function of F$_{2-10keV}$/F$_{isotropic}$ and Fe K$\alpha$ EW for the 12$\mu$m sample (see LaMassa et al. 2009 for a discussion of fitted N$_H$ for the [OIII] sample); as the [OIV] line is the least affected by the host galaxy contaminations mentioned above, we use F$_{[OIV]}$ as F$_{isotropic}$. With the exception of several sources, the fitted N$_H$ values approximately trace the degree of absorption implied by the obscuration diagnostics. However, a handful of sources lay several orders of magnitude below this trend, and are labeled in Figure \ref{nh_cthick}. This result is consistent with the findings of Cappi et al. (2006), where several Sy2s have fitted N$_H$ values an order of magnitude below that suggested by F$_{2-10 keV}$/F$_{[OIII],obs}$. Both F$_{2-10keV}$/F$_{isotropic}$ and the Fe K$\alpha$ EW diagnostics point to the same sources as being anomalous, with NGC 3982 and NGC 4501 missing from the Fe K$\alpha$ plot due to having an unconstrained EW or upper limit on the EW, respectively. All five of these sources required only a single power law model (with a thermal component in many cases) to adequately fit the spectrum. The low observed X-ray fluxes and high Fe K$\alpha$ EW values indicate that the emission is primarily seen in scattering and/or reflection, rather than transmission through the obscuring medium. Hence such fitted N$_H$ values are likely associated with the line of sight absorption to the scattered/reflected component, suggesting that simple models of a foreground screen extincting the central source do not always recover the intrinsic absorption.

Partial covering models, parametrized in this work by a double absorbed power law with the two photon indices tied together, can also misrepresent the inherent column density. For example, such a model fairly fit the spectra for NGC 1068 ($\chi^2$=450.4 with 247 degrees of freedom), yet the best fit N$_H$ was $\sim$9$\times10^{22}$ cm$^{-2}$ whereas the lower limit on this column density from higher energy observations is 10$^{25}$ cm$^{-2}$ \citep{M97}. Though a partial covering model could more realistically represent the geometry of the system, assuming a certain percentage of transmitted light through the obscuring medium with the rest scattered into the line of sight, it could be subject to the same limitations discussed above for single absorbed power law models. 

Published N$_H$ distributions could potentially be biased, skewed to lower values, though checks based on obscuration diagnostics can help mitigate this problem. For example, Akylas et al. (2006) analyzed the X-ray spectra for 359 sources from \textit{XMM-Newton} and the \textit{Chandra} Deep Field - South (CDFS), deriving intrinsic column densities from fitted N$_H$ values though adopting a column density of 5$\times$ 10$^{24}$ cm$^{-2}$ for the cases where $\Gamma <$ 1, a signature of Compton-thick obscuration. However, as Cappi et al. (2006) note, this criterion could indicate a Compton-thick source while the Fe K$\alpha$ EW and flux diagnostics suggest Compton-thin (e.g. NGC 4138 and NGC 4258) or vice versa (e.g. NGC 3079). Tozzi et al. (2006) use a reflection model (PEXRAV in XSpec) for Compton-thick sources in the CDFS, which are defined as those AGN with a better fit statistic using PEXRAV than an absorbed power law model. However, as Murphy \& Yaqoob point out (2009), such a model describes reflection off of an accretion disk, which is not physically relevant for the putative torus obscuration. Derived N$_H$ values could then potentially be suspect for some sources. Other diagnostics are therefore crucial in checking the reliability of fitted N$_H$ values. For example, Krumpe et al. (2008) find the ratio of X-ray to optical flux, as well as the non-detection of an Fe K$\alpha$ line in the stacked spectrum of 14 type II QSOs (AGN with intrinsic L$_{2-10keV} \geq$ 10$^{44}$ erg/s), to verify their distribution of moderately absorbed, but not Compton-thick, column densities.

\subsubsection{Variable Sources}
It is intriguing to note that all X-ray variable sources in this study are on the high end of the obscuration flux diagnostics (see Figures \ref{cthick_hist} and \ref{lx_l_iso}). These high F$_{2-10 keV}$/F$_{isotropic}$ flux ratios may indicate that the X-ray emission from the central source is seen directly. However, the high Fe K$\alpha$ EW for the \textit{Chandra} and July 2002 observations of NGC 4388 (0.29$^{+0.11}_{-0.08}$ keV and 0.62$^{+0.10}_{-0.10}$ keV, respectively) and for the \textit{XMM-Newton} observations of NGC 7582 (0.58$^{+0.04}_{-0.04}$ keV and 0.31$^{+0.05}_{-0.05}$ keV) are higher than predicted for transmission-dominated spectra, where the EW with respect to the primary transmitted emission is $<$0.18 keV \citep{M02}. \citet{n7582x} propose a double absorption geometry to account for the variability in NGC 7582: a ``thick'' absorber which attenuates just the central source, attributed to the putative torus, and a ``thin'' absorber which enshrouds the primary and reflected emission and is located externally to the torus. They postulate that this inner, ``thick'' absorber is inhomogeneous, accounting for the observed X-ray variability. A similar scenario may be present for NGC 4388 and be responsible for both sources switching from transmitted-dominated to reflection-dominated states (or vice versa). The Fe K$\alpha$ EWs for the two other variable sources, NGC 5506 and NGC 7172, as well as the flux ratio diagnostics are consistent with Compton-thin sources, implying the central source is consistently viewed directly.

\subsection{Are Compton-Thick Sources Unique?}
Here we investigate whether Compton-thick sources differ from their Compton-thin counterparts in terms of host galaxy  and AGN properties. In particular, we examined whether systematic differences exist in intrinsic AGN power, Eddington ratio (L$_{bolometric}$/L$_{Eddington}$), central black hole mass (M$_{BH}$), the AGN contribution to the ionization field, and star formation activity. The results are summarized in Figures \ref{liso_c_thick} through \ref{alpha_c_thick} and in Table \ref{host_prob}, where we utilized survival analysis to calculate Spearman $\rho$ values and the associated probabilities that the obscuration diagnostics are uncorrelated with host galaxy properties: P$<$0.05 indicates that the quantities are significantly correlated ($\geq$2$\sigma$ level). The values of the relevant host galaxy parameters used in this analysis are presented in \citet{me2}.

To test whether Compton-thick sources have unique AGN properties, we searched for correlations between toroidal obscuration and intrinsic AGN luminosity, accretion rate and central black hole mass (M$_{BH}$\footnote{M$_{BH}$ measured using velocity dispersion and the M-$\sigma$ relation \citep{m-sigma}. See LaMassa et al. 2010 for literature references to M$_{BH}$ for the 12$\mu$m sample; velocity dispersions for the [OIII]-sample were derived from SDSS.}). As discussed previously, the [OIV] 25.89$\mu$m line serves as a robust proxy of intrinsic AGN flux as it is mainly ionized by the central engine and not affected by host galaxy reddening as the [OIII] line is (e.g. Mel\'endez et al. 2008, Diamond-Stanic et al. 2009, Rigby et al 2009). We therefore utilize L$_{[OIV]}$ as L$_{isotopic}$ in Figures \ref{liso_c_thick} and \ref{edd_c_thick} and Table \ref{host_prob}. According to survival analysis, a marginal statistically significant correlation exists between L$_{isotropic}$ and two of the Compton-thick flux ratios (F$_{2-10keV}$/F$_{[OIII],obs}$ and F$_{2-10keV}$/F$_{[OIV]}$), with a marginal significant anticorrelation between L$_{isotropic}$ and Fe K$\alpha$ EW. Figure \ref{liso_c_thick}, however, shows these dependencies to be weak with a wide scatter, especially considering the error bars which can not be accommodated in the survival analysis calculation. We find no correlations between implied column density and accretion rate (using L$_{[OIV]}$/M$_{BH}$ as a proxy for the Eddington ratio) and M$_{BH}$ (Figures \ref{edd_c_thick} and \ref{mbh_c_thick}); survival analysis does indicate a weak significant relationship between Eddington parameter and F$_{2-10keV}$/F$_{MIR}$, but this is likely driven by the dependence on L$_{[OIV]}$. As the weak correlation between luminosity and obscuration is tenuous at best, we conclude that Compton-thick sources do not have systematically different AGN properties from their less obscured counterparts.

Could there be a relation between the obscuration shrouding the central engine and the large amounts of dust and gas necessary for starburst activity? As Levenson et al. (2004, 2005) point out, NGC 5135 and NGC 7130 (both members of the 12$\mu$m sample) are starburst galaxies that likely harbor Compton-thick AGN. The combined [OIII] and 12$\mu$m samples provide us an opportunity to test if such a relation is generic. We use infrared quantities to illuminate the relative importance of starburst versus AGN activity: F$_{[OIV]}$/F$_{[NeII]}$, which probes the hardness of the ionization field as F$_{[OIV]}$ is largely ionized by the AGN whereas [NeII] 12.81$\mu$m is excited by star formation processes; the EW of the 17 $\mu$m polycyclic aromatic hydrocarbon (PAH) feature \citep{Genzel}, which includes the emission features between 16.4-17.9 $\mu$m; and the MIR spectral index $\alpha_{20-30\mu m}$\footnote{$\alpha_{\lambda_1 - \lambda_2}=log(f_{\lambda_1}/f_{\lambda_2})/log(\lambda_1/\lambda_2)$} \citep{Deo}.\footnote{We note that we only have these data for 27 of the 28 12$\mu$m sources presented in this work as NGC 1068 had saturated low-resolution \textit{Spitzer} data. We were therefore unable to obtain a PAH 17 $\mu$m EW value or $\alpha_{20-30\mu m}$ value for this source.} A higher value of F$_{[OIV]}$/F$_{[NeII]}$ indicates the dominance of AGN activity whereas larger PAH EW at 17$\mu$m and $\alpha_{20-30\mu m}$ values denote higher levels of starburst activity. As Figures \ref{oiv_neii_c_thick} - \ref{alpha_c_thick} and Table \ref{host_prob} illustrate, a correlation between column density and hardness of incident ionization field/star formation activity do not exist. These results suggest that the gas responsible for starburst processes likely originates in regions of the galaxy not associated with the putative torus, and similarly that gas from the interstellar medium does not contribute significantly to toroidal AGN obscuration in hard (2-10 keV) X-rays.

\subsection{NuSTAR: Detection at Higher Energies}
In order to confirm a source as Compton-thick, observations at higher energies ($>$10 keV) are necessary. The spectral characteristic of heavily obscured AGN is the so-called ``Compton-hump'', a peak in the spectrum between 20-30 keV which is caused by the competing effects of absorption on the low-energy end and Compton down-scattering on the high energy range. The Nuclear Spectroscopic Telescope Array (NuSTAR), to be launched in 2012, is sensitive in the 5 - 80 keV band, and could thus confirm our obscured candidates as Compton-thick sources, if they are detected.

To test if these sources would be detectable by NuSTAR, we simulated higher energy spectra, using the XSpec command \textit{fakeit}, based on the best-fit model and the associated response and background files provided by the NuSTAR team (http://www.nustar.caltech.edu/for-astronomers/simulations). For the three non-detections in the 12$\mu$m sample, we simulated spectra using a model that takes into account Compton scattering assuming a spheroidal obscuration geometry (PLCABS in XSpec), with N$_H$=$10^{24}$ cm$^{-2}$, $\Gamma$=1.8 and the maximum number of scatterings set to 5; the normalization was adjusted such that model 2-10 keV flux equaled the upper limits we calculated via Bayesian analysis. Using the simulated observed source and background count rate, we estimated the exposure time necessary for a source to be detected at the 5$\sigma$ level over the background. We find that all but five sources from the 12$\mu$m sample (NGC 1386, NGC 1667, Tololo 1238-364, NGC 4968 and NGC 6890) and four from the [OIII]-sample (2MASX 08035923+2345201, 2MASX J10181928+3722419, 2MASX J13463217+6423247 and NGC 5695) will be detected with exposure times less than 100 ks (see Appendix). However, though our simulations indicate the three non-detected 2-10 keV sources will be observable by NuSTAR, this is an optimistic estimate, and should be treated with caution.

In \citet{me2}, we noted that the majority of these Sy2s are undetected by the Swift BAT Surveys, indicating that these sources are heavily absorbed. However, as NuSTAR probes to a much deeper flux level in a million second observation than the BAT surveys ($\sim$2$\times 10^{-14}$ erg s$^{-1}$ cm$^{-2}$ vs. the limiting BAT flux of $\sim 3.1 \times 10^{-11}$ erg s$^{-1}$ cm$^{-2}$), the majority of these heavily obscured sources will likely be detected if observed by NuSTAR.

\section{Conclusions}
We have analyzed archival \textit{Chandra} and \textit{XMM-Newton} observations for two nearly complete homogeneous samples of Sy2 galaxies: one selected from the SDSS on the basis of observed [OIII] flux and a MIR sample from the original \textit{IRAS} 12$\mu$m sample. The combined sample provided us with 45 Sy2s with existing \textit{Chandra} and/or \textit{XMM-Newton} data. Of these, three were not detected above the background (F08572+3915, NGC 5953 and NGC 7590) and four exhibited evidence of variability among multiple X-ray observations (NGC 4388, NGC 5506, NGC 7172 and NGC 7582).

We probed the amount of absorption present in these sources by comparing the 2-10 keV X-ray flux with optical and MIR proxies of intrinsic AGN power (F$_{intrinsic}$: the fluxes of the [OIII]$\lambda$ 5007 \AA\ and [OIV]25.89 $\mu$m emission lines and the MIR continuum flux at 13.5 $\mu$m) and by investigating X-ray spectral signatures of obscuration (i.e. Fe K$\alpha$ EW). We compared such obscuration diagnostics with fitted column densities and explored the implications of these diagnostics on the AGN geometry.  We also investigated whether a connection exists between the column density of the obscuring medium and host galaxy characteristics. Our results are summarized as follows:

\begin{enumerate}

\item The majority of the combined sample has F$_{2-10keV}$/F$_{intrinsic}$ values an order of magnitude or lower than the mean values for Sy1s. The statistically significant anti-correlation between F$_{2-10keV}$/F$_{intrinsic}$ and Fe K$\alpha$ EW indicates that these lower diagnostic flux ratios are due to obscuration rather than inherent X-ray weakness in Sy2s. Thus a majority of these sources are potentially Compton-thick, consistent with the results of previous studies (e.g. Risaliti 1999).

\item A wide range of obscuration diagnostic values are present, indicating a continuum of column densities and/or inclination angles, rather than a clear segregation into Compton-thick and Compton-thin sub-populations. Though the diagnostics do generally point to the same sources as likely heavily absorbed, disagreement does exist for a handful of Sy2s. Such a discrepancy is to be expected based on the various biases affecting the observed intrinsic flux proxies and the inherent spread in such isotropic flux indicators (e.g. see \citet{me2}). Hence, nominal Compton-thick boundaries should be considered approximate.

\item Though recent work (\citet{ga10}, \citet{Trouille}) shows the luminosity functions for X-ray selected and [OIII]-selected AGN to be consistent, the various selection techniques favor differ classes of objects. Heavily obscured sources, present in optically selected samples, are missing from 2-10 keV X-ray samples. Sample selection based on intrinsic flux proxies are therefore necessary to include the Compton-thick population, especially since highly absorbed sources constitute the majority of our homogeneously selected samples.

\item Though fitted column densities generally tend to trace the absorption implied by obscuration diagnostics, several glaring inconsistencies are present. Such discrepancies are most extreme when the hard X-ray spectrum is best fit by a single absorbed power law, implying that the simple geometry of a foreground screen attenuating the central source does not recover the intrinsic absorption. This could result from scattering and/or reflected emission dominating over the transmitted continuum, where the fitted column density reflects line of sight absorption rather than the obscuration enshrouding the AGN. Such a result indicates that published N$_H$ distributions derived from single absorbed power law models can be similarly biased, systematically under-representing the intrinsic column density of type 2 AGN. Other diagnostics are therefore crucial in checking the validity of fitted column densities.

\item The X-ray variable Sy2s populate the less obscured range of the flux ratio obscuration diagnostics. Two of these sources (NGC 4388 and NGC 7582) do show evidence of switching to a reflection-dominated state, as indicated by the change in the Fe K$\alpha$ EWs. As Piconcelli et al. (2007) suggest, this change could reflect an inhomogeneous thick absorber covering the central source, with a thin absorber attenuating both the reflected and transmitted emission. The other two variable sources (NGC 5506 and NGC 7172) show signs of Compton-thin absorption, suggesting that the central source is viewed directly. 

\item We do not find compelling evidence that Compton-thick sources have unique AGN properties (intrinsic AGN luminosity, accretion rate and central black hole mass) or star formation activity. Though three out of four obscuration diagnostics are significantly correlated with intrinsic AGN luminosity, the significance is marginal and the relationships display a wide scatter. Evidence linking more obscured sources to more luminous central engines is therefore tenuous at best. No correlation exists between toroidal AGN obscuration and the relative amount of ionization due to the central engine compared to star formation processes (F$_{[OIV]}$/F$_{[NeII]}$, EW of the 17$\mu$m PAH feature, $\alpha_{20-30\mu m}$) and AGN absorption. Though several starburst galaxies do seem to host Compton-thick AGN (e.g. Levenson et al. 2004, 2005), such a relation is not present globally. Hence, we conclude that the gas responsible for star formation processes is not associated with the toroidal obscuration hiding the central engine.

\item Based on simulated high-energy (10-40 keV) spectra using the best-fit modeling of the 2-10 keV spectra, we estimate that the majority of this sample (36 out of 45) will be detected if observed by NuSTAR. The more heavily obscured sources which have not been detected by BAT surveys could likely be identified by NuSTAR as this future mission will probe to lower flux levels ($\sim$2$\times 10^{-14}$ erg s$^{-1}$ cm$^{-2}$ vs. $\sim 3.1 \times 10^{-11}$ erg s$^{-1}$ cm$^{-2}$). These observations would confirm whether the heavily absorbed sources are indeed Compton-thick.

\end{enumerate}

\acknowledgments{}


\begin{figure}[ht]
\subfigure[Black - XMM PN spectrum, red - Chandra spectrum]{\includegraphics[scale=0.33,angle=-90]{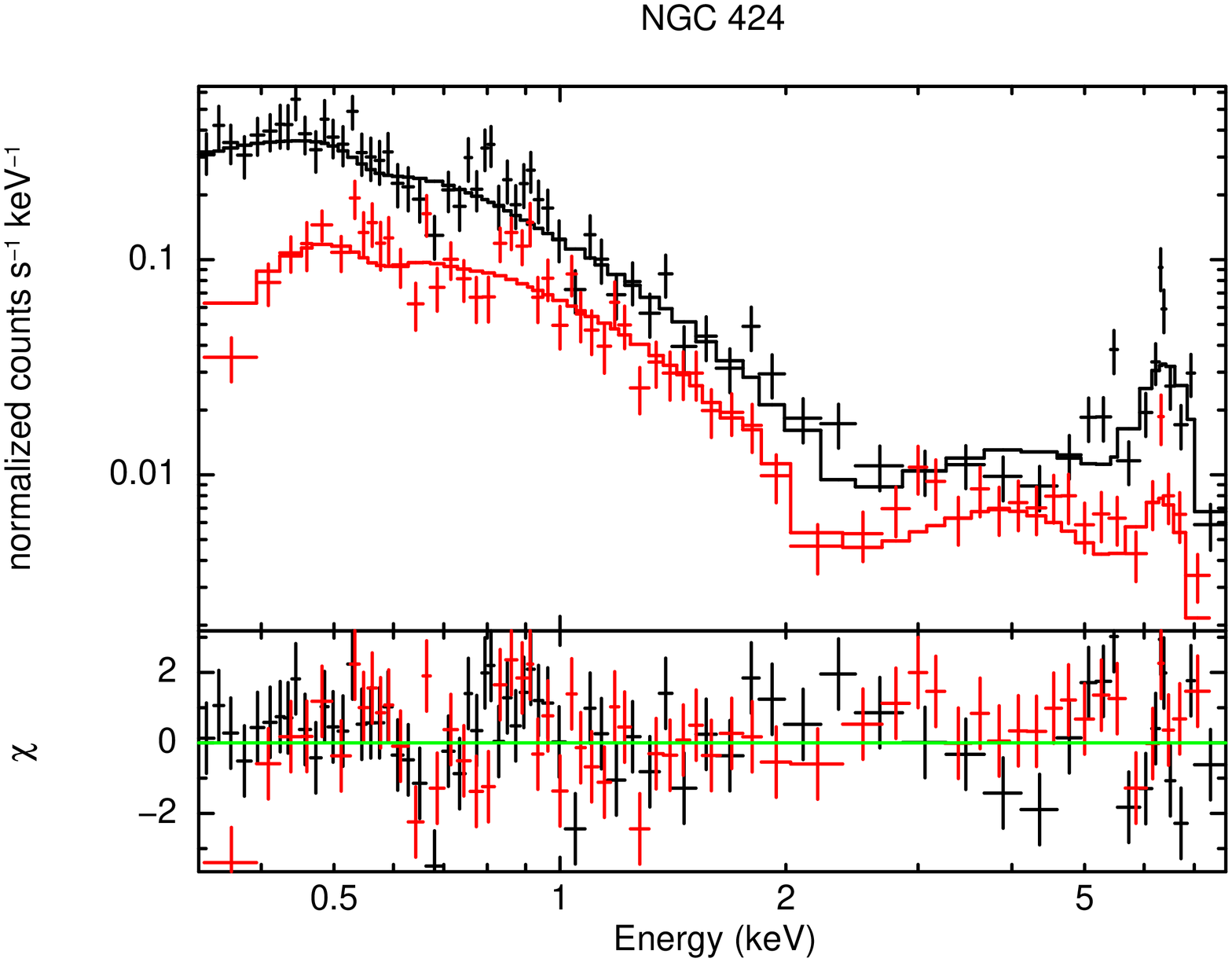}}
\hspace{30pt}
\subfigure[Black - XMM MOS2 spectrum from Jul 29, 2000 observation, red - XMM MOS2 spectrum from July 30, 2000 observation]{\includegraphics[scale=0.33,angle=-90]{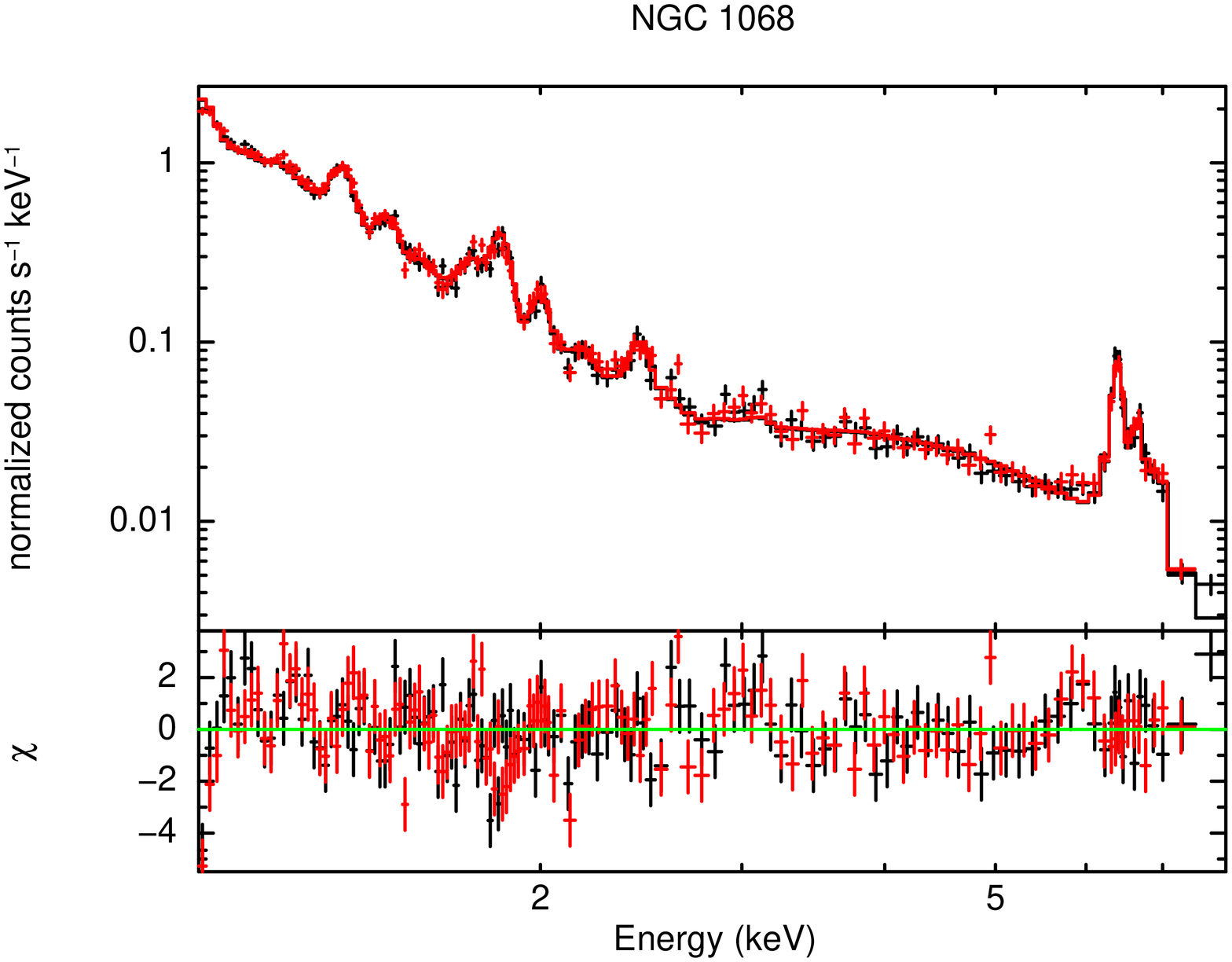}}

\subfigure[XMM PN spectrum.]{\includegraphics[scale=0.33,angle=-90]{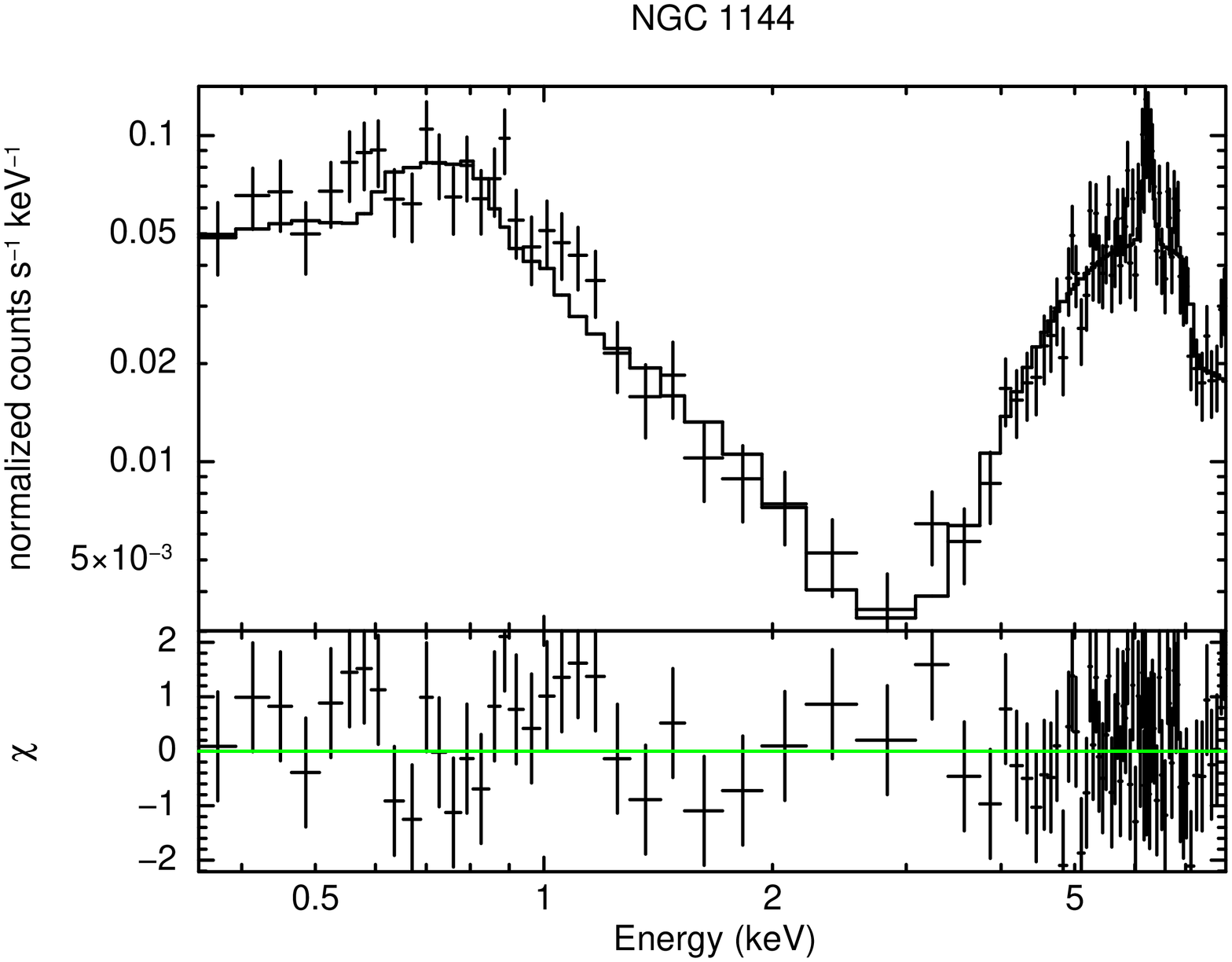}}
\hspace{30pt}
\subfigure[XMM PN spectrum.]{\includegraphics[scale=0.33,angle=-90]{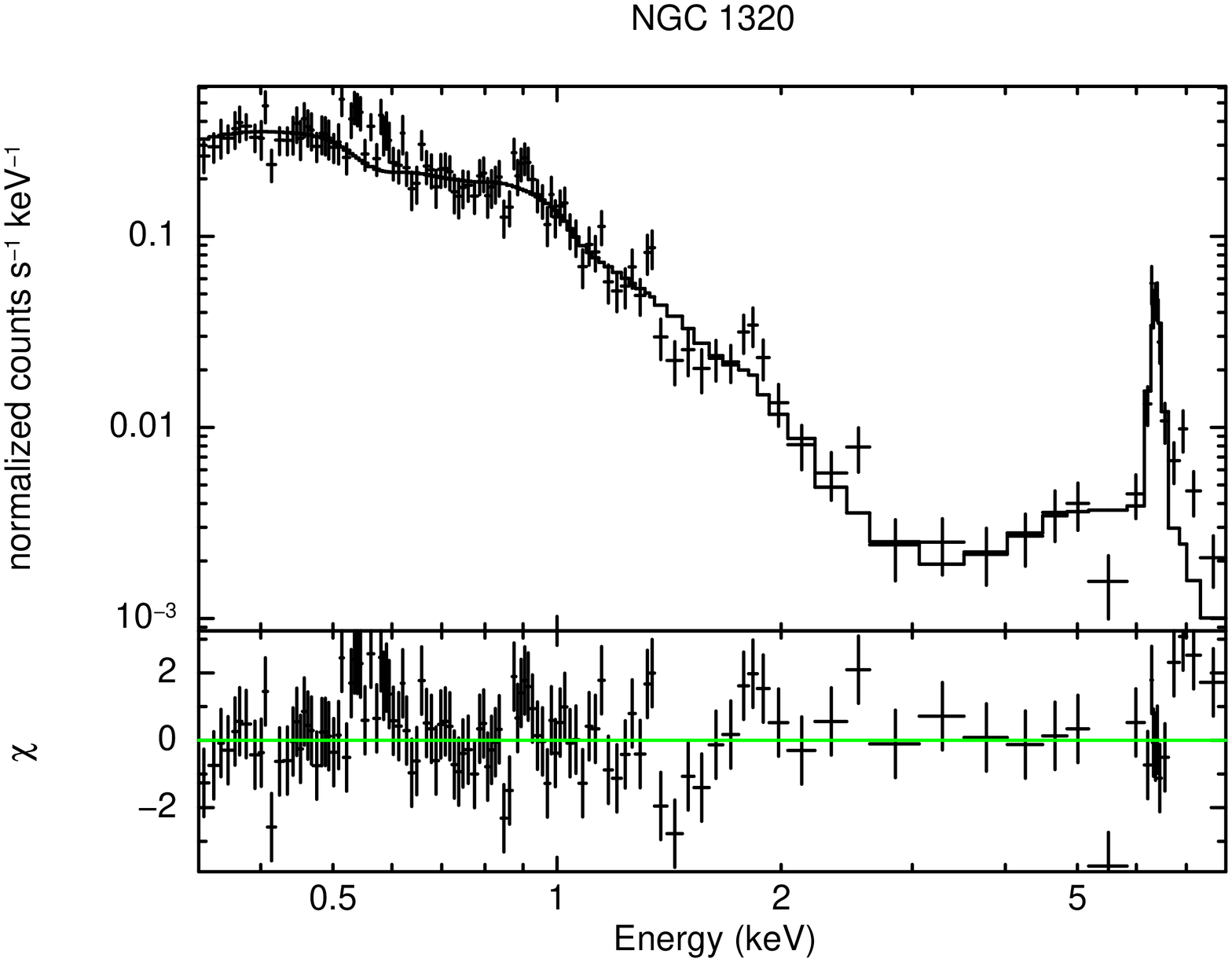}}

\subfigure[Black - XMM PN spectrum, red - Chandra spectrum.]{\includegraphics[scale=0.33,angle=-90]{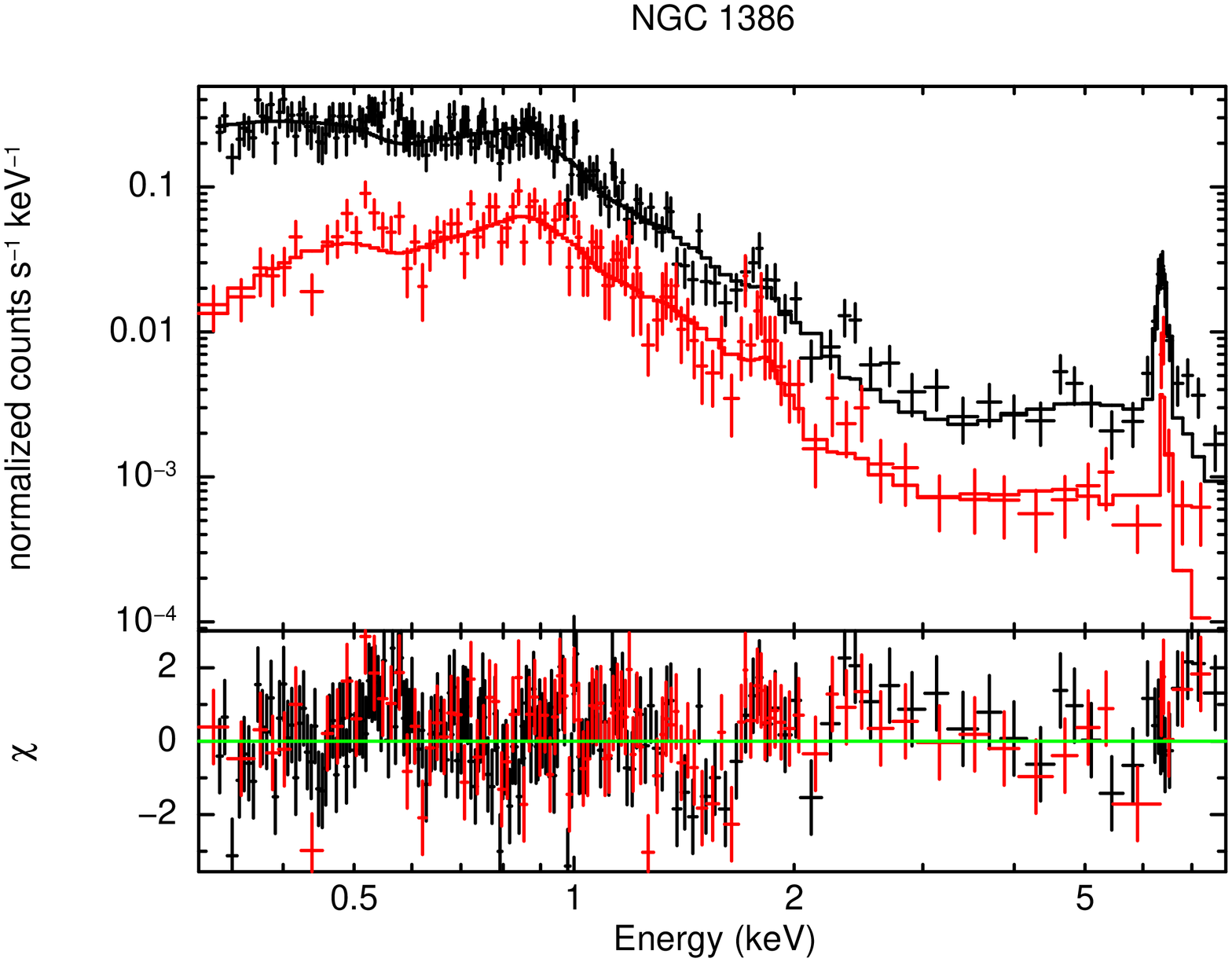}}
\hspace{30pt}
\subfigure[XMM PN spectrum.]{\includegraphics[scale=0.33,angle=-90]{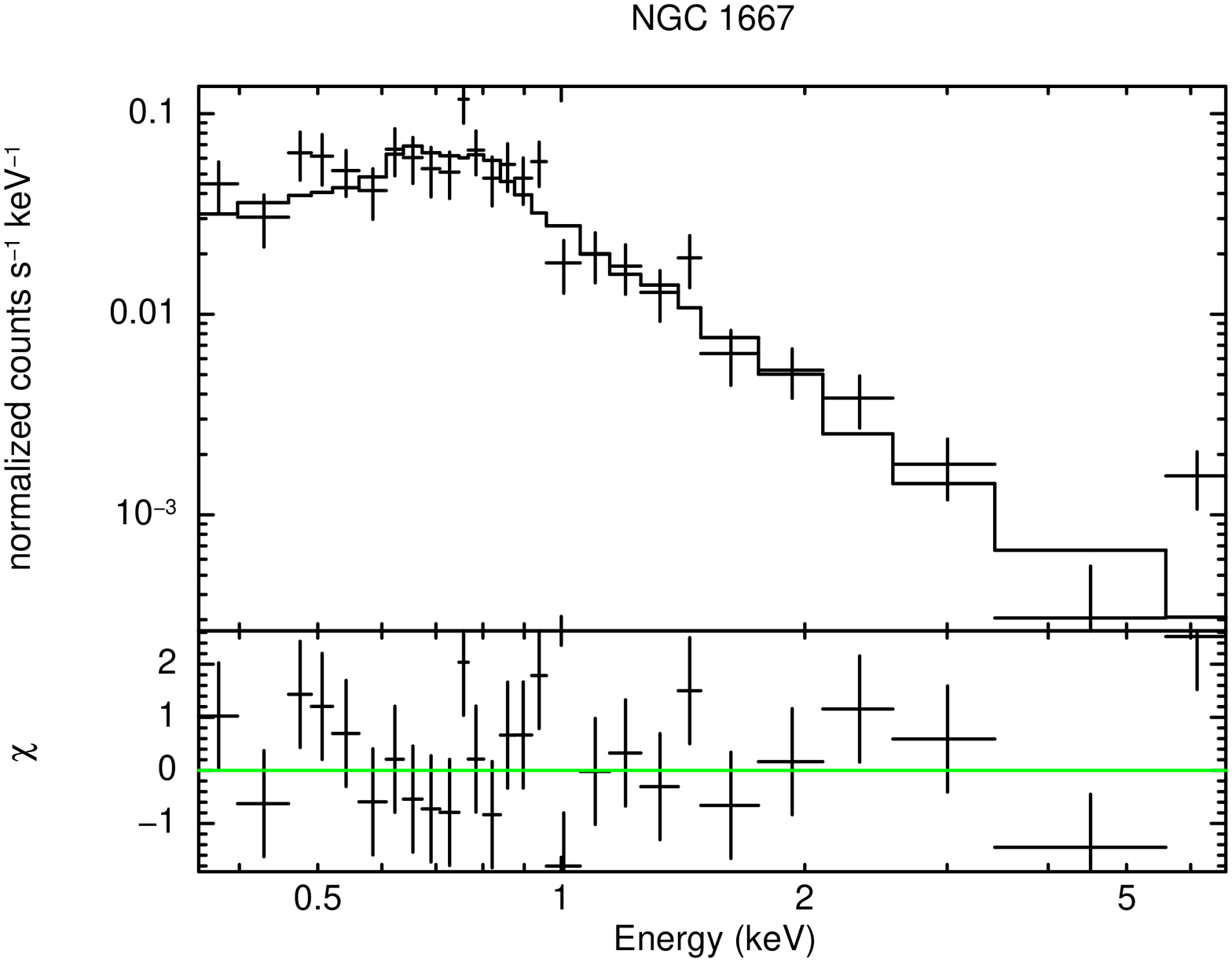}}
\end{figure}

\begin{figure}[ht]
\subfigure[Black - XMM PN spectrum, red - Chandra spectrum from 2001 observation, green - Chandra spectrum from 2002 observation.]{\includegraphics[scale=0.33,angle=-90]{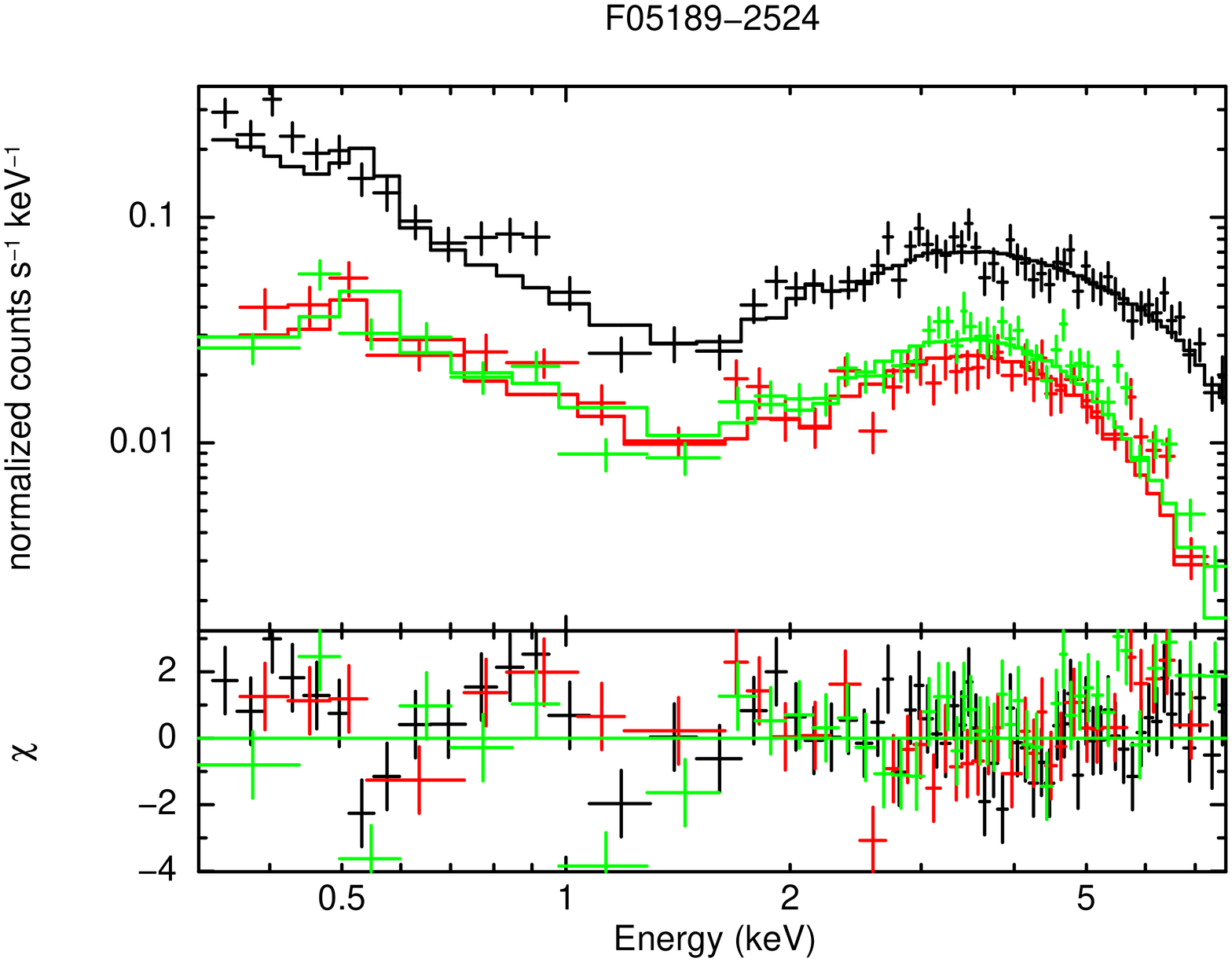}}
\hspace{30pt}
\subfigure[Black - Chandra spectrum.]{\includegraphics[scale=0.33,angle=-90]{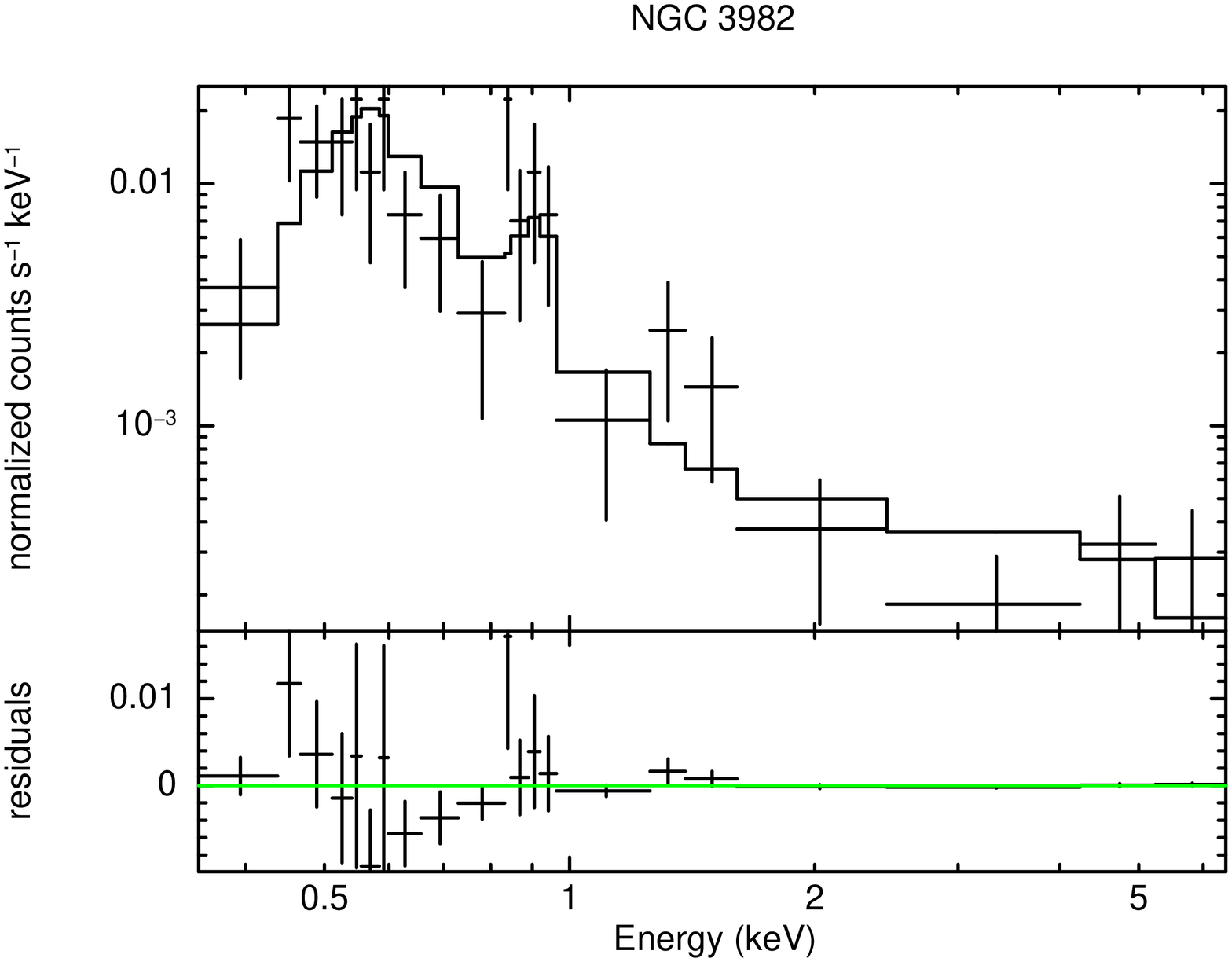}}

\subfigure[Black - XMM PN spectrum from December 2002 observation, red - XMM PN spectrum from July 2002 observation.]{\includegraphics[scale=0.33,angle=-90]{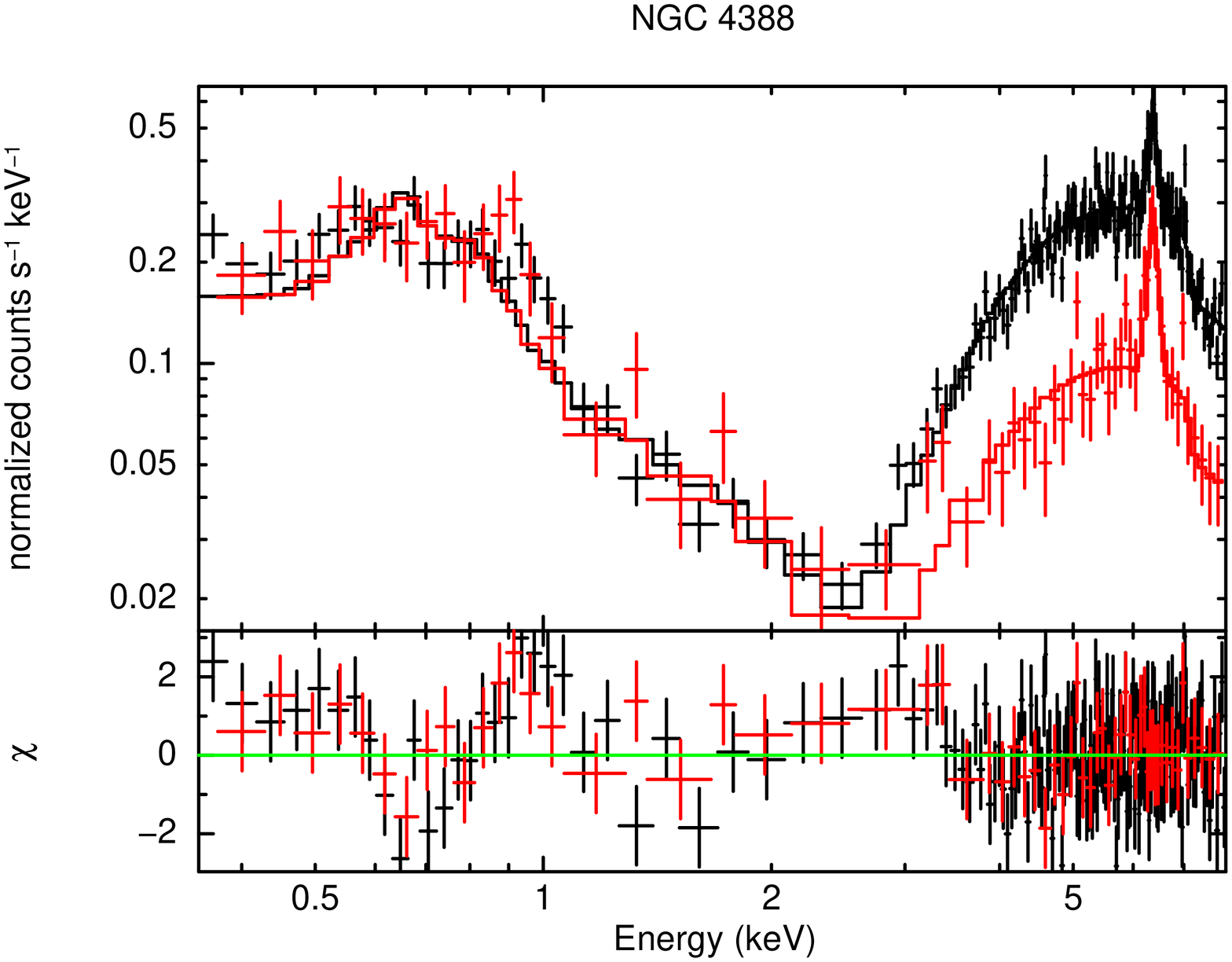}}
\hspace{30pt}
\subfigure[Chandra spectrum.]{\includegraphics[scale=0.33,angle=-90]{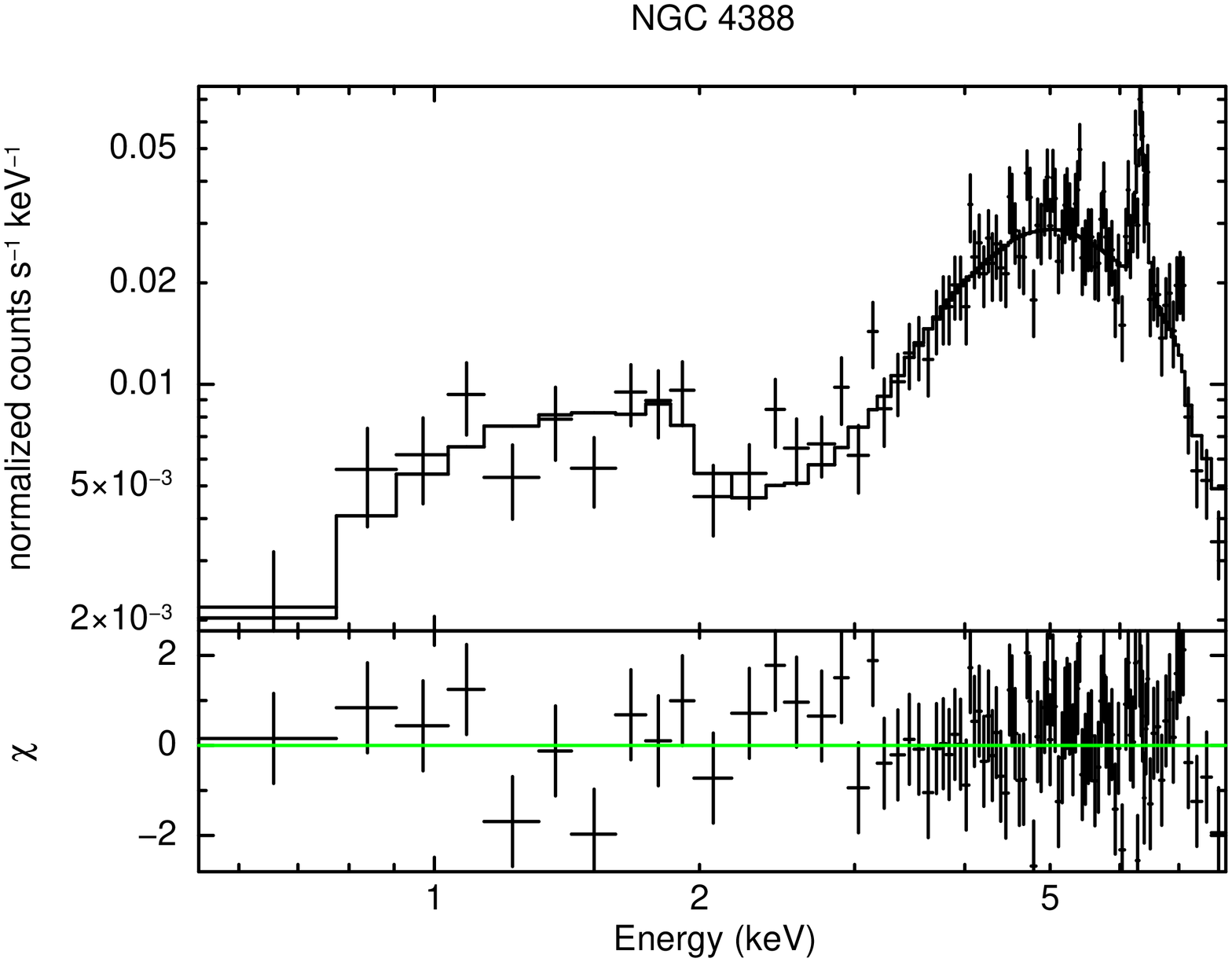}}

\subfigure[Chandra spectrum.]{\includegraphics[scale=0.33,angle=-90]{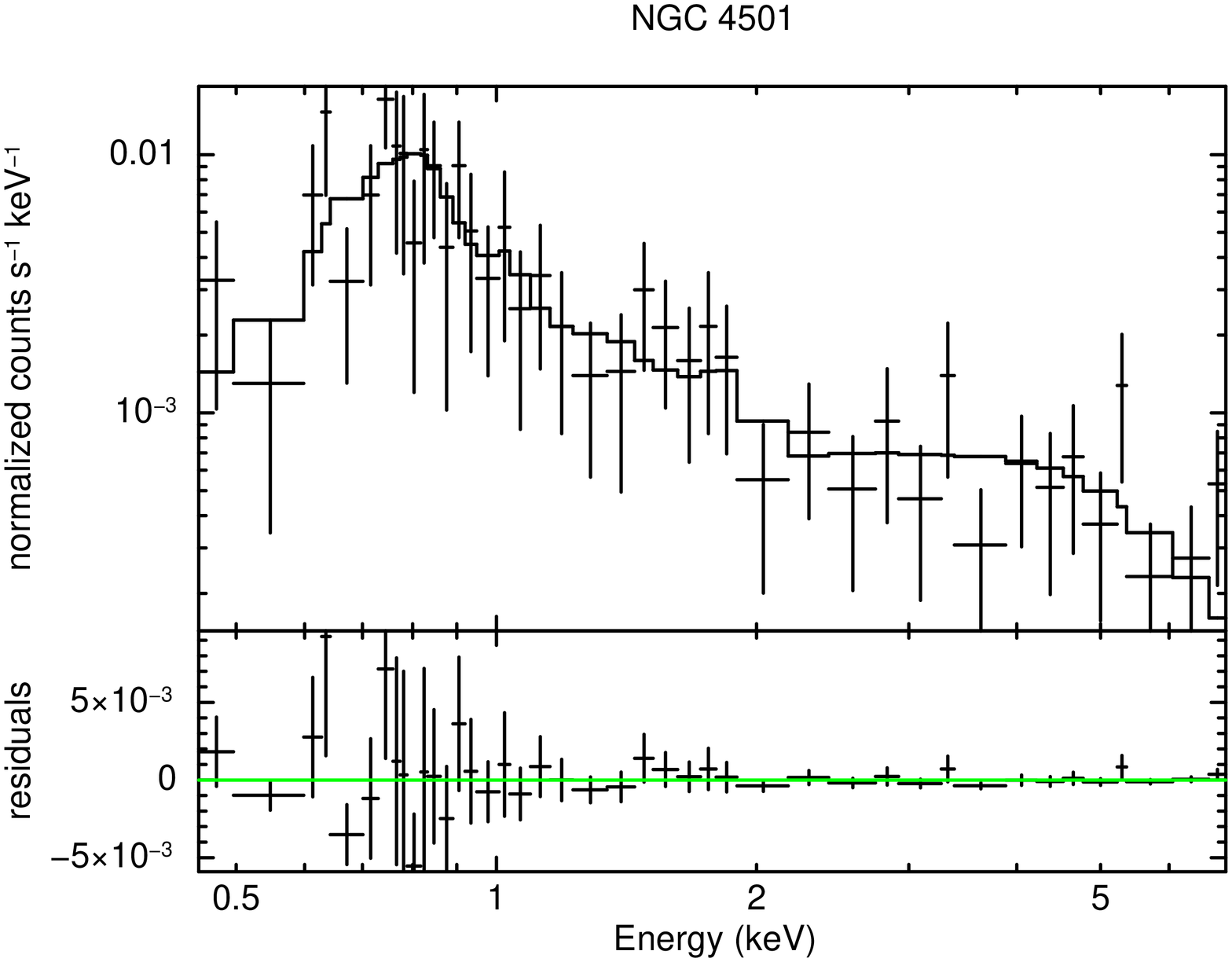}}
\hspace{30pt}
\subfigure[Chandra spectrum.]{\includegraphics[scale=0.33,angle=-90]{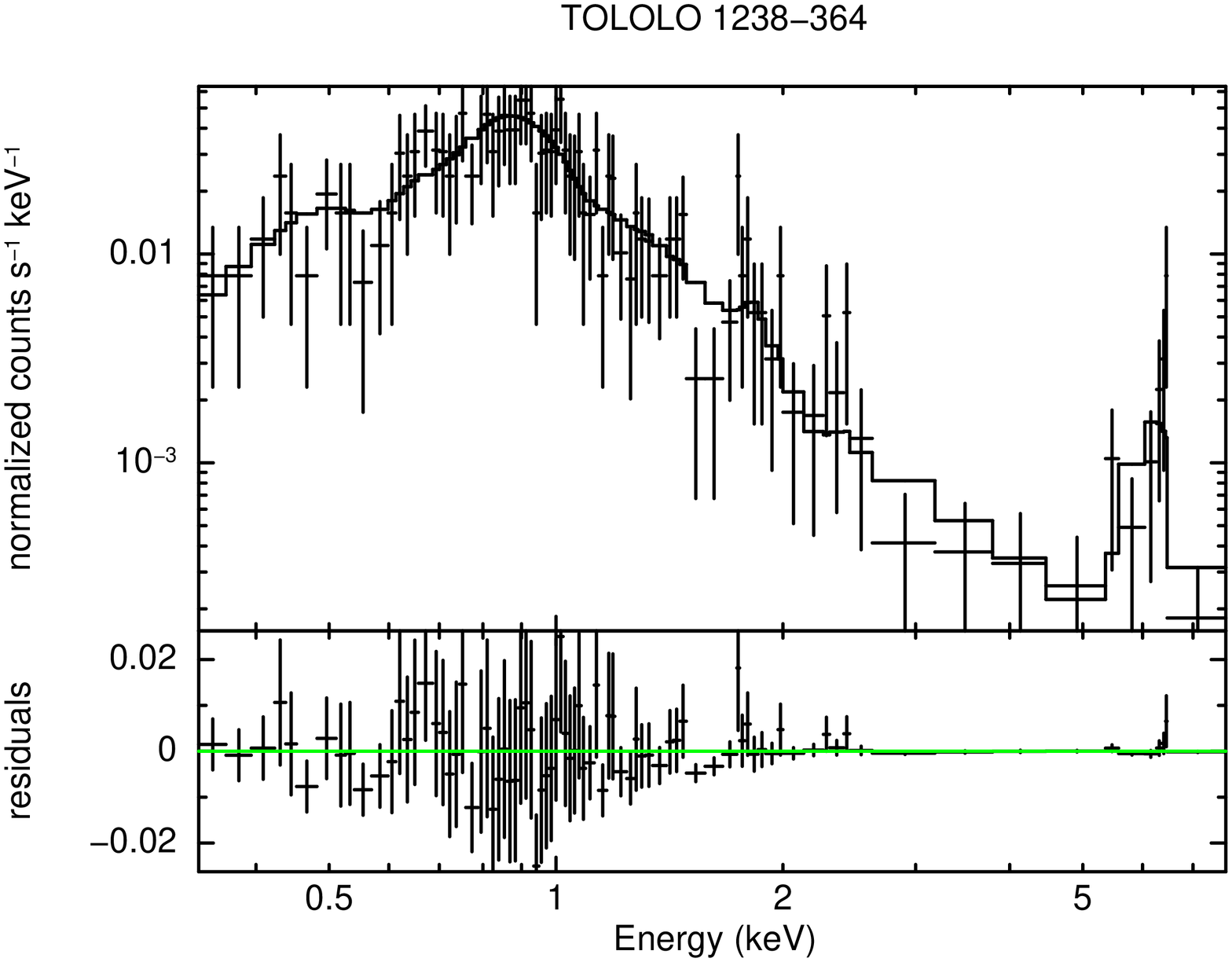}}

\end{figure}

\begin{figure}
\subfigure[Black - XMM PN spectrum from 2001 observation, red - XMM PN spectrum from 2004 observation.]{\includegraphics[scale=0.33,angle=-90]{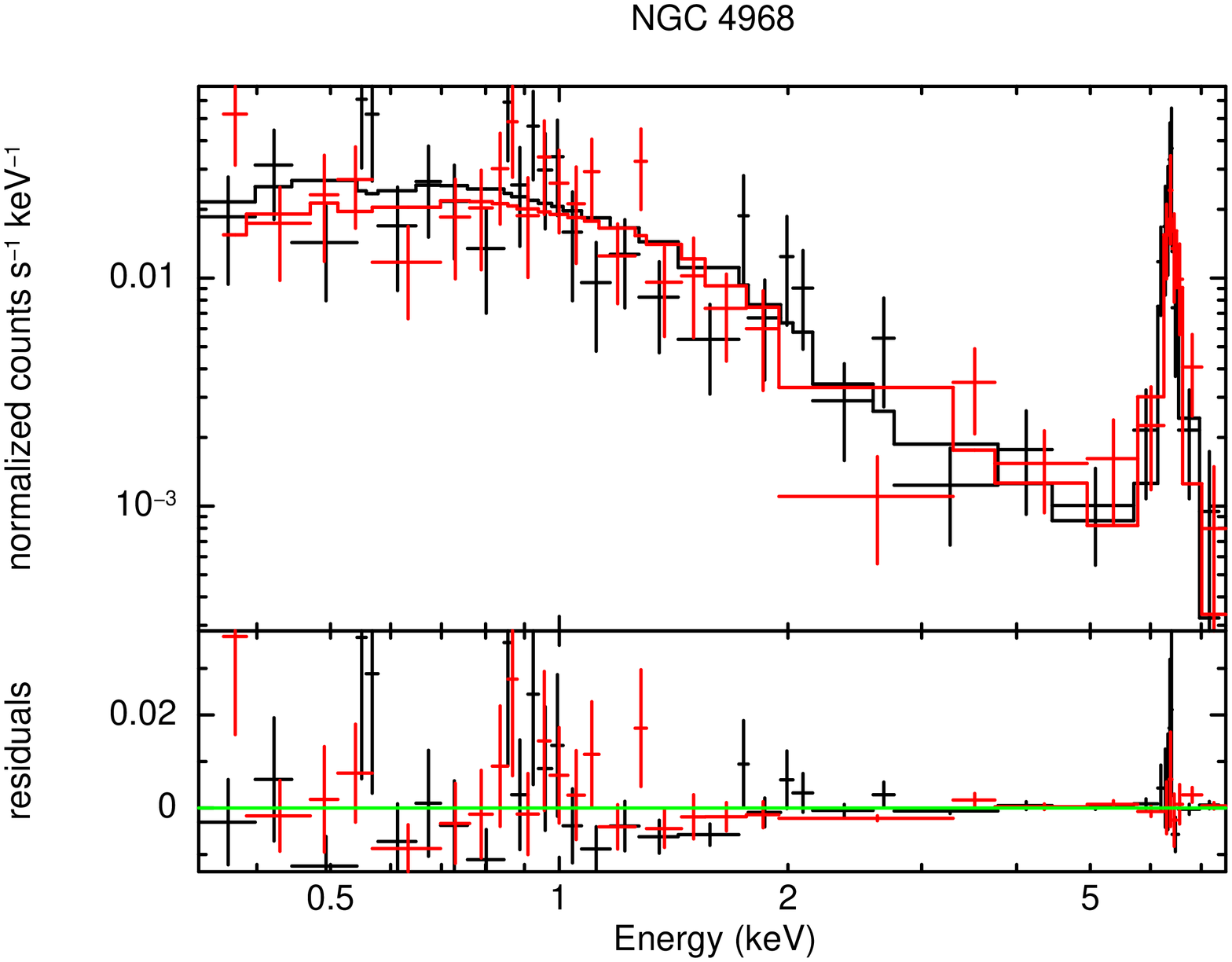}}
\hspace{30pt}
\subfigure[XMM PN spectrum.]{\includegraphics[scale=0.33,angle=-90]{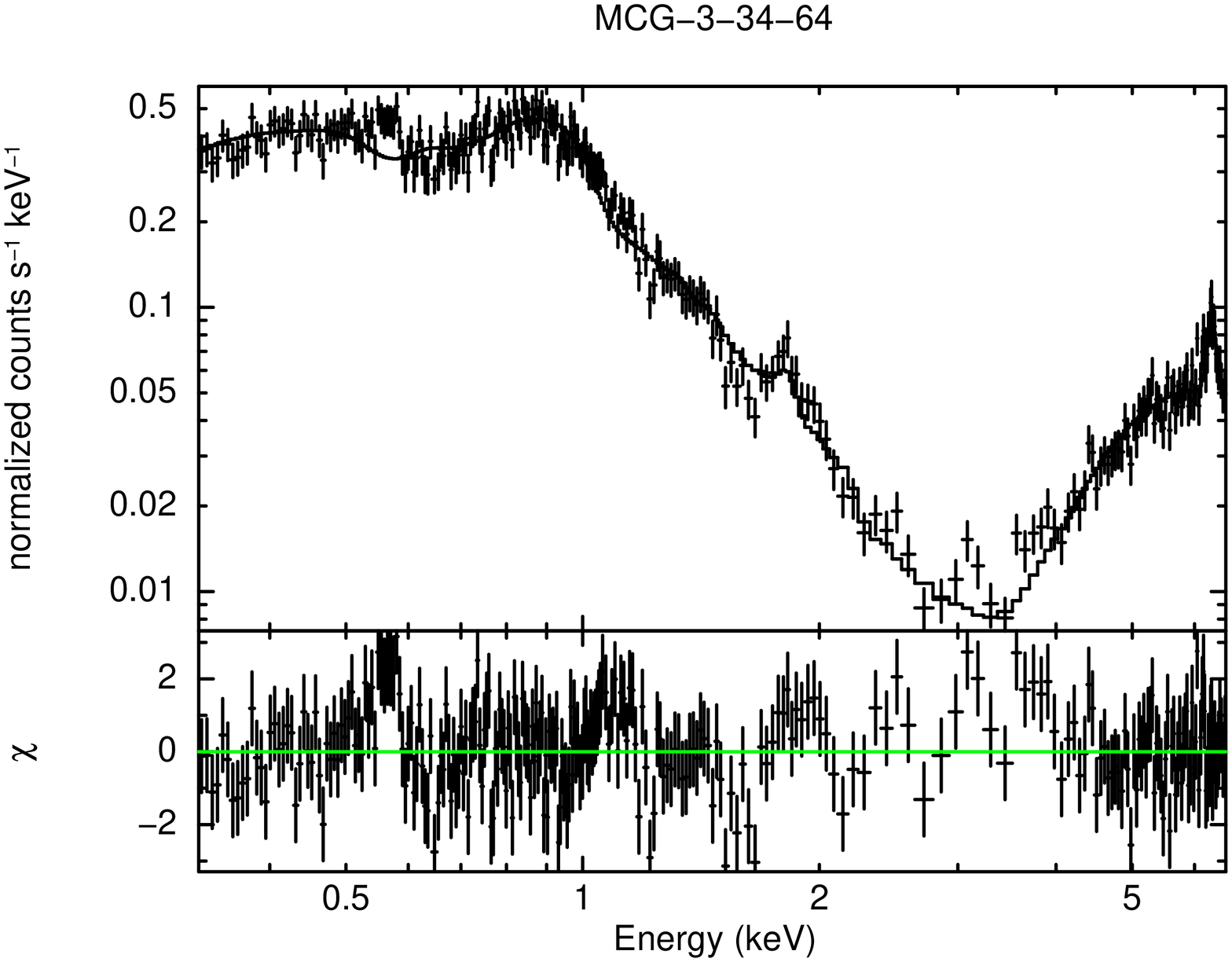}}

\subfigure[Chandra spectrum.]{\includegraphics[scale=0.33,angle=-90]{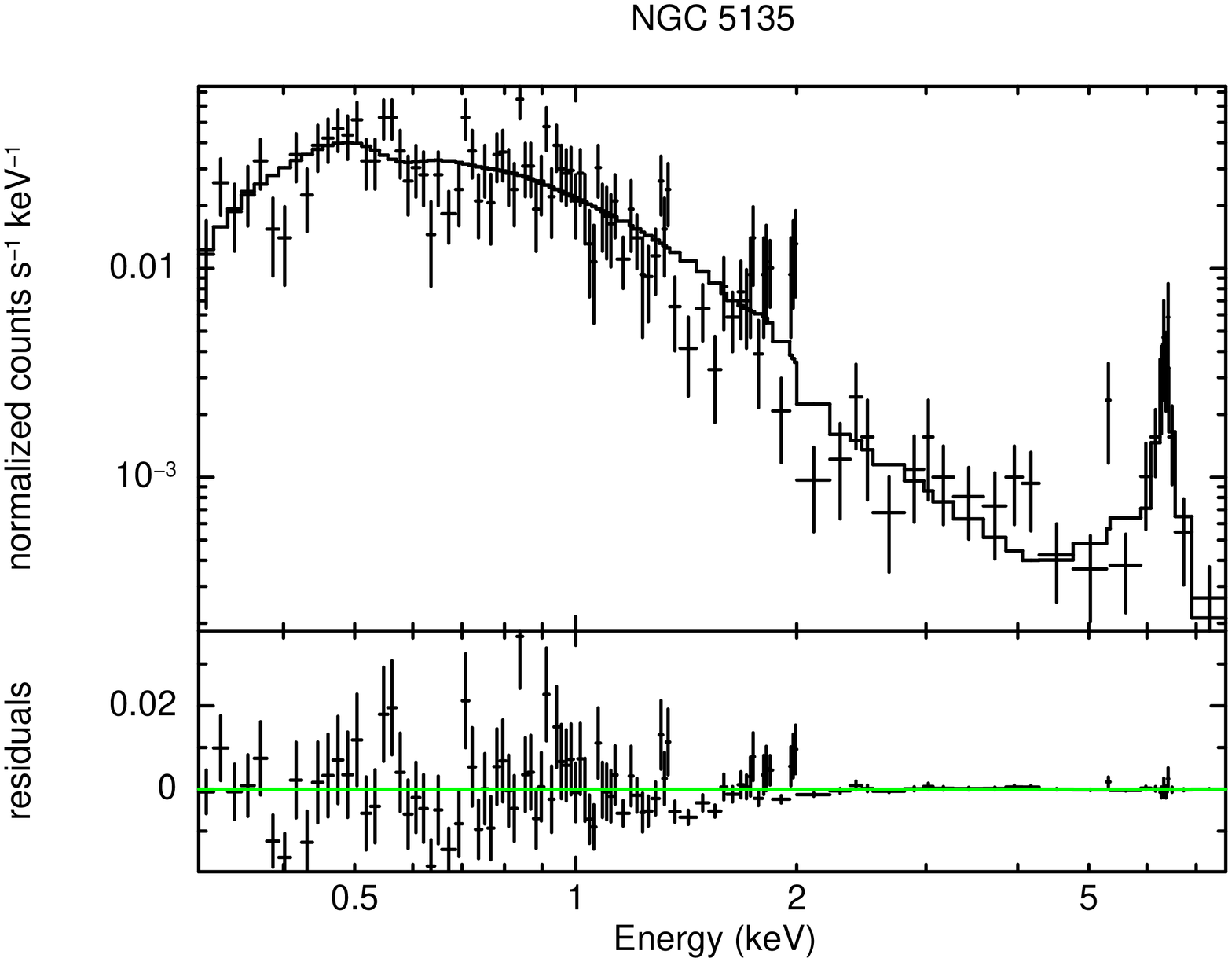}}
\hspace{30pt}
\subfigure[Black -Chandra spectrum from 2000 observation , red - Chandra spectrum from 2001 observation, green - Chandra spectrum from 2003 observation.]{\includegraphics[scale=0.33,angle=-90]{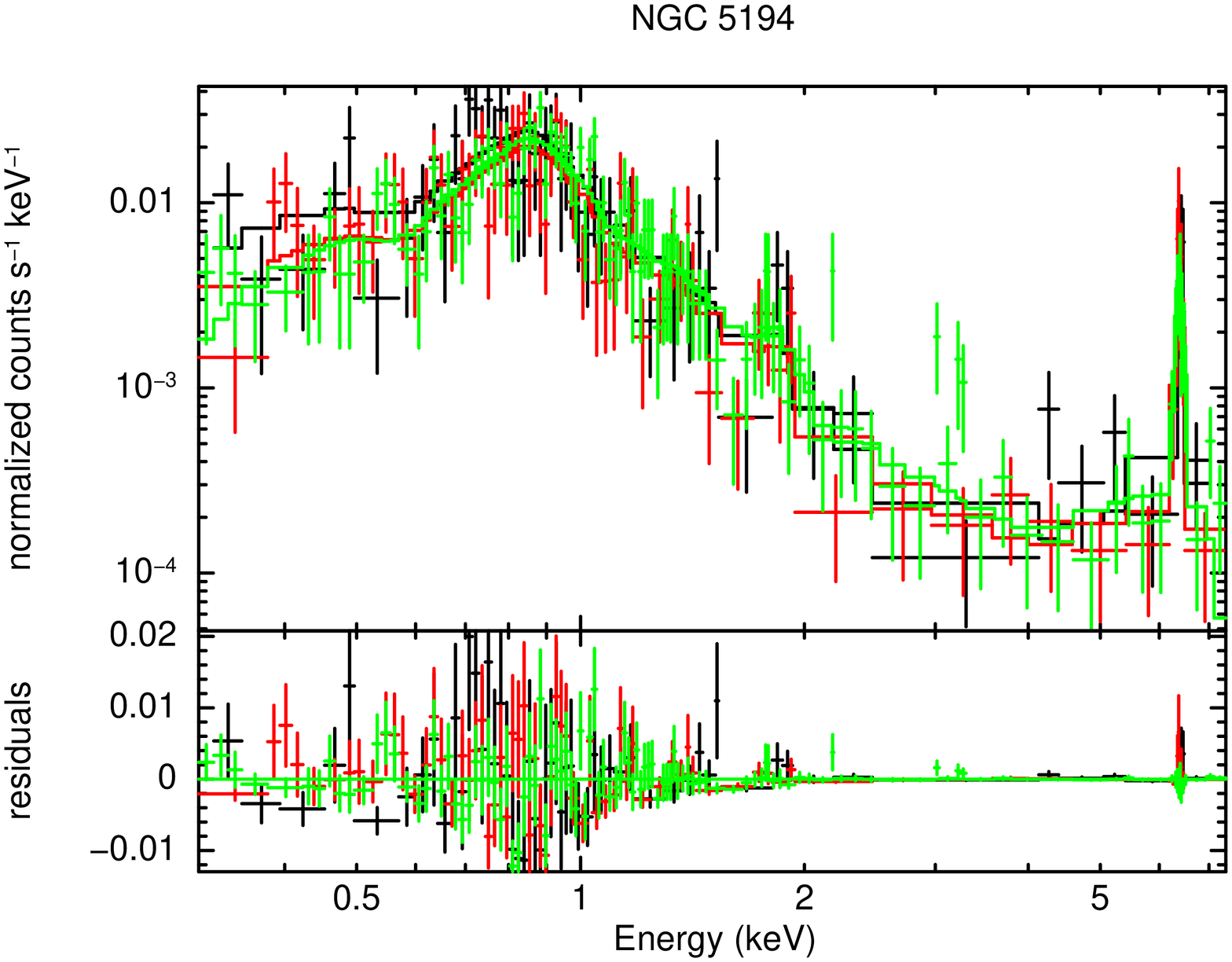}}

\subfigure[Chandra spectrum.]{\includegraphics[scale=0.33,angle=-90]{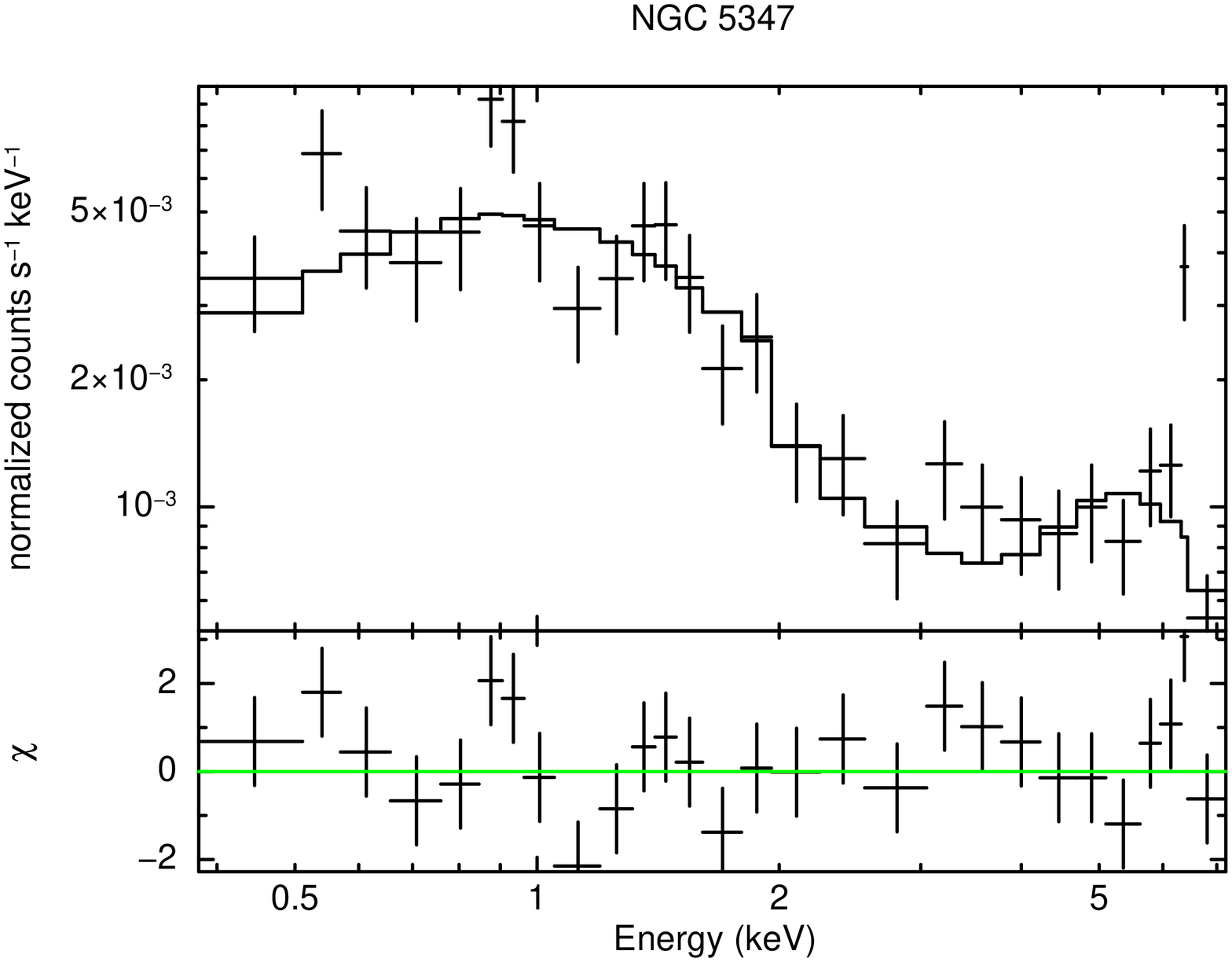}}
\hspace{30pt}
\subfigure[Black - XMM PN spectrum, red - Chandra spectrum.]{\includegraphics[scale=0.33,angle=-90]{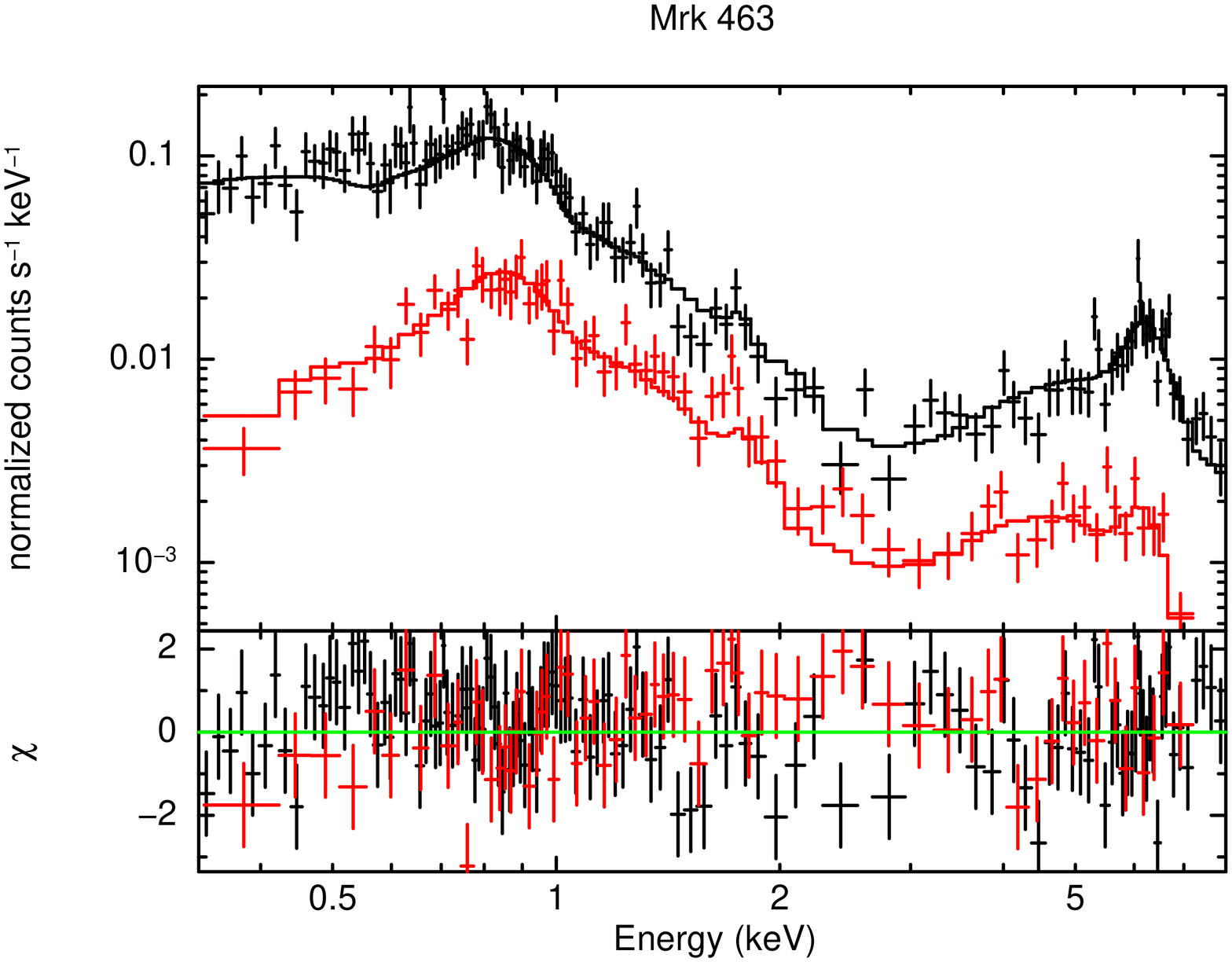}}

\end{figure}

\begin{figure}[ht]
\subfigure[Black - XMM PN spectrum from Jul 11, 2004 observation, red - XMM PN spectrum from Feb 2, 2001 observation, green - XMM PN spectrum from Jul 14, 2004 observation, blue - XMM PN spectrum from Jul 22, 2004 observation.]{\includegraphics[scale=0.33,angle=-90]{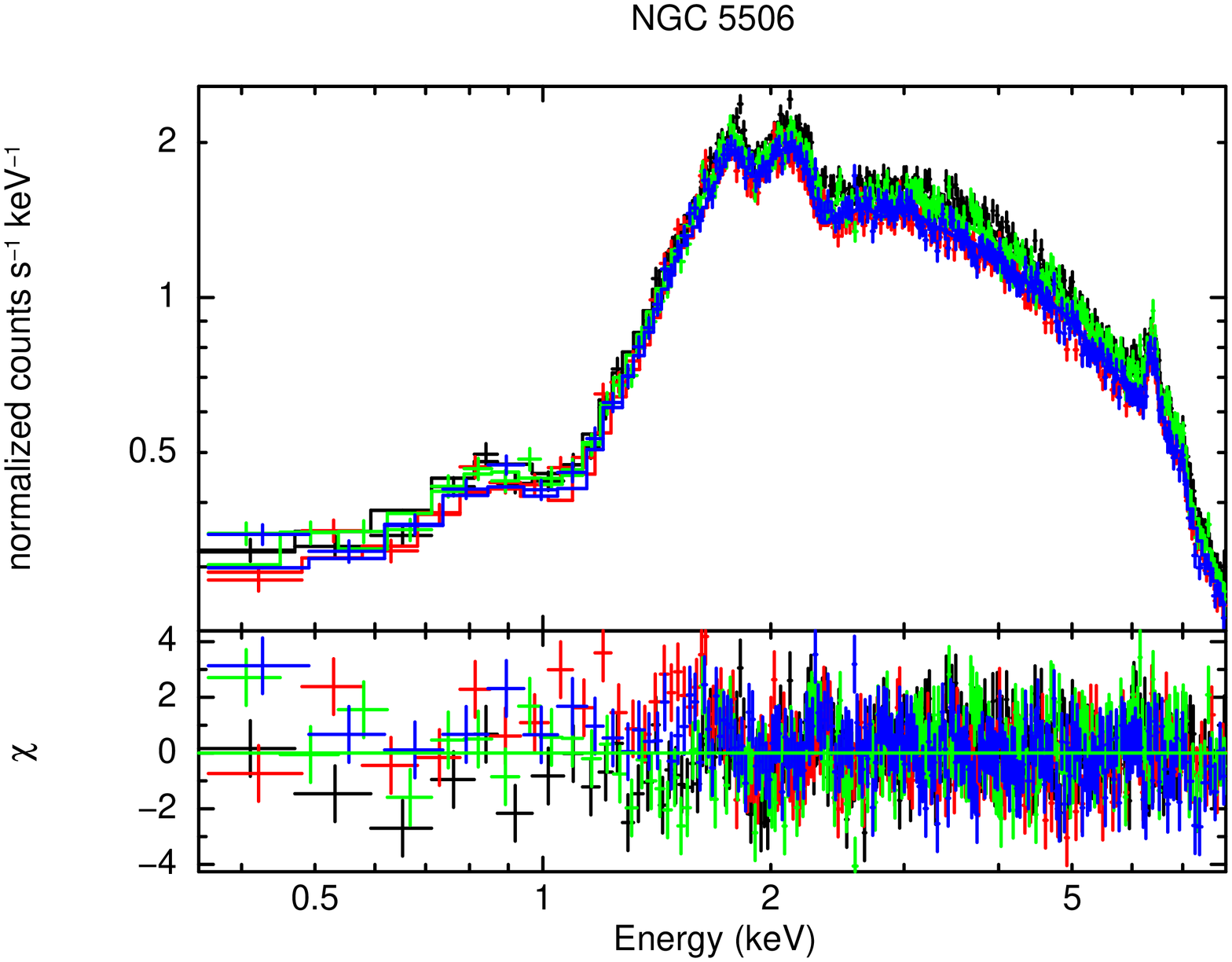}}
\hspace{30pt}
\subfigure[Black - XMM PN spectrum from 2001 observation, red - XMM PN spectrum from Aug 2004 observation, green - XMM PN spectrum from 2008 observation, blue - XMM PN spectrum from 2009 observation.]{\includegraphics[scale=0.33,angle=-90]{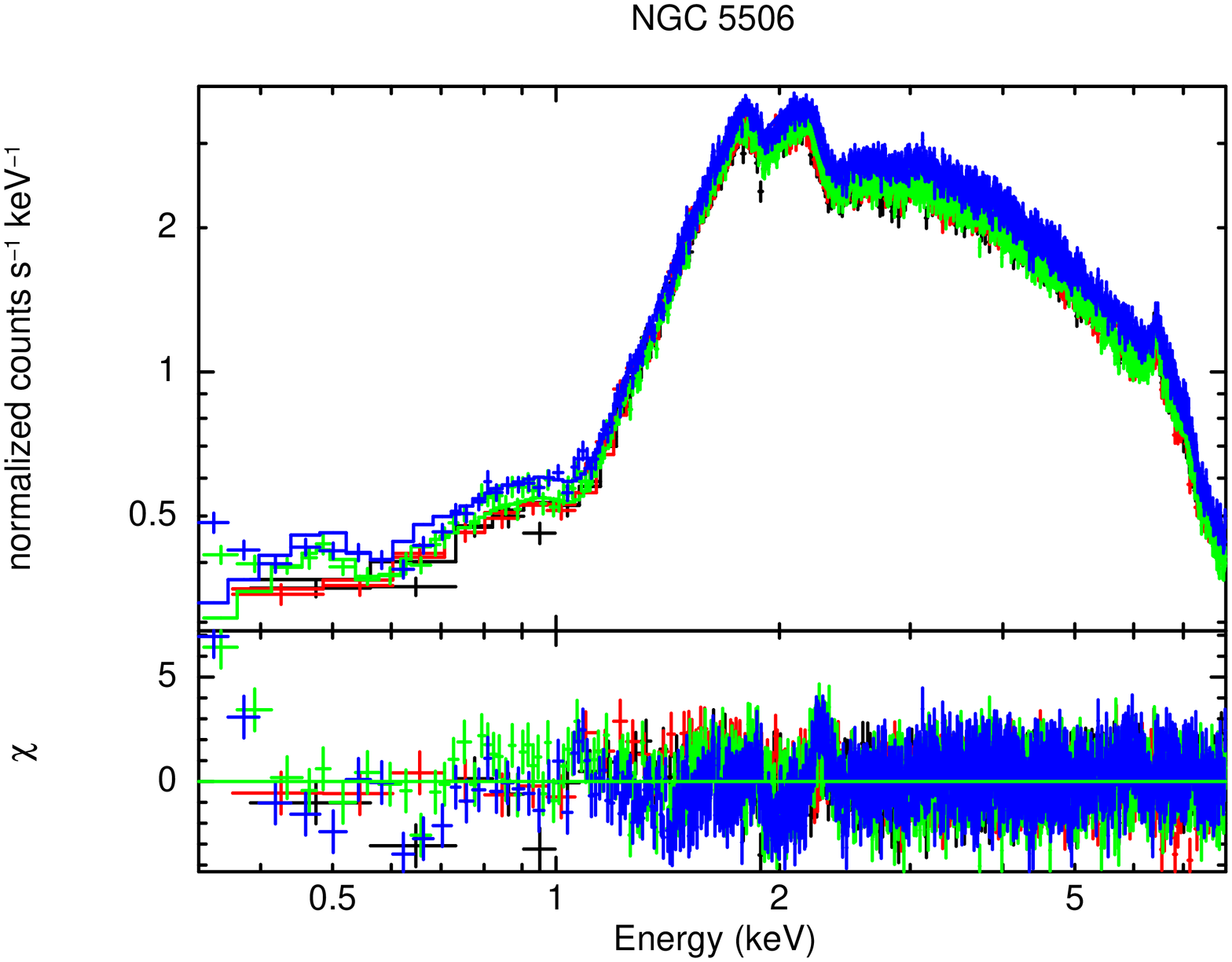}}

\subfigure[Chandra spectrum.]{\includegraphics[scale=0.33,angle=-90]{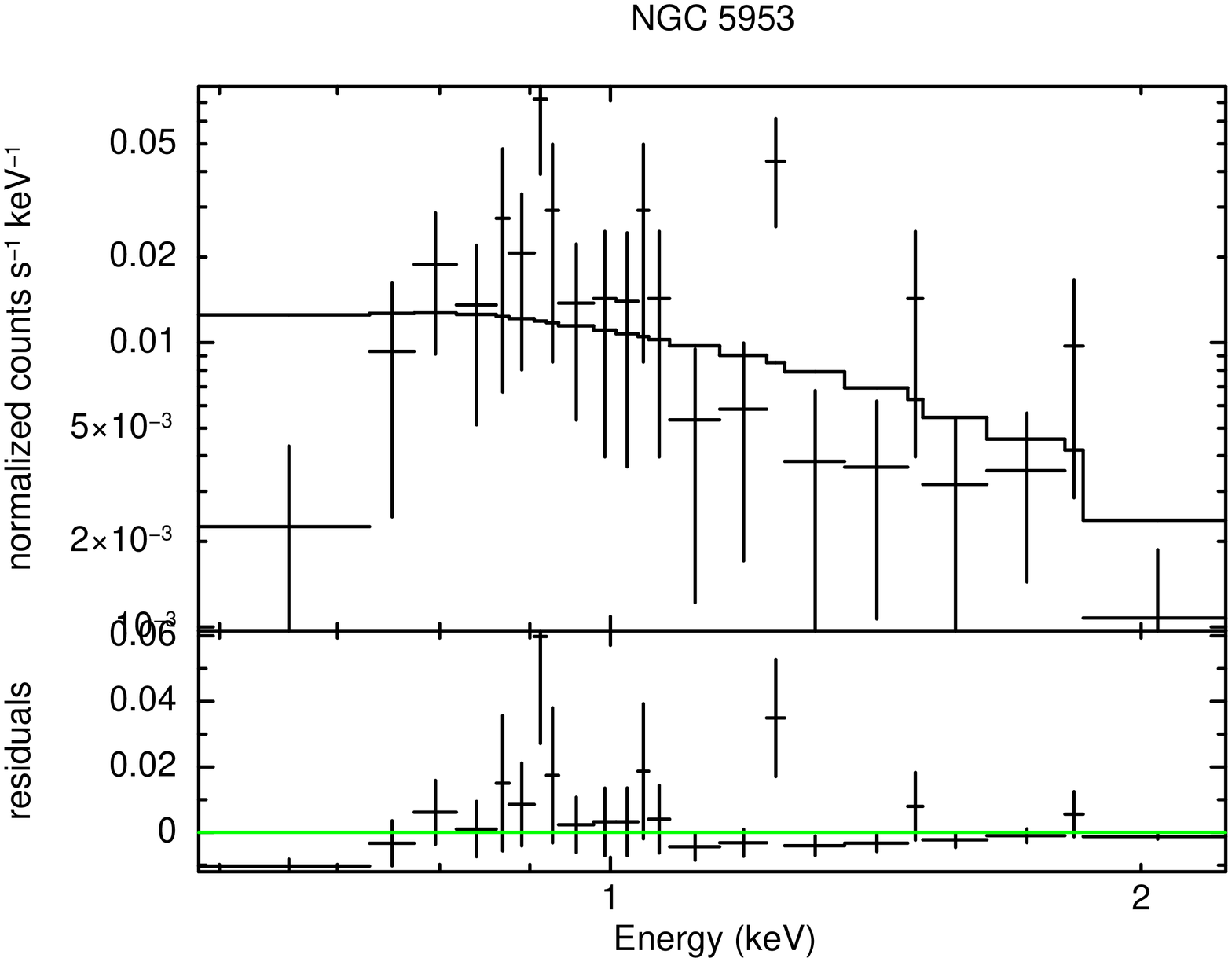}}
\hspace{30pt}
\subfigure[Black - XMM PN spectrum from 2002 observation, red - XMM PN spectrum from 2003 observation, green - XMM MOS1 spectrum from Jan 2005 observation, blue -  Chandra spectrum.]{\includegraphics[scale=0.33,angle=-90]{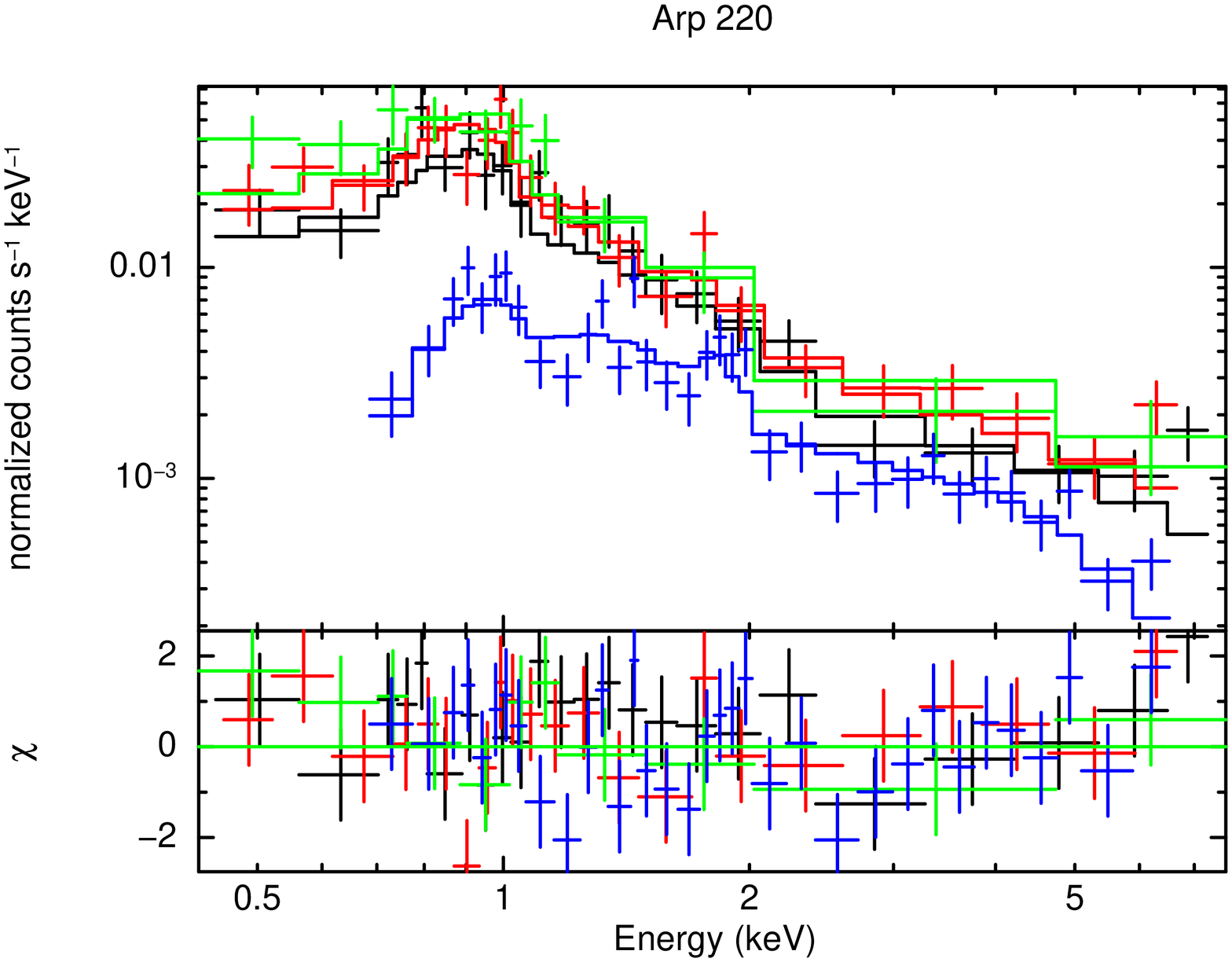}}

\subfigure[XMM PN spectrum.]{\includegraphics[scale=0.33,angle=-90]{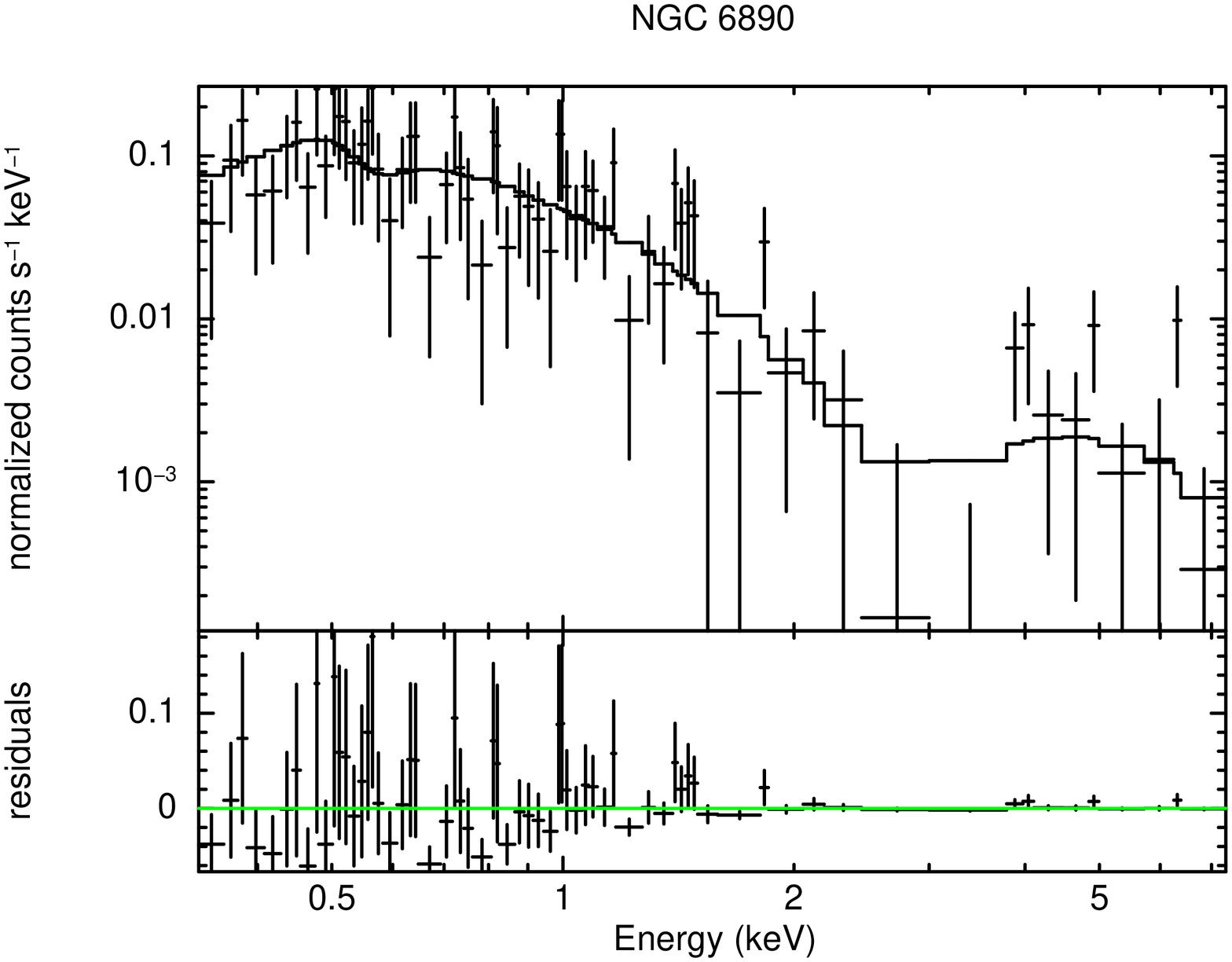}}
\hspace{30pt}
\subfigure[Chandra spectrum.]{\includegraphics[scale=0.33,angle=-90]{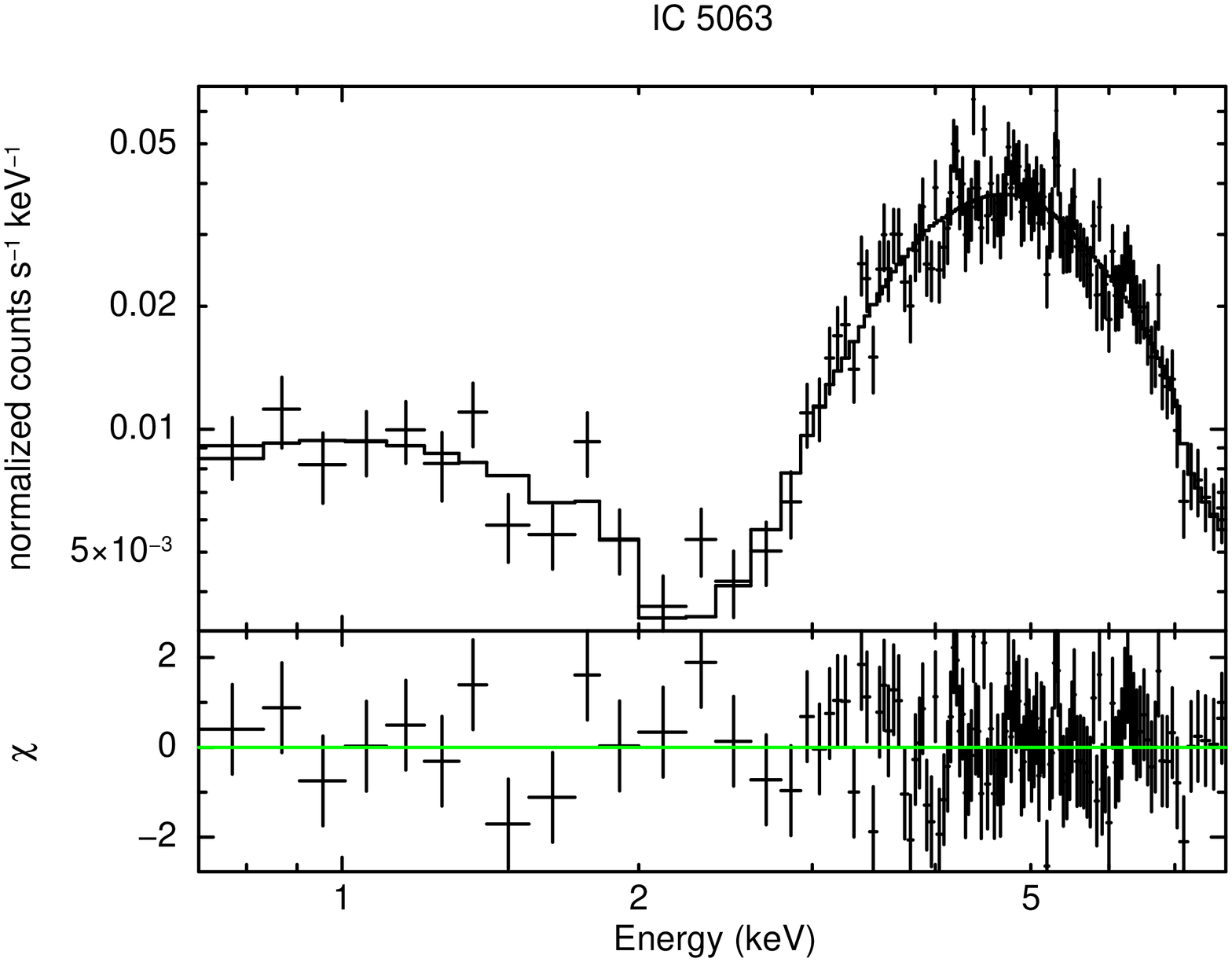}}

\end{figure}

\begin{figure}
\centering
\subfigure[]{\includegraphics[scale=0.30,angle=90]{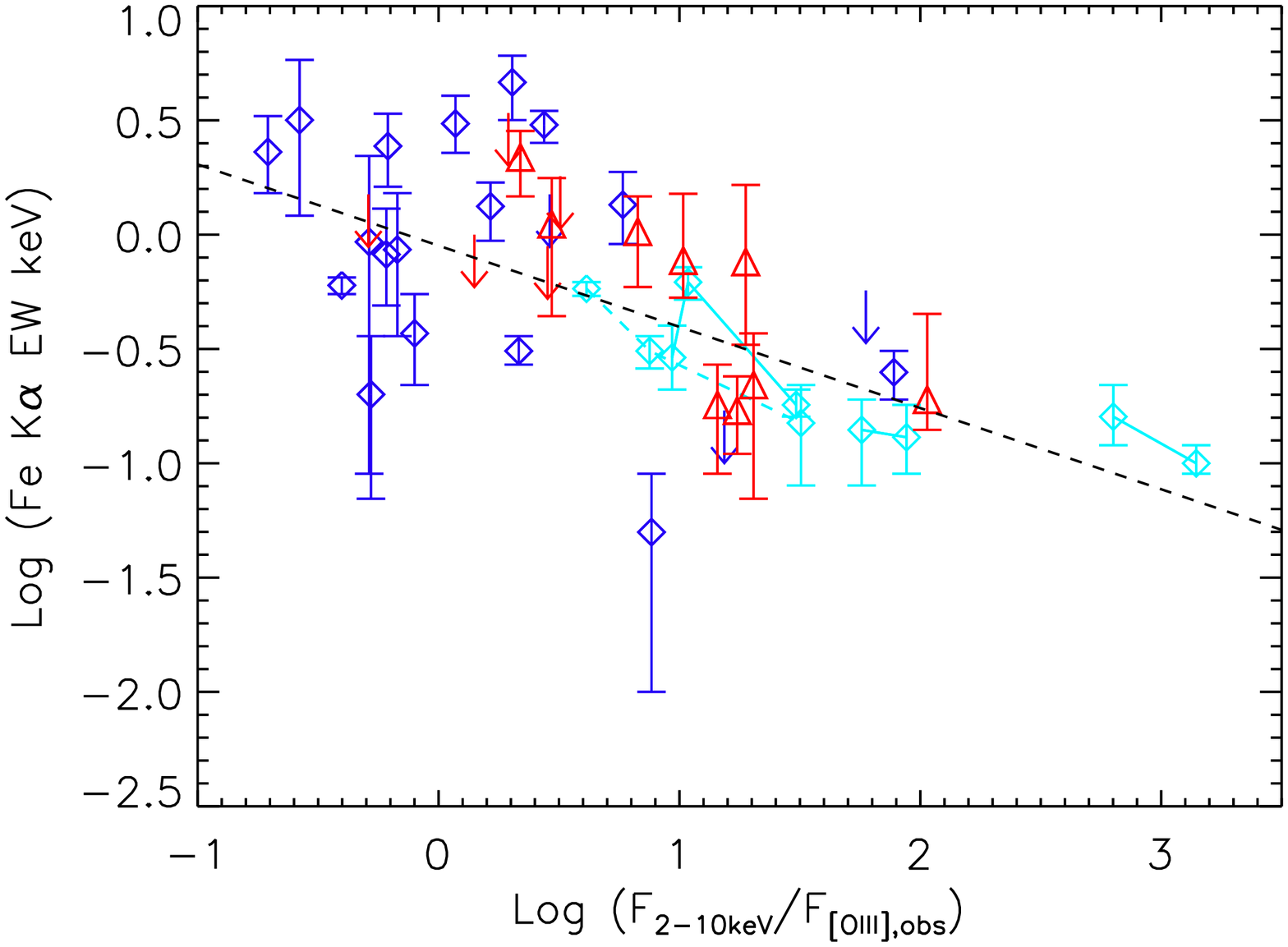}}
\subfigure[]{\includegraphics[scale=0.30,angle=90]{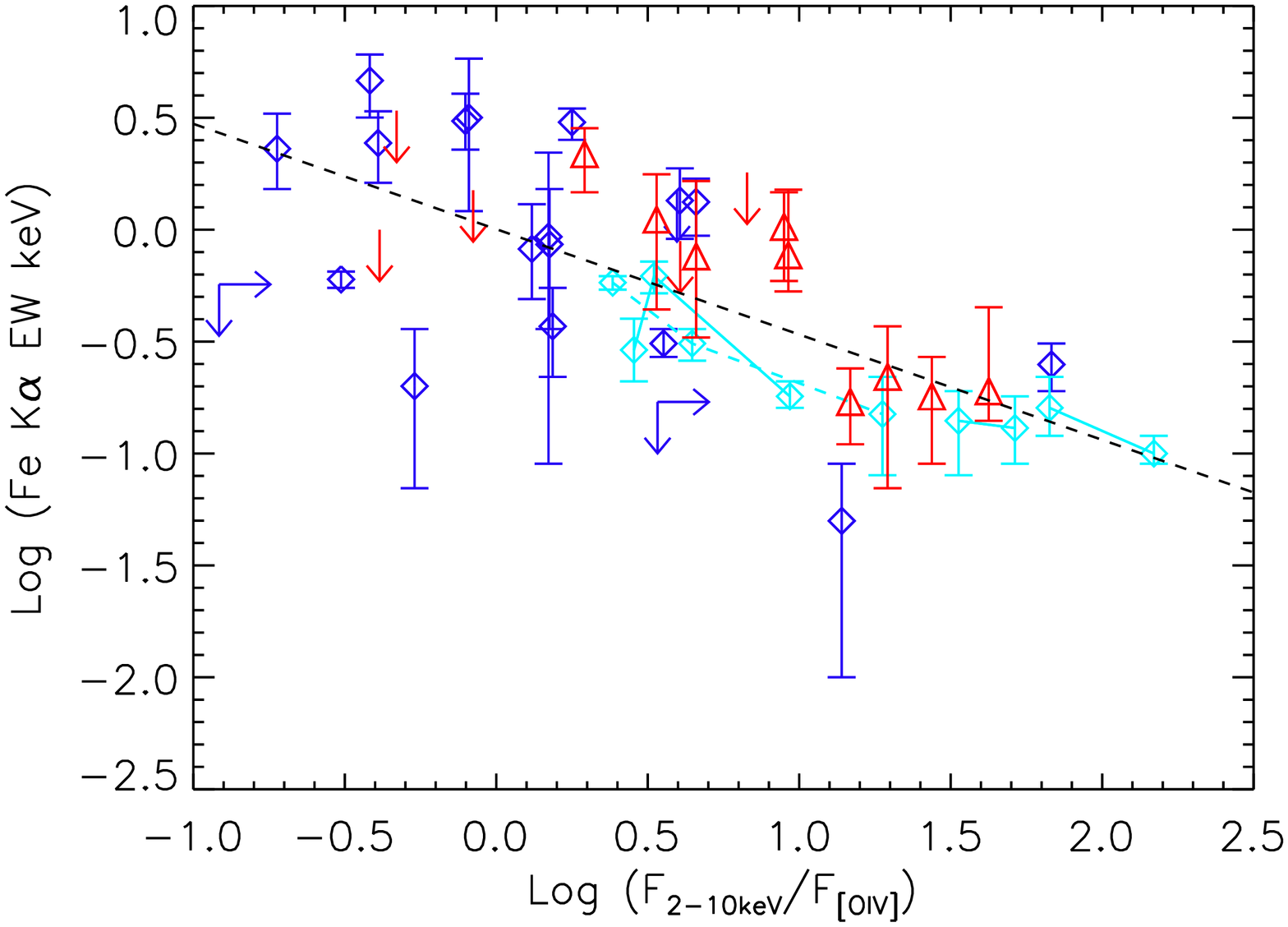}}
\subfigure[]{\includegraphics[scale=0.30,angle=90]{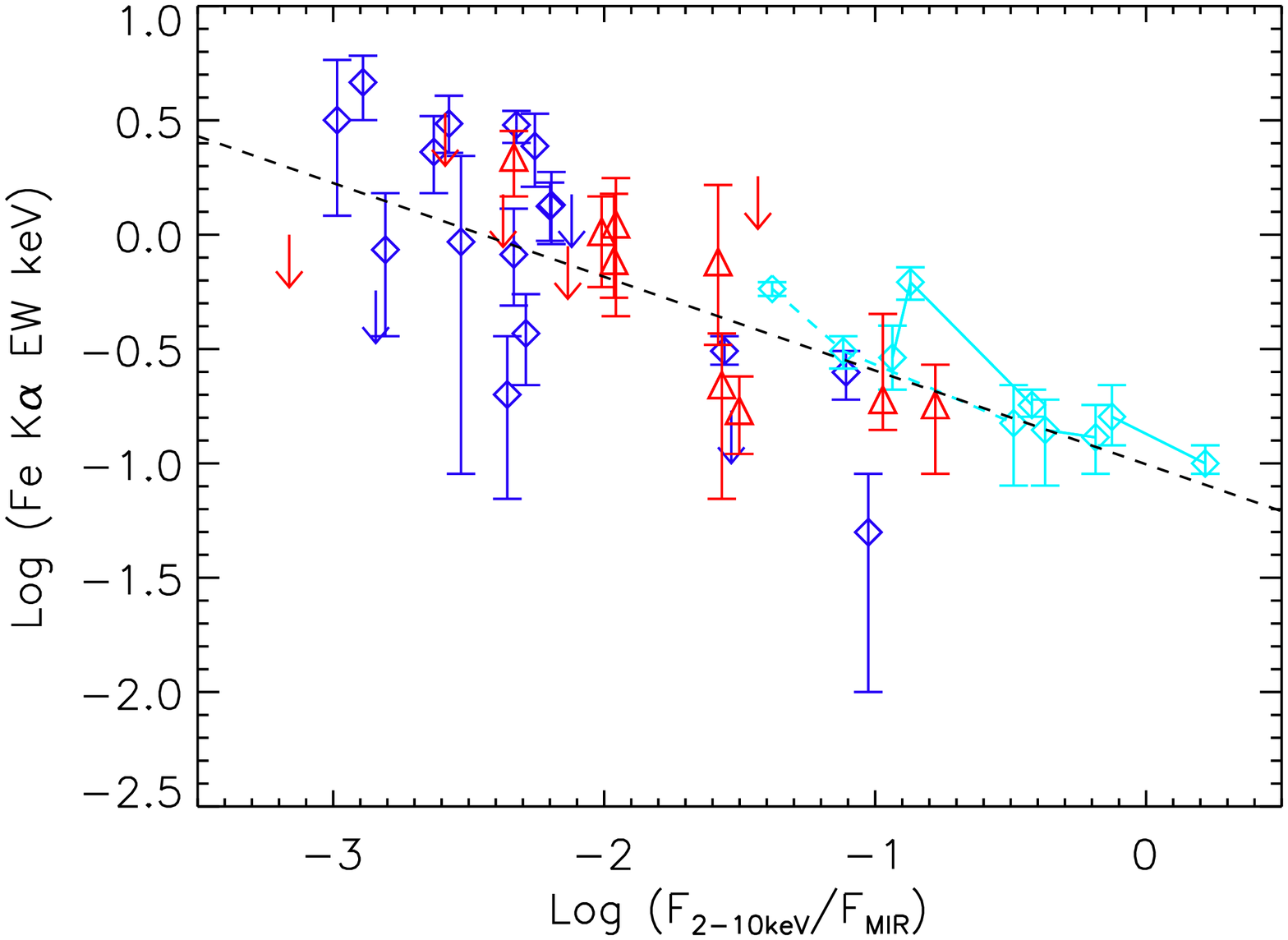}}
\caption{\label{ew_ratios}Fe K$\alpha$ EW as a function of obscuration diagnostic ratios. Color coding is similar to \ref{lx_l_iso}, though a dashed line is used for variable source NGC 7582 to avoid confusion with other variable sources having similar values. The statistically significant anti-correlations among all three relationships (Spearman's $\rho$= -0.647, -0.657, -0.645, respectively, calculated from survival analysis) indicate obscuration is primarily responsible for X-ray attenuation. In b), the two sources with lower limits on F$_{2-10keV}$/F$_{[OIV]}$ are plotted for illustrative purposes and were not included in the survival analysis calculations. The fitted relationships are a) slope = -0.36 $\pm$0.07 dex with $\sigma$=0.37 dex and intercept of -0.05 dex, b) slope = -0.47 $\pm$0.08 dex with $\sigma$=0.34 dex and intercept of 0.002 dex and c) slope = -0.41 $\pm$ 0.06 dex with $\sigma$=0.32 dex and intercept of -1.00 dex.}
\end{figure}


\begin{figure}
\epsscale{1.2}
\plottwo{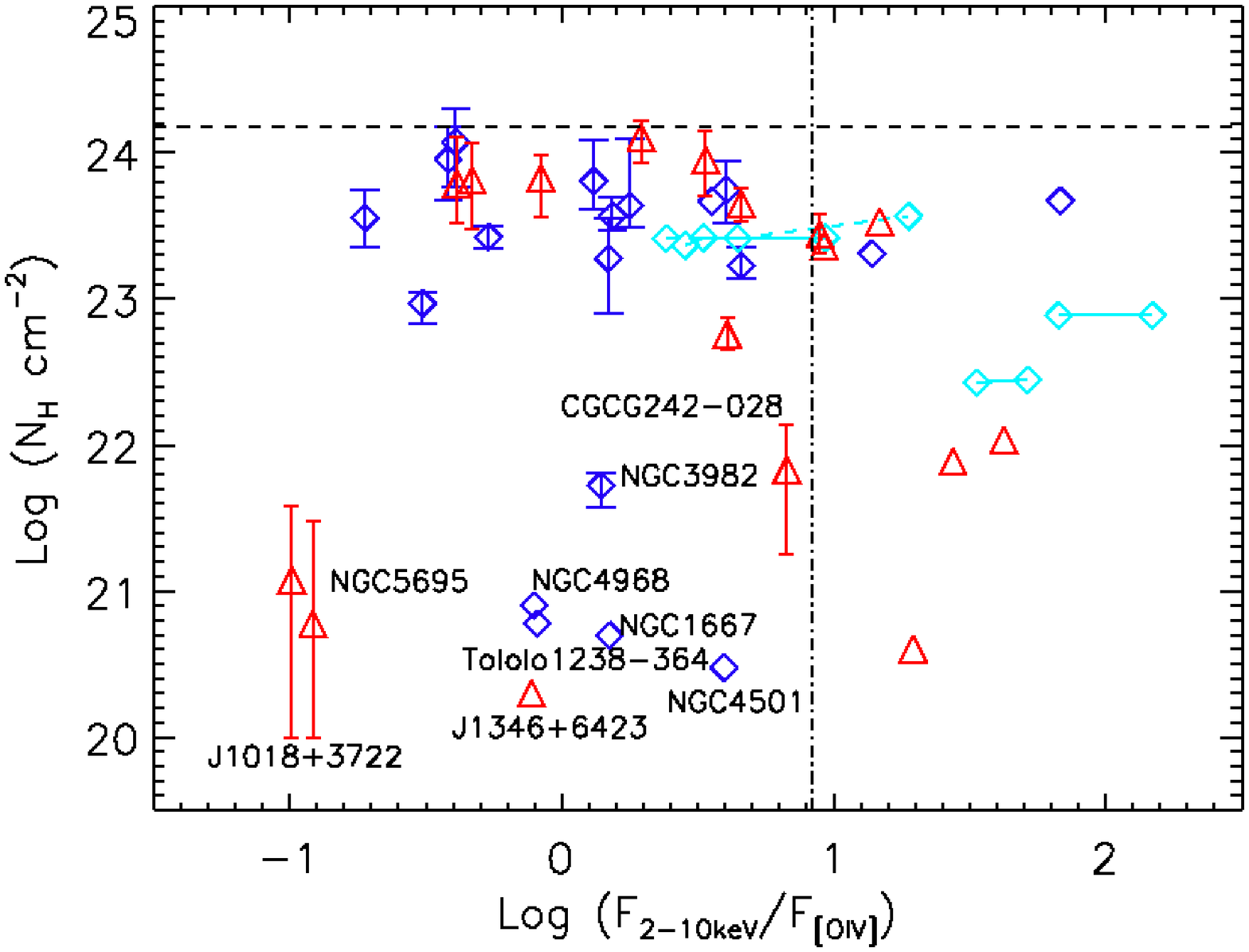}{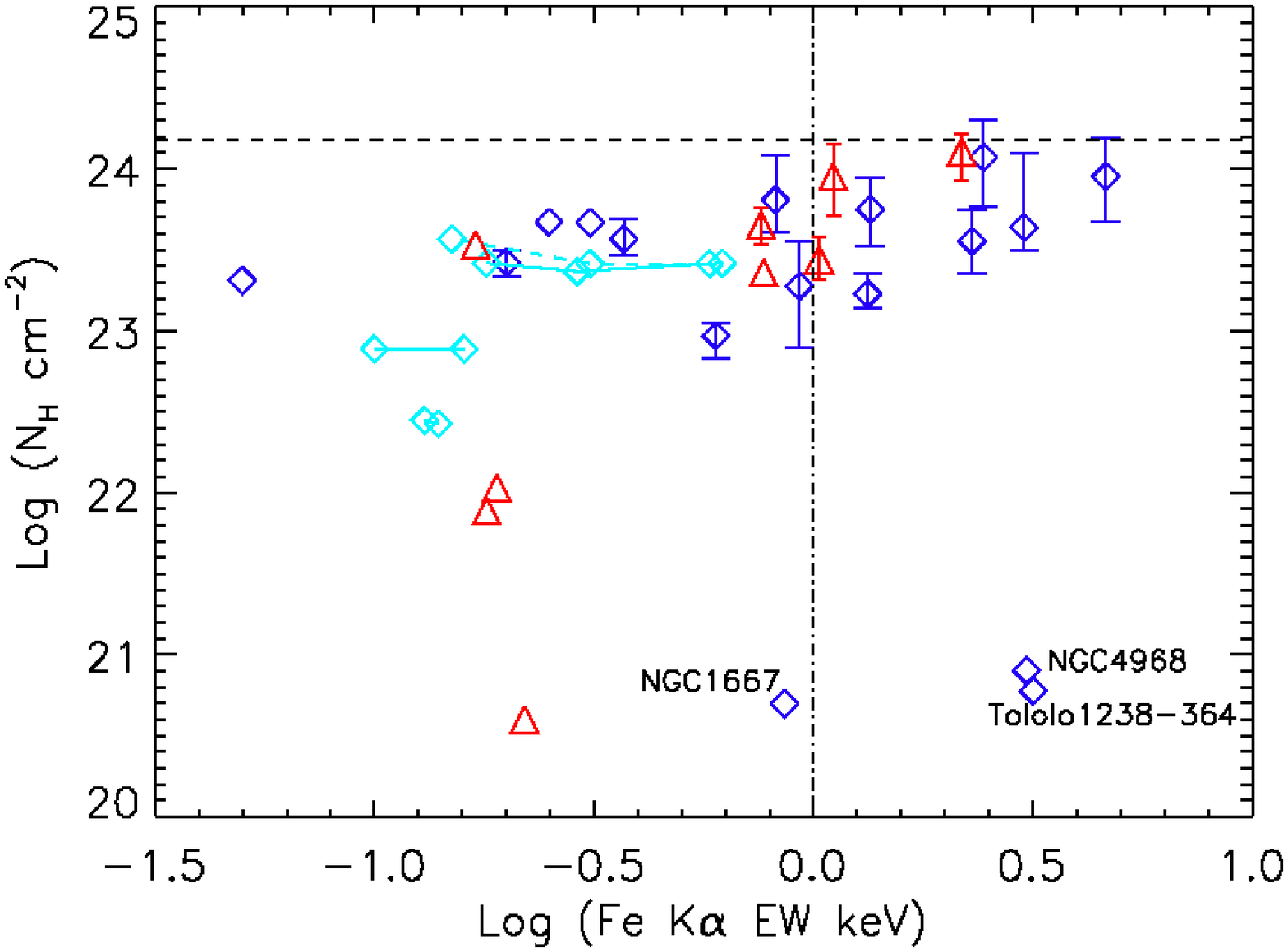}
\caption{\label{nh_cthick}Fitted N$_H$ as a function of obscuration diagnostics for the 12$\mu$m sample. The sources without error bars either had the best-fit absorption equal to the Galactic value, and therefore frozen at this value during fitting, or had N$_H$ error bars smaller than the symbol size. The dashed lines indicate the boundary for a Compton-thick column density (N$_H \geq 1.5 \times 10^{24}$ cm$^{-2}$) and the dashed-dotted line indicates nominal Compton-thick boundaries based on obscuration diagnostics (log (F$_{2-10keV}$/F$_{[OIV]}$) $\leq$ 0.9 dex, an order of magnitude less than the average value for Sy1s, and Fe K$\alpha$ EW $\geq$ 1 keV). Sources that are likely heavily obscured according to the obscuration diagnostics, yet have low fitted column densities, are labeled. Color coding same as Figure \ref{ew_ratios}.}
\end{figure}

\begin{figure}
\centering
\subfigure[]{\includegraphics[scale=0.30,angle=90]{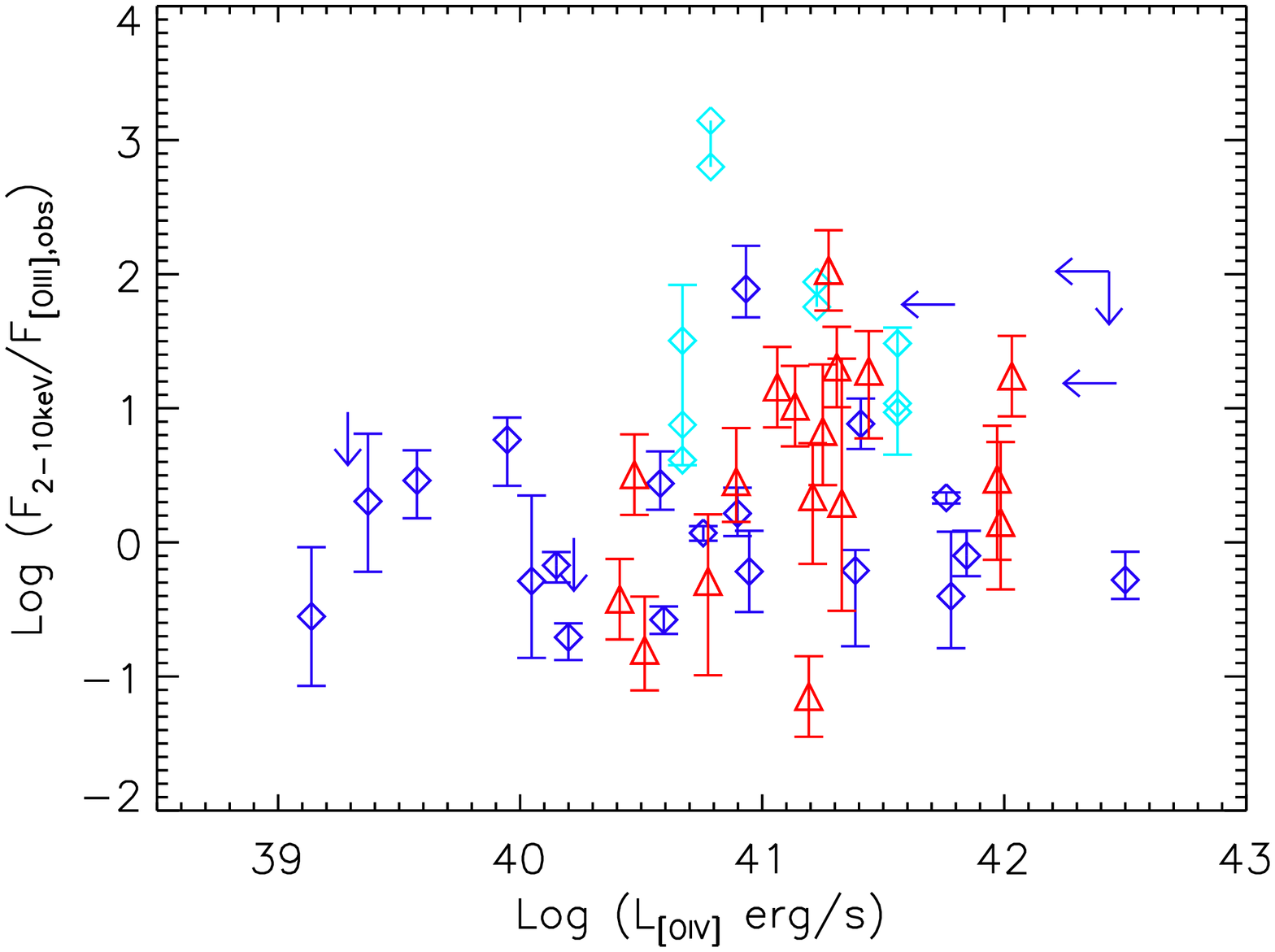}}
\subfigure[]{\includegraphics[scale=0.30,angle=90]{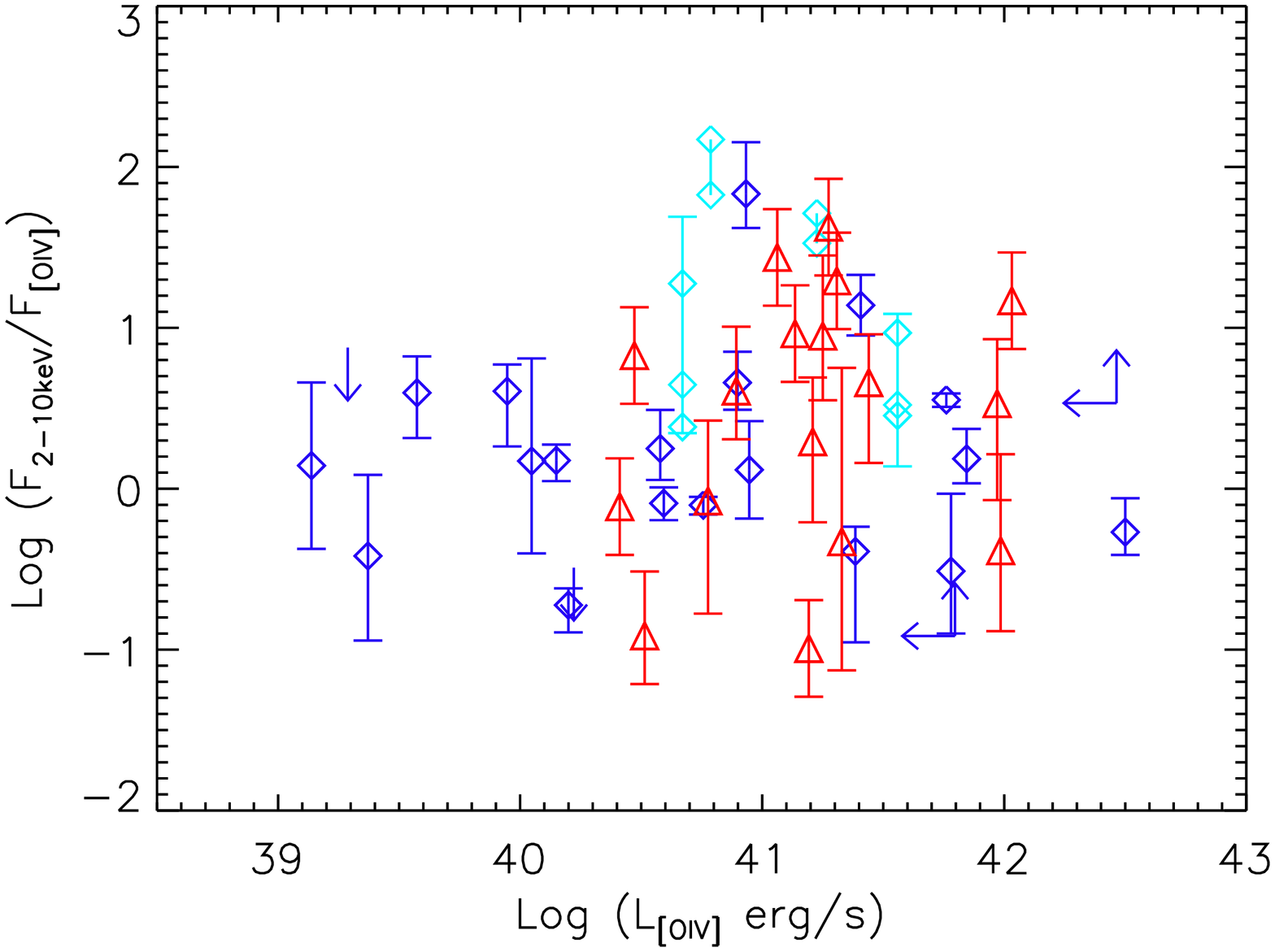}}
\subfigure[]{\includegraphics[scale=0.30,angle=90]{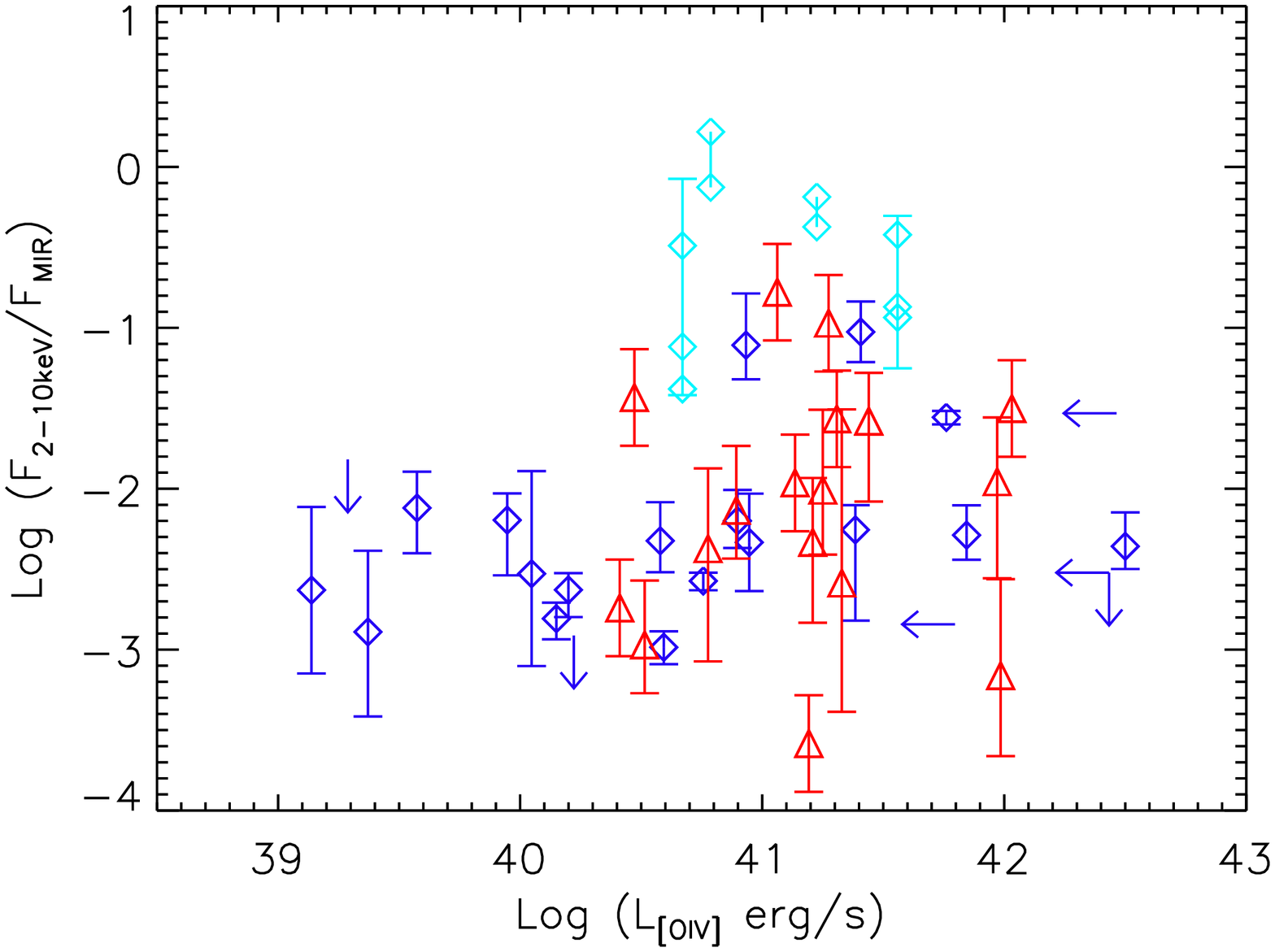}}
\subfigure[]{\includegraphics[scale=0.30,angle=90]{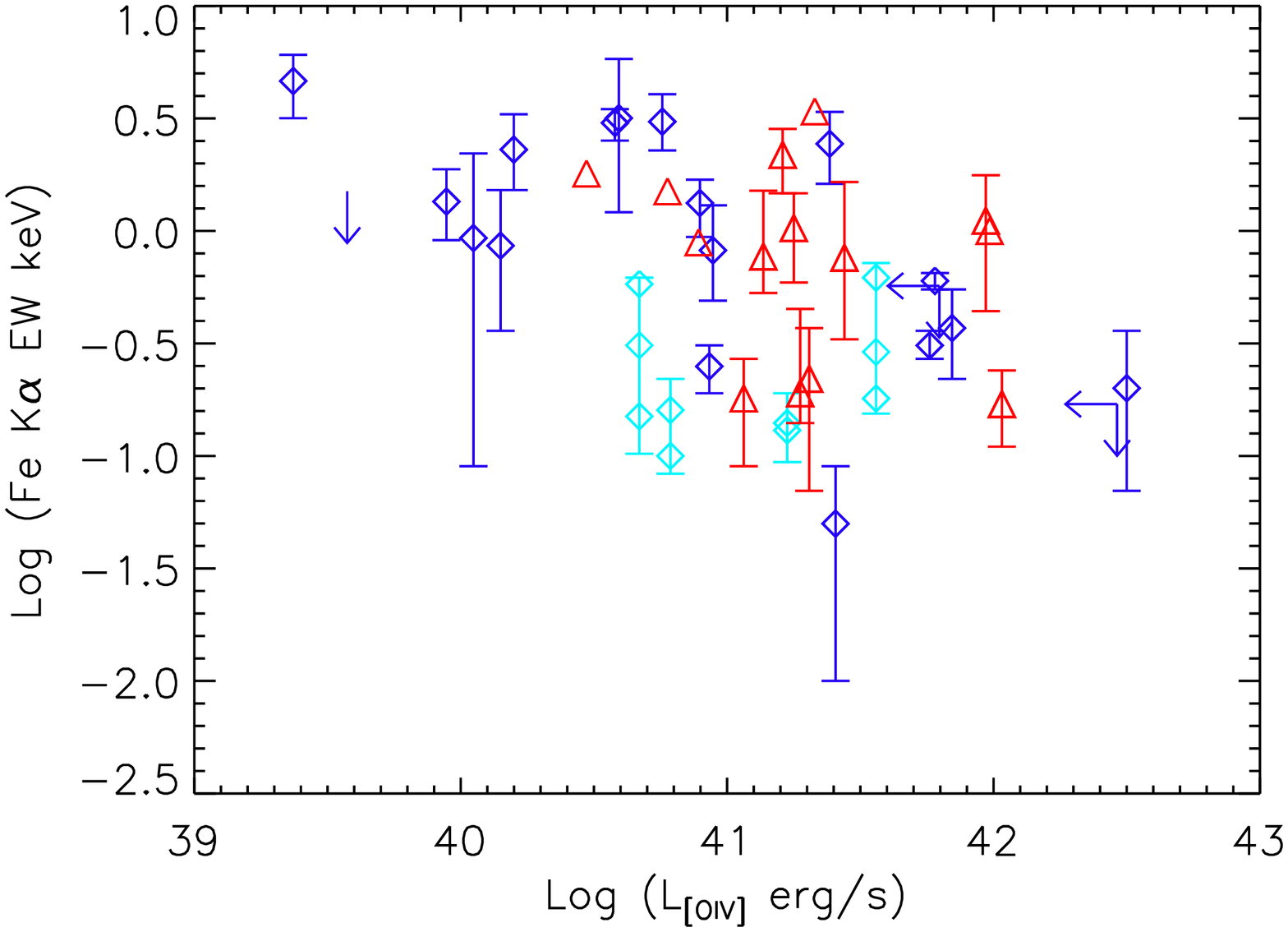}}
\caption{\label{liso_c_thick}Obscuration diagnostics vs. intrinsic AGN luminosity, parametrized by L$_{[OIV]}$ The lower limits on F$_{2-10keV}$/F$_{[OIV]}$ are displayed for illustrative purposes and not included in the survival analysis calculation. Survival analysis indicates a marginal statistically significant correlation among three of these relationships ($\rho$=0.273, 0.185, 0.349 and -0.302, respectively), however a wide scatter is evident. Color coding same as Figure \ref{lx_l_iso}.}
\end{figure}

\begin{figure}
\subfigure[]{\includegraphics[scale=0.30,angle=90]{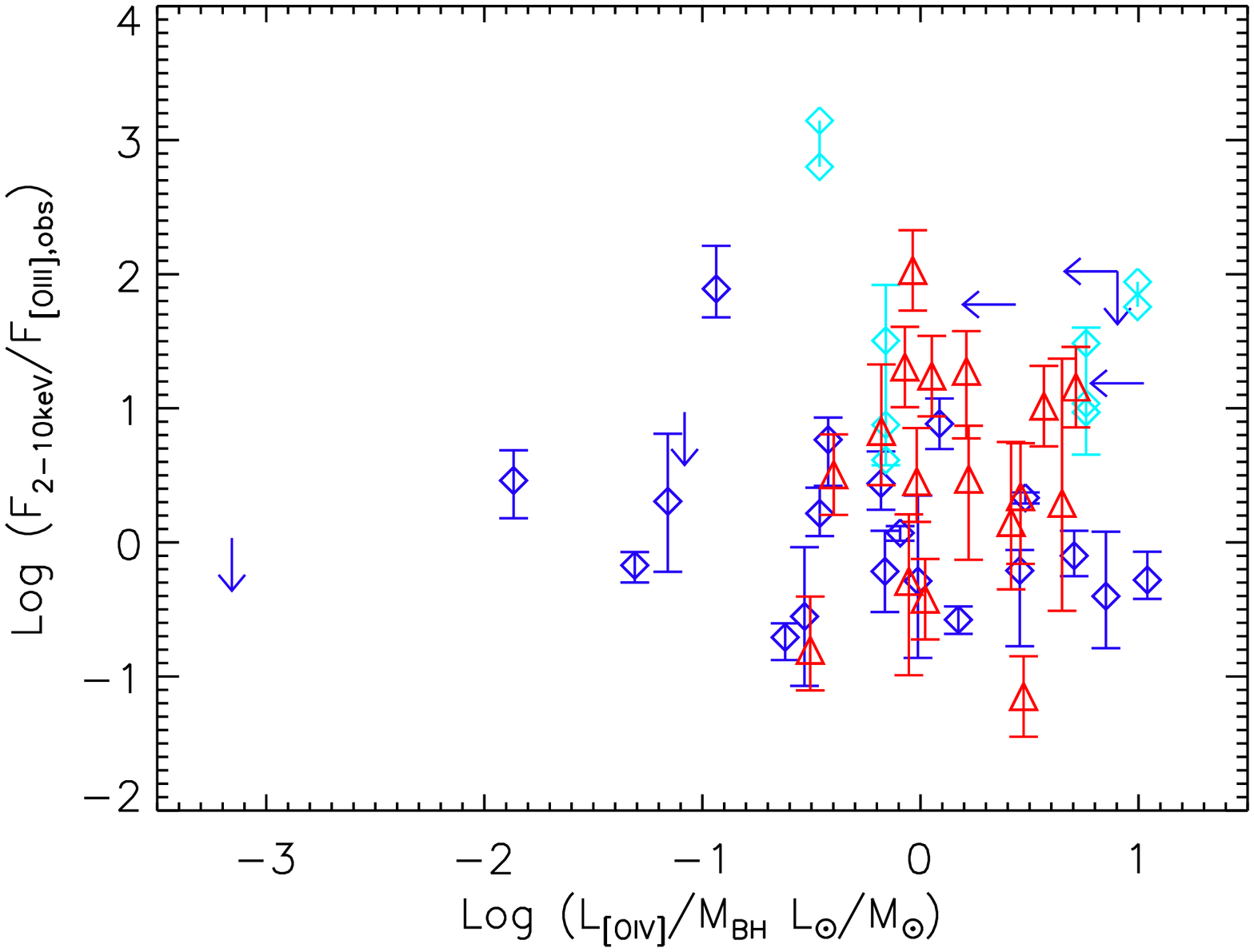}}
\subfigure[]{\includegraphics[scale=0.30,angle=90]{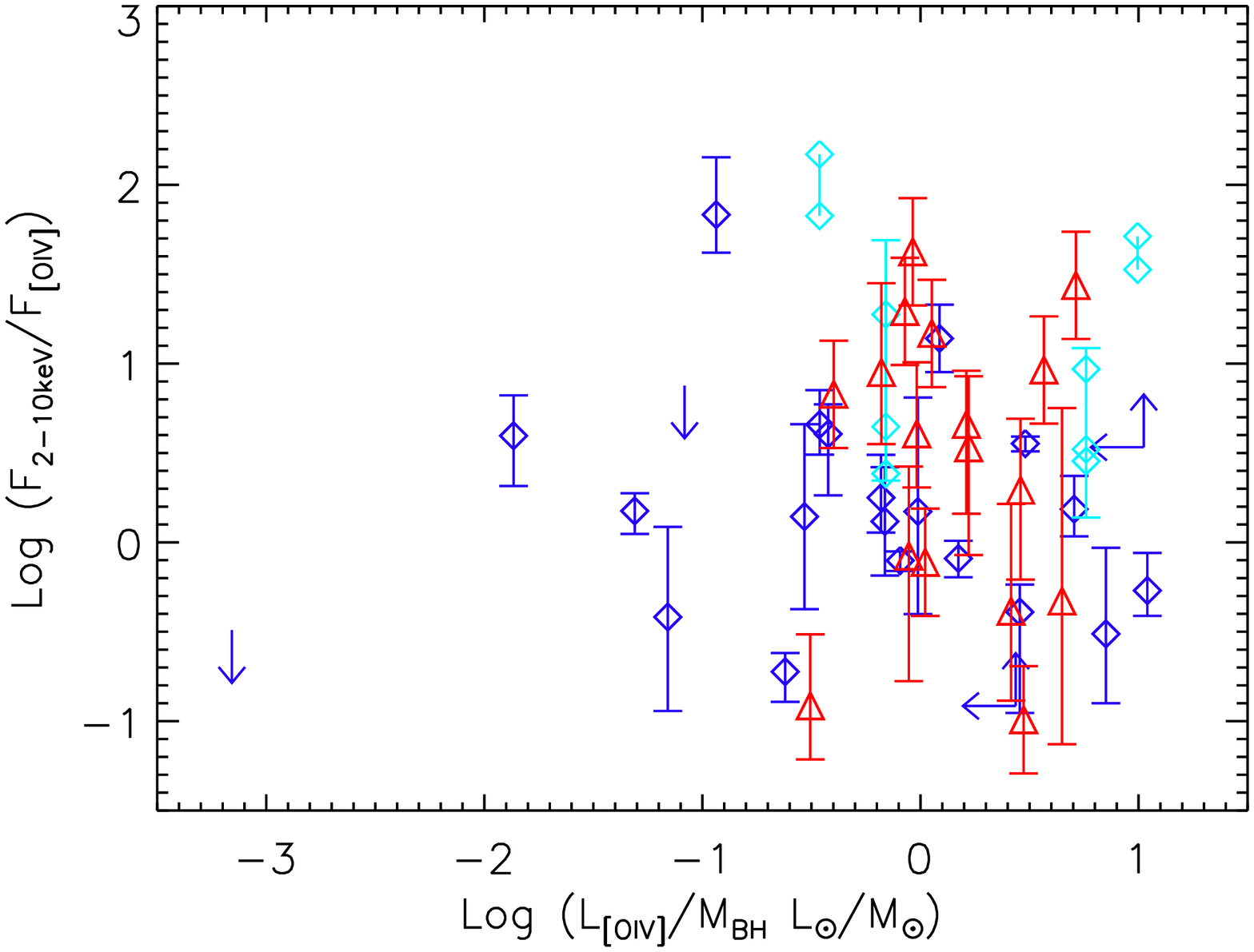}}
\subfigure[]{\includegraphics[scale=0.30,angle=90]{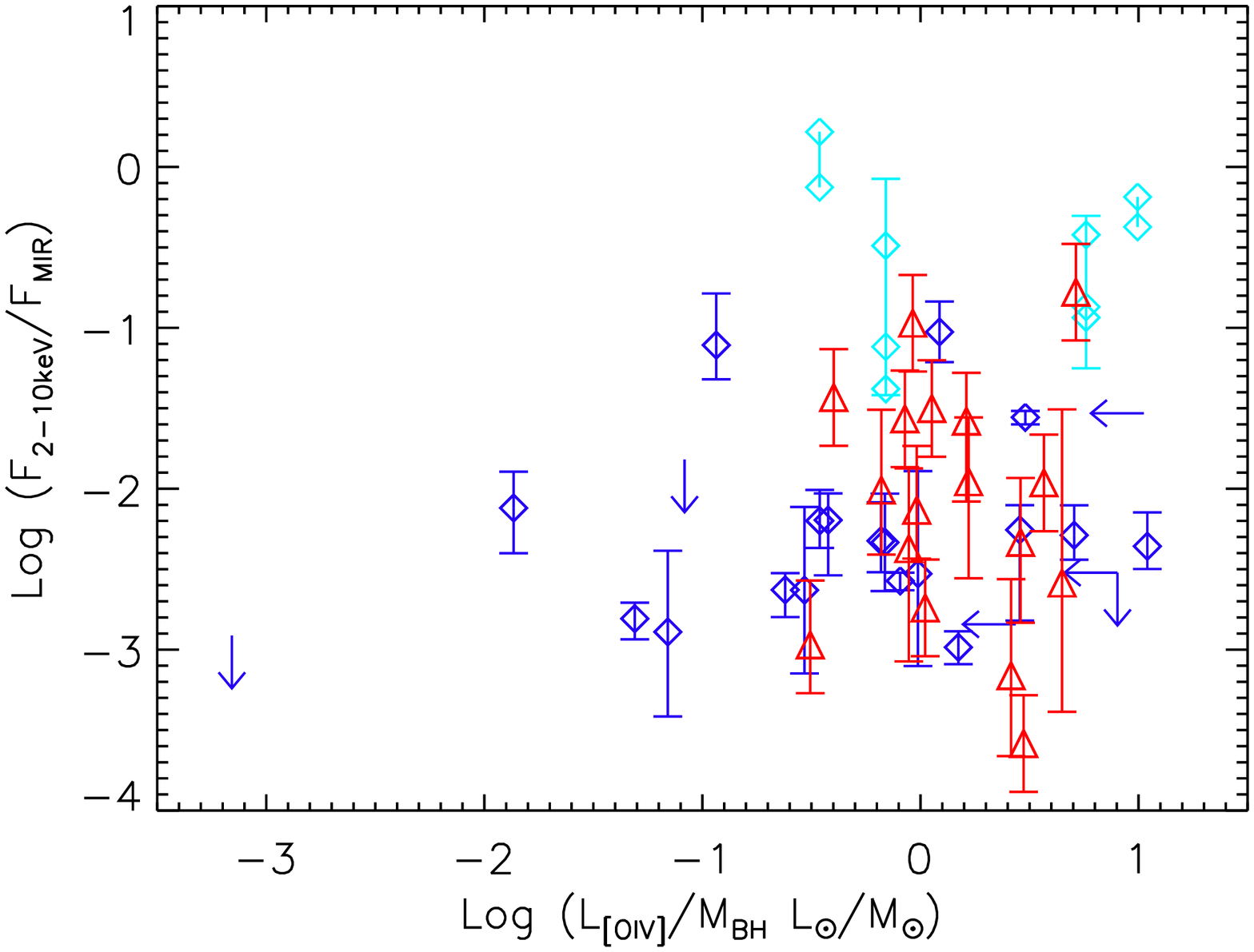}}
\subfigure[]{\includegraphics[scale=0.30,angle=90]{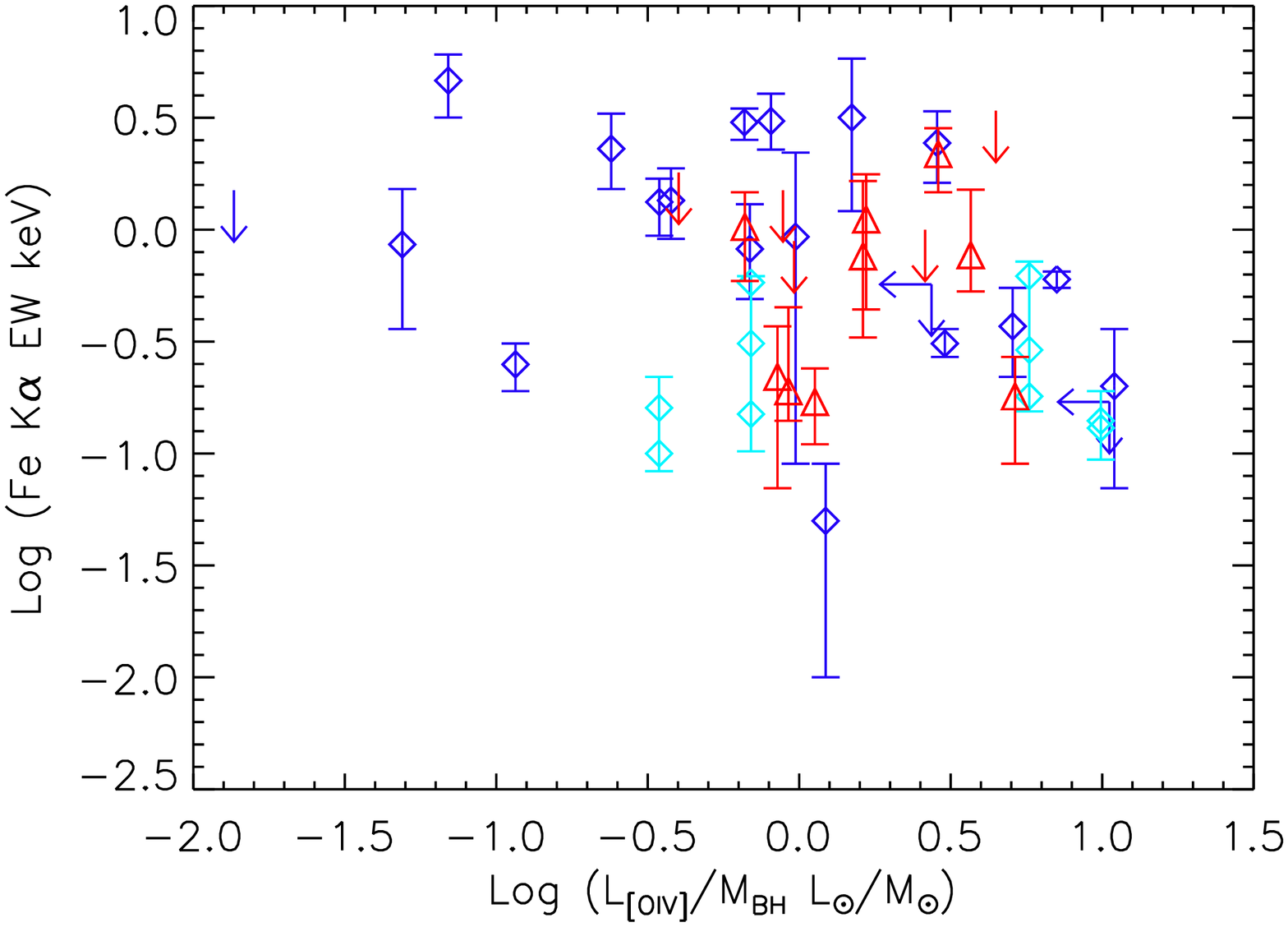}}
\caption{\label{edd_c_thick}Obscuration diagnostics vs. Eddington ratio. The lower limits on F$_{2-10keV}$/F$_{[OIV]}$ are displayed for illustrative purposes and not included in the survival analysis. With the exception of F$_{2-10keV}$/F$_{MIR}$, which survival analysis suggests is marginally significantly correlated with Eddington parameter, no trends are present: $\rho$=0.110, 0.056, 0.280 and -0.219, respectively. Color coding same as Figure \ref{lx_l_iso}.}
\end{figure}

\begin{figure}
\subfigure[]{\includegraphics[scale=0.30,angle=90]{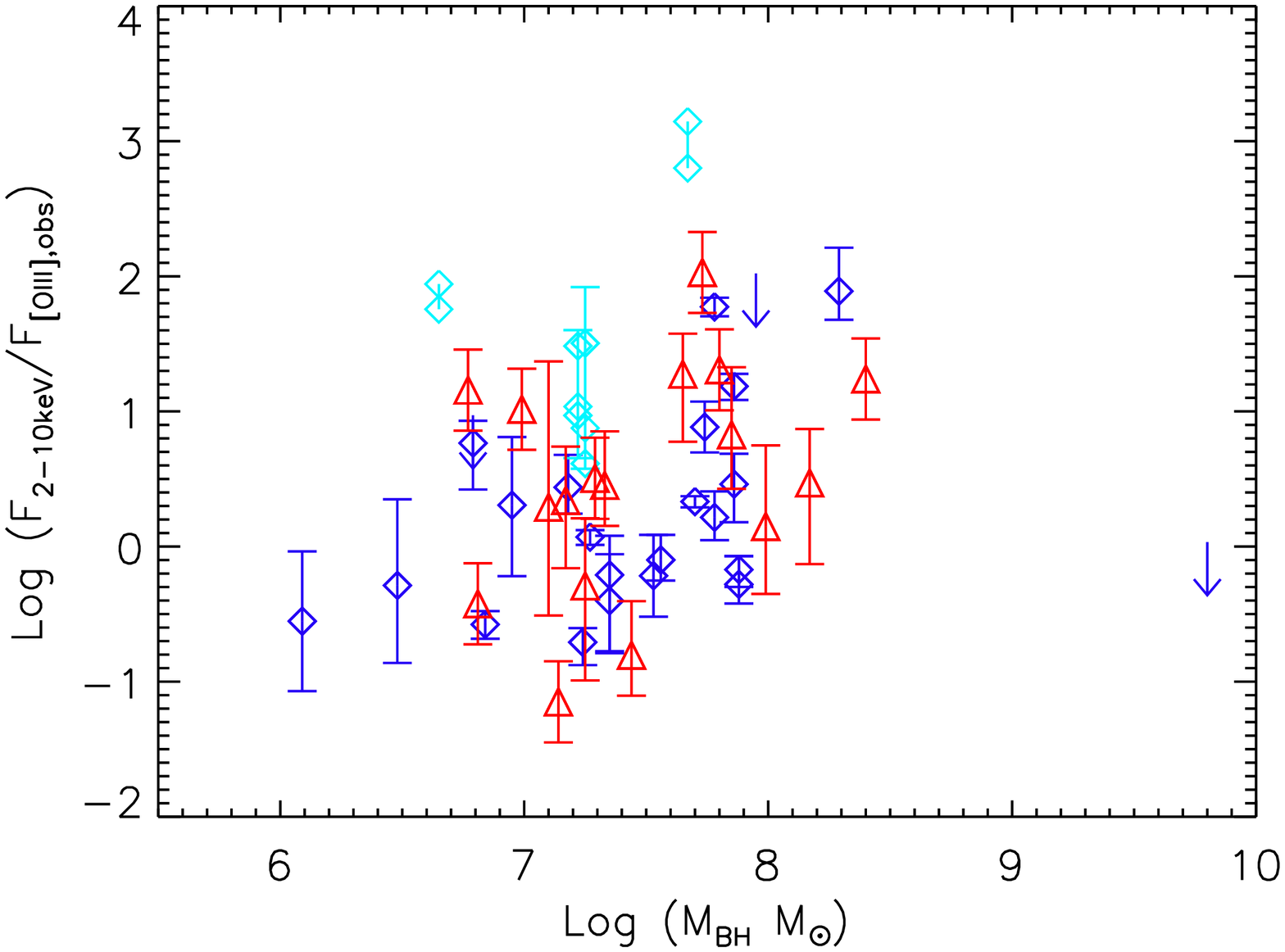}}
\subfigure[]{\includegraphics[scale=0.30,angle=90]{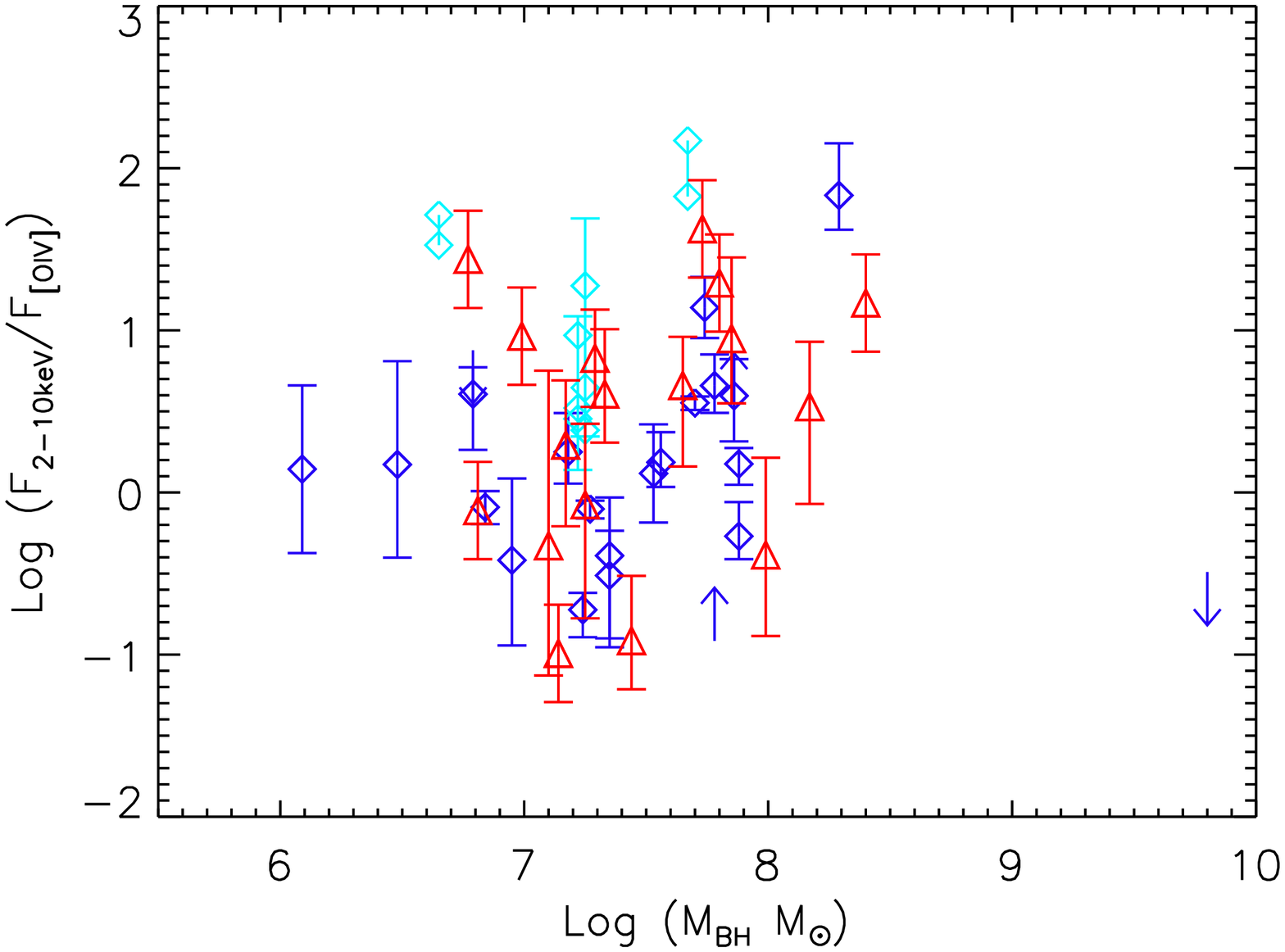}}
\subfigure[]{\includegraphics[scale=0.30,angle=90]{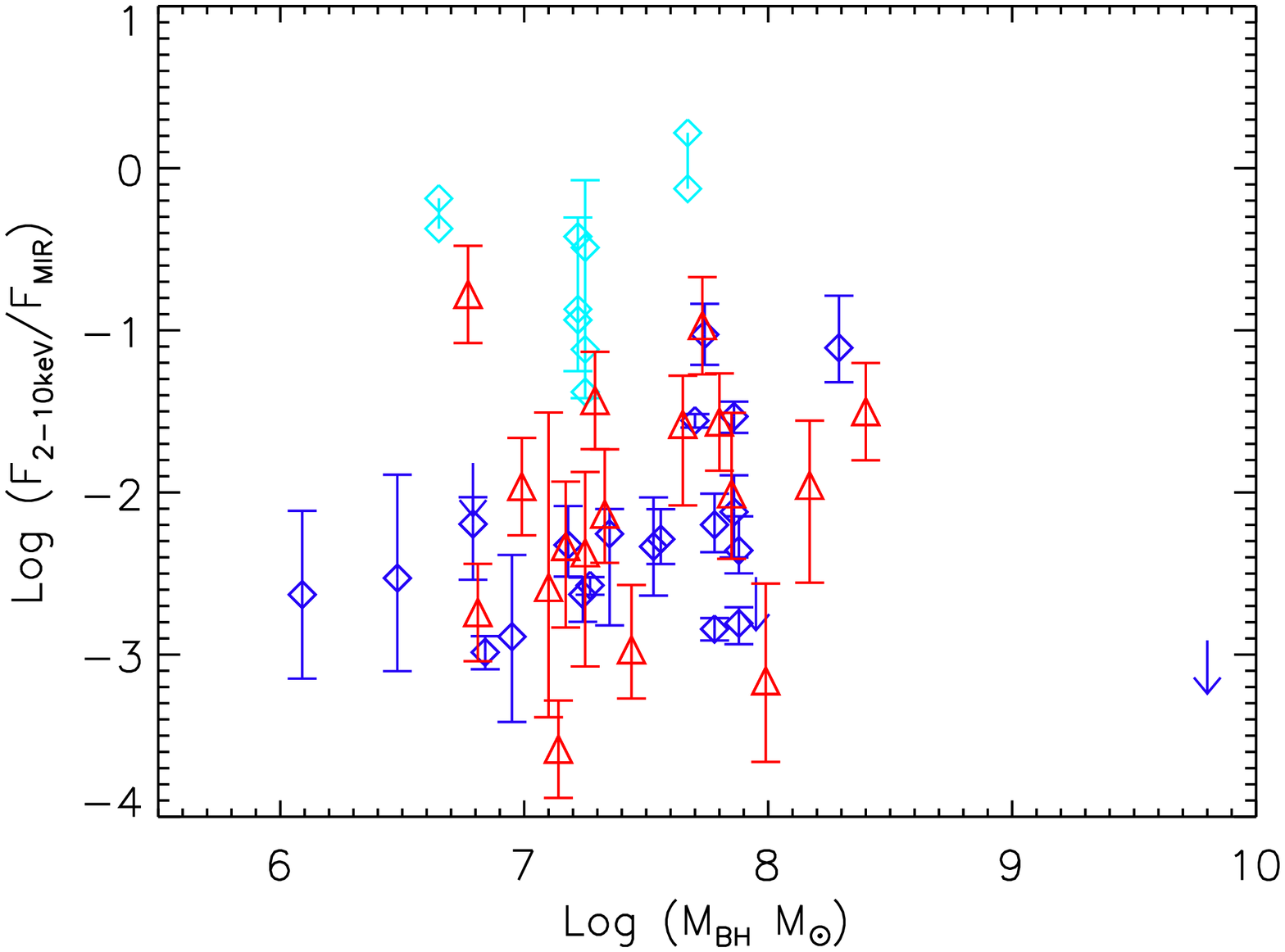}}
\subfigure[]{\includegraphics[scale=0.30,angle=90]{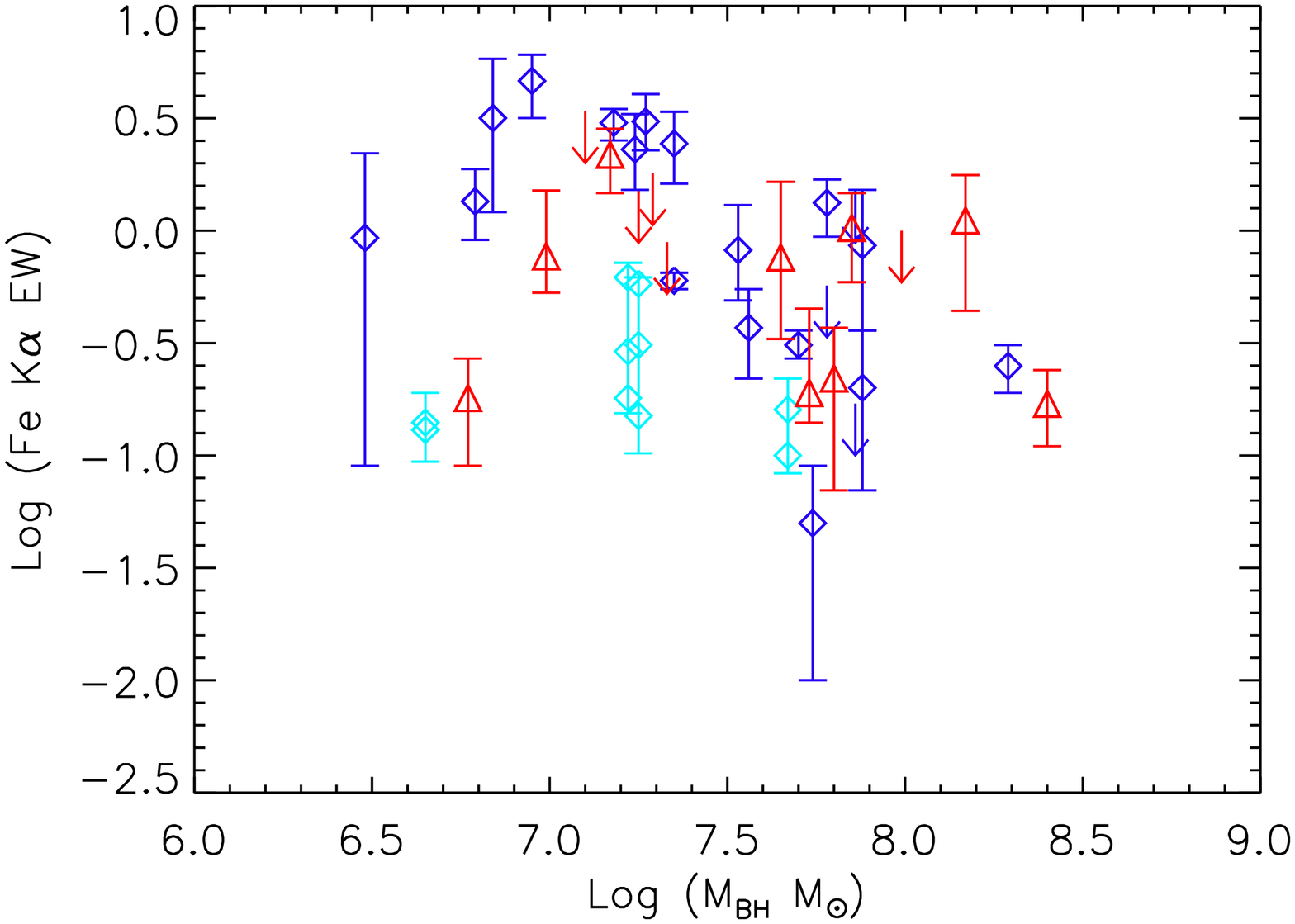}}
\caption{\label{mbh_c_thick}Obscuration diagnostics vs. M$_{BH}$. The lower limits on F$_{2-10keV}$/F$_{[OIV]}$ are displayed for illustrative purposes and not included in the survival analysis. No statistically significant trends are present: $\rho$=0.148, 0.115, -0.012  and -0.234, respectively. Color coding same as Figure \ref{lx_l_iso}.}
\end{figure}

\begin{figure}
\centering
\subfigure[]{\includegraphics[scale=0.30,angle=90]{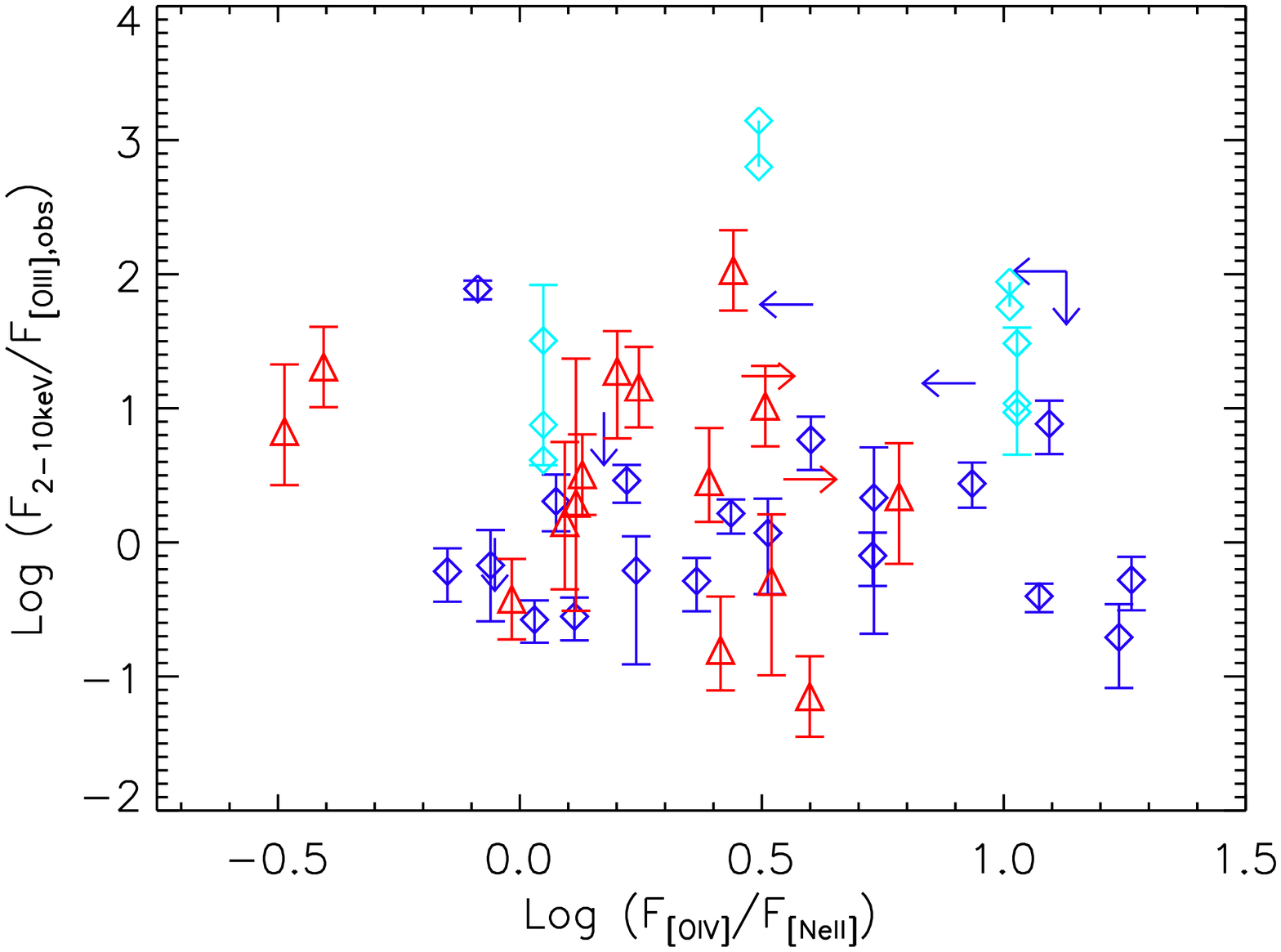}}
\subfigure[]{\includegraphics[scale=0.30,angle=90]{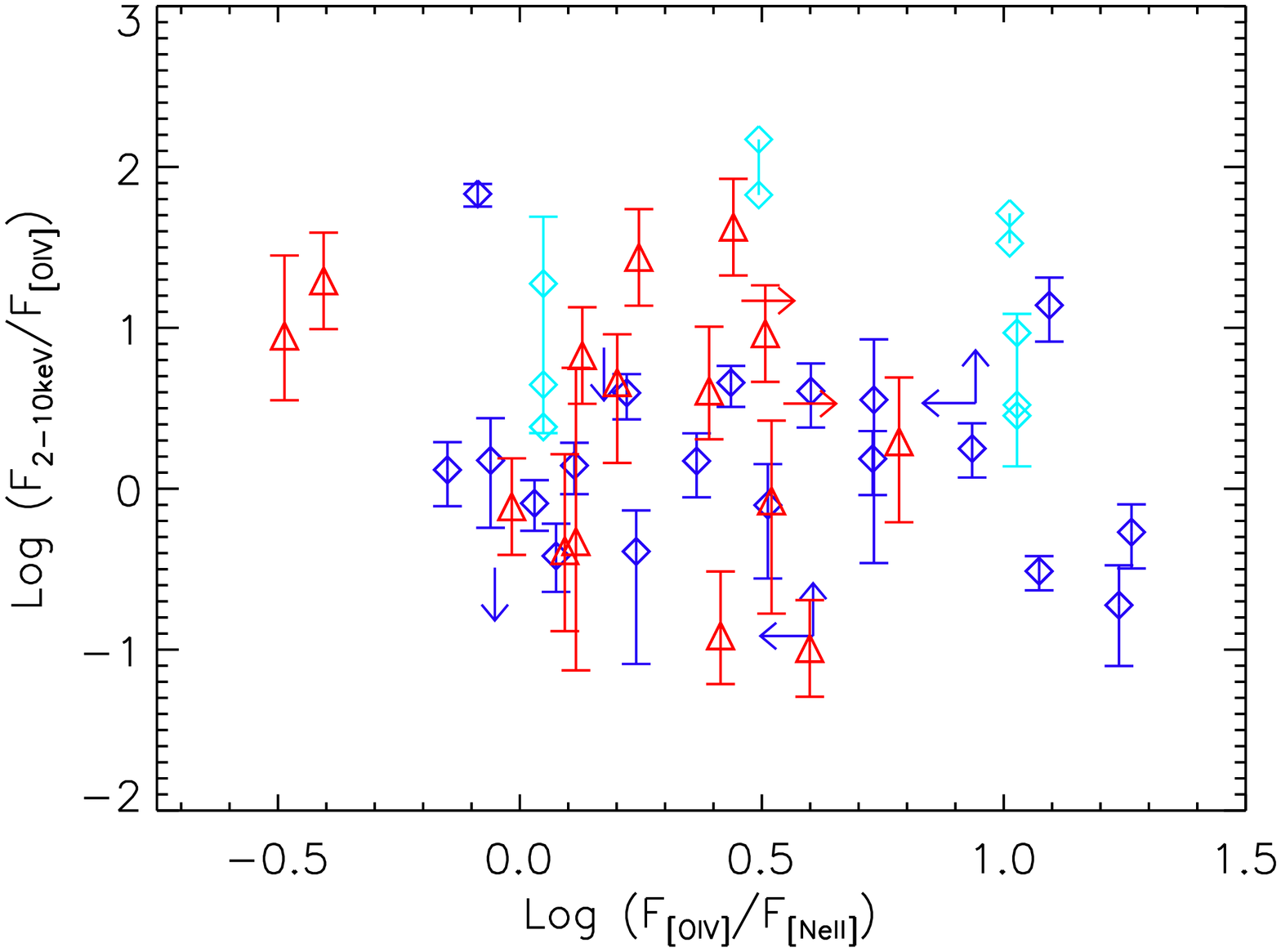}}
\subfigure[]{\includegraphics[scale=0.30,angle=90]{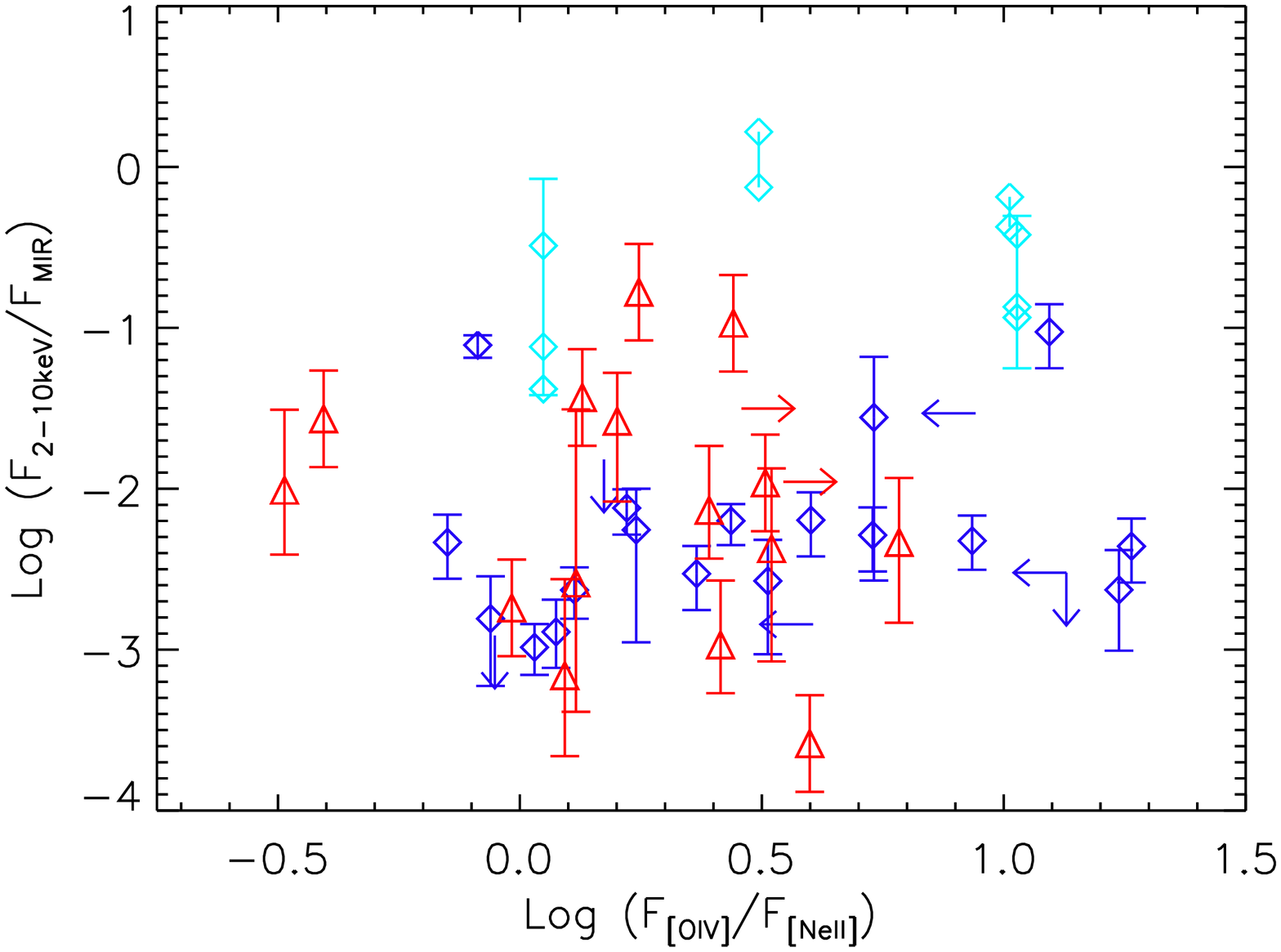}}
\subfigure[]{\includegraphics[scale=0.30,angle=90]{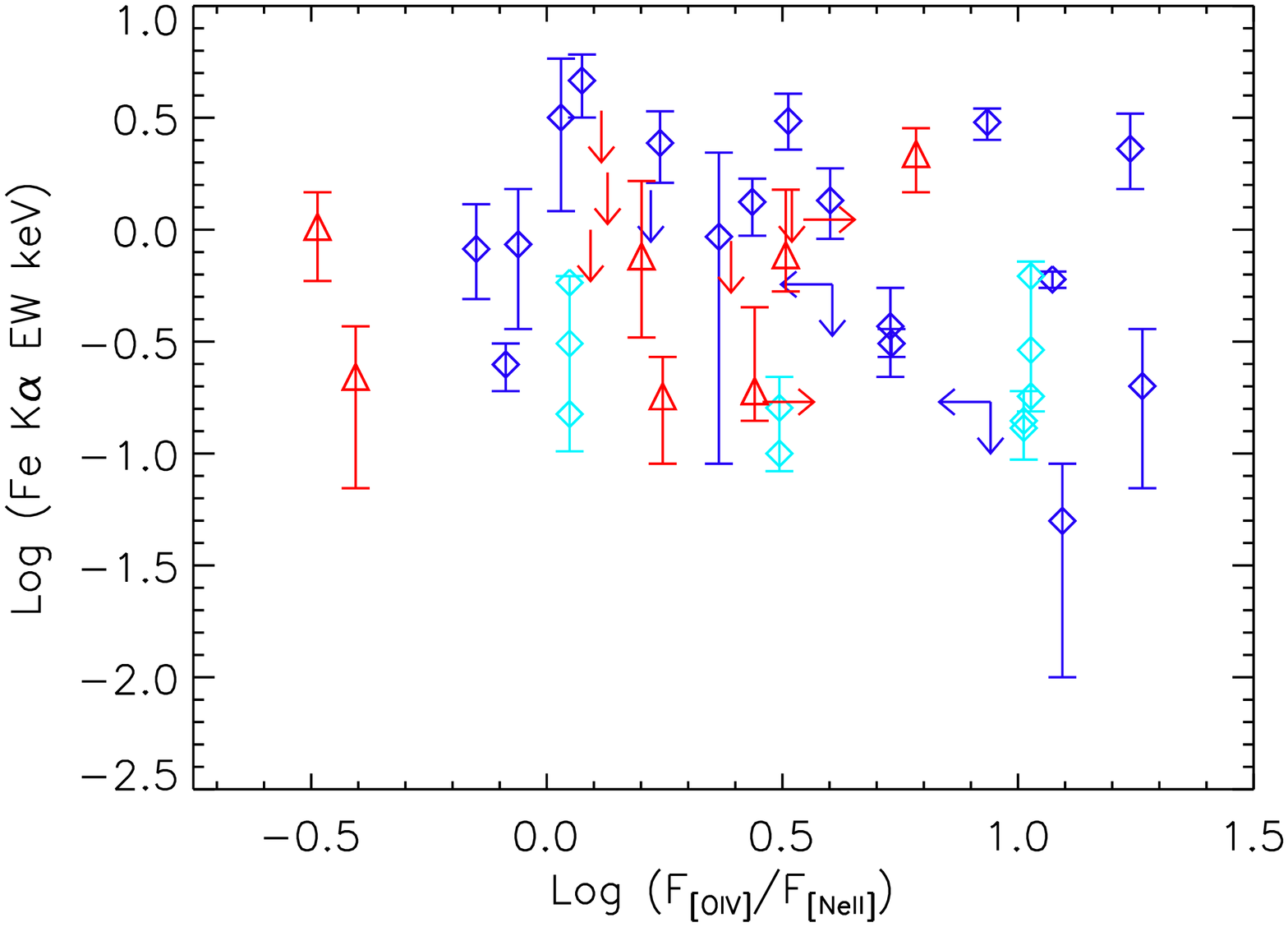}}
\caption{\label{oiv_neii_c_thick}Obscuration diagnostics vs. F$_{[OIV]}$/F$_{[NeII]}$, a proxy for the relative strength of the ionizing continuum from the AGN versus starburst activity. The lower limits on F$_{[OIV]}$/F$_{[NeII]}$ and F$_{2-10keV}$/F$_{[OIV]}$ are displayed for illustrative purposes and not included in the survival analysis. No statistically significant trends are apparent: $\rho$=0.036, 0.032, 0.269 and -0.130, respectively. Color coding same as Figure \ref{lx_l_iso}.}
\end{figure}

\begin{figure}
\centering
\subfigure[]{\includegraphics[scale=0.30,angle=90]{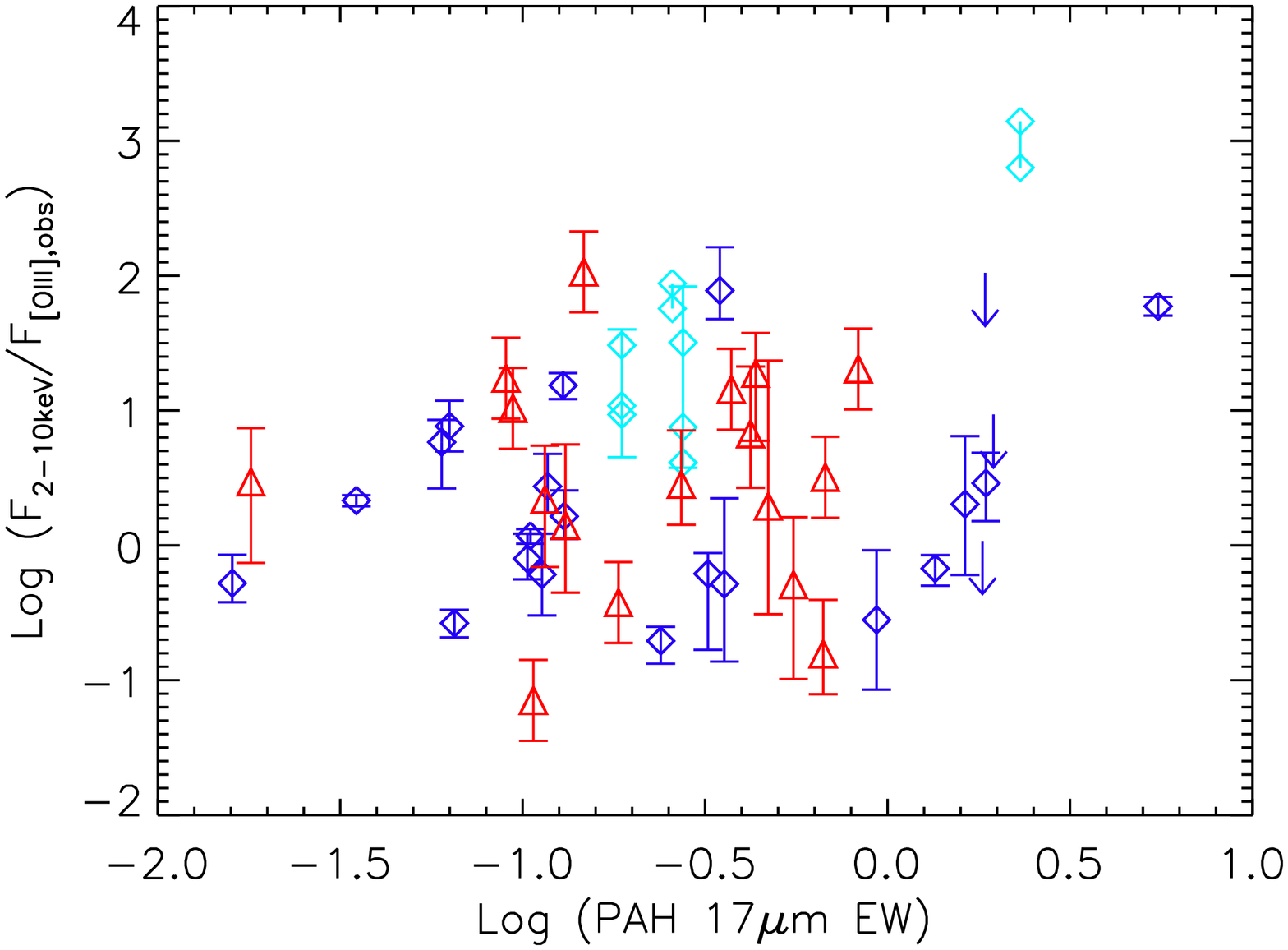}}
\subfigure[]{\includegraphics[scale=0.30,angle=90]{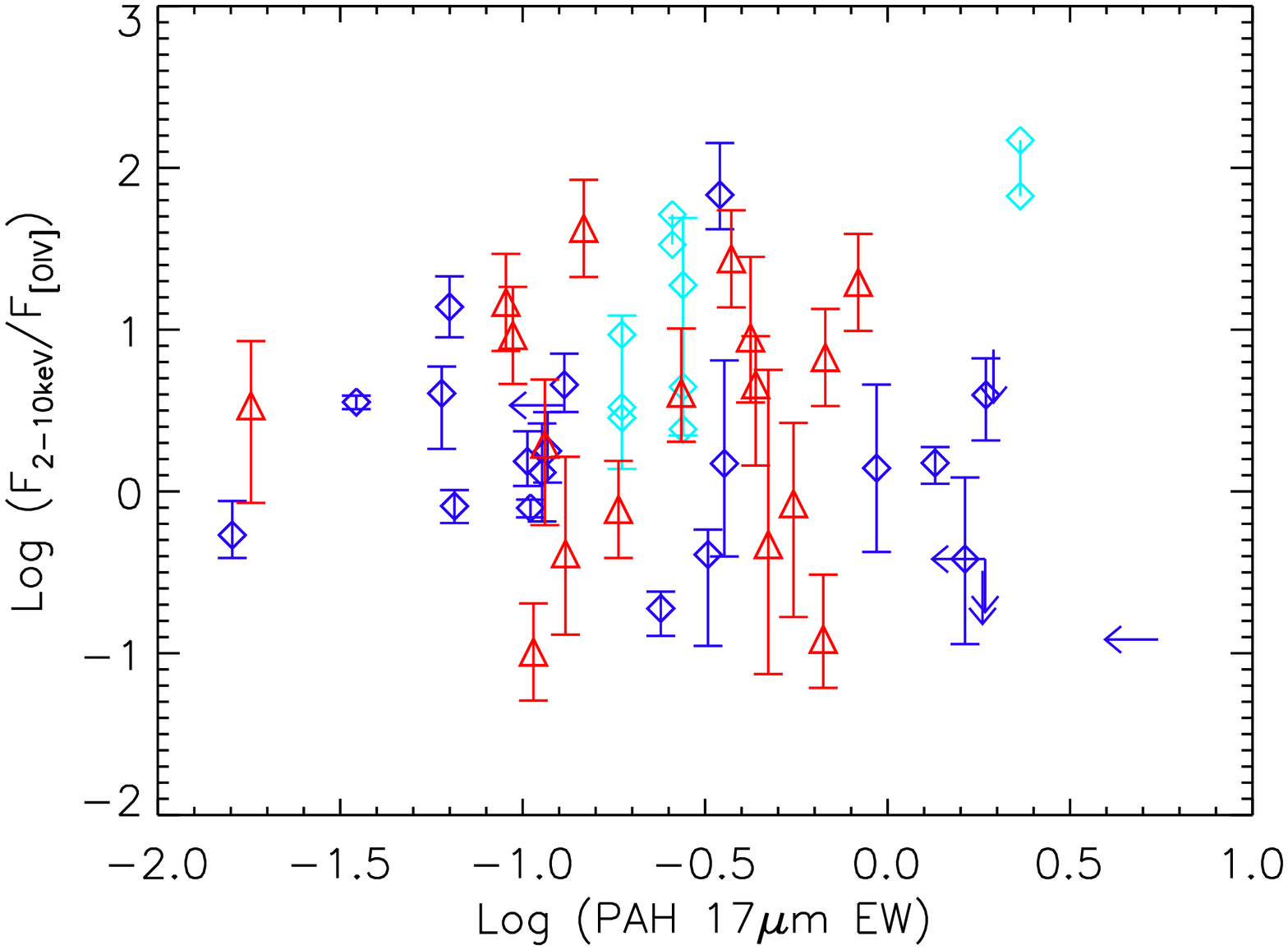}}
\subfigure[]{\includegraphics[scale=0.30,angle=90]{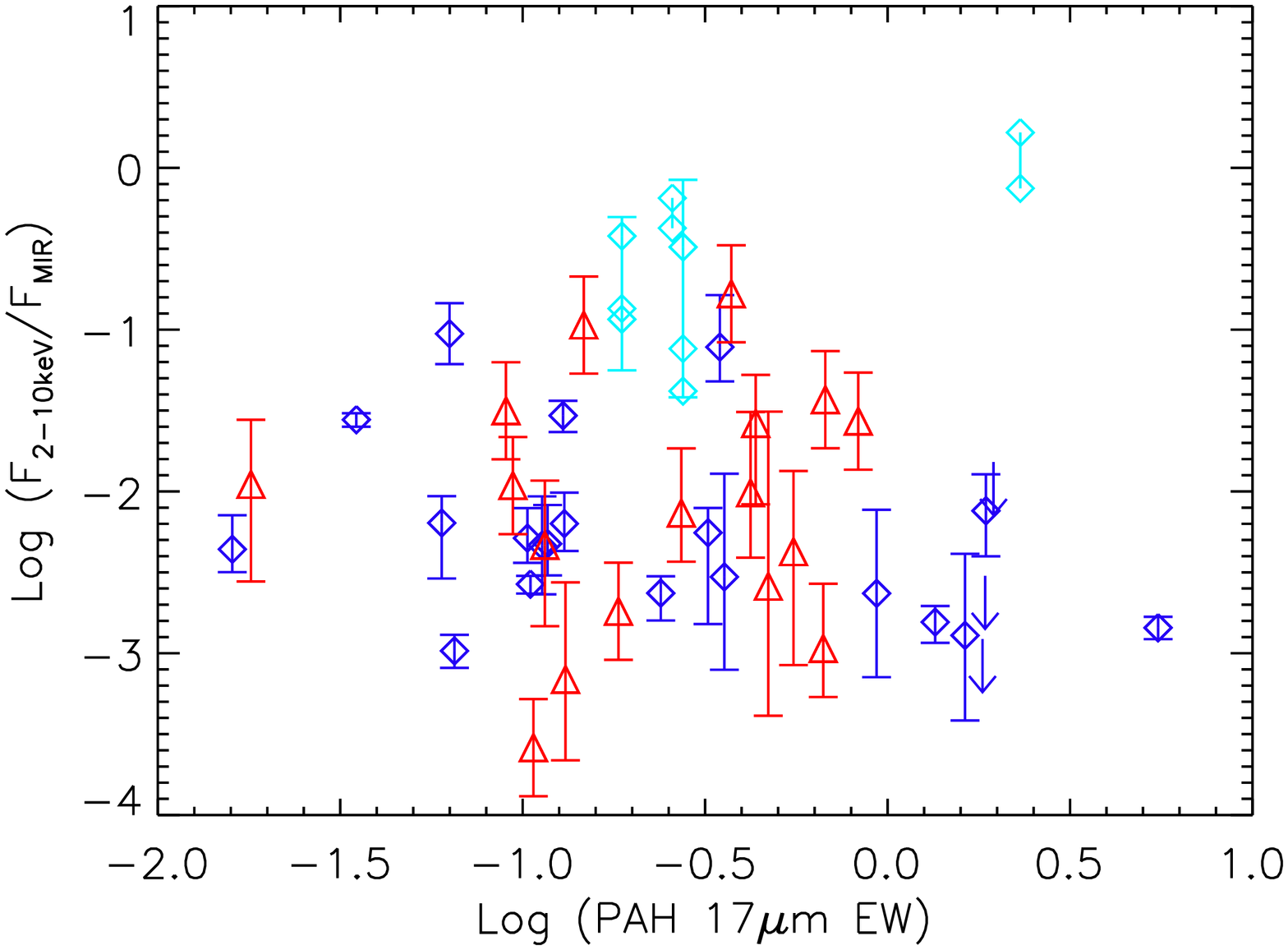}}
\subfigure[]{\includegraphics[scale=0.30,angle=90]{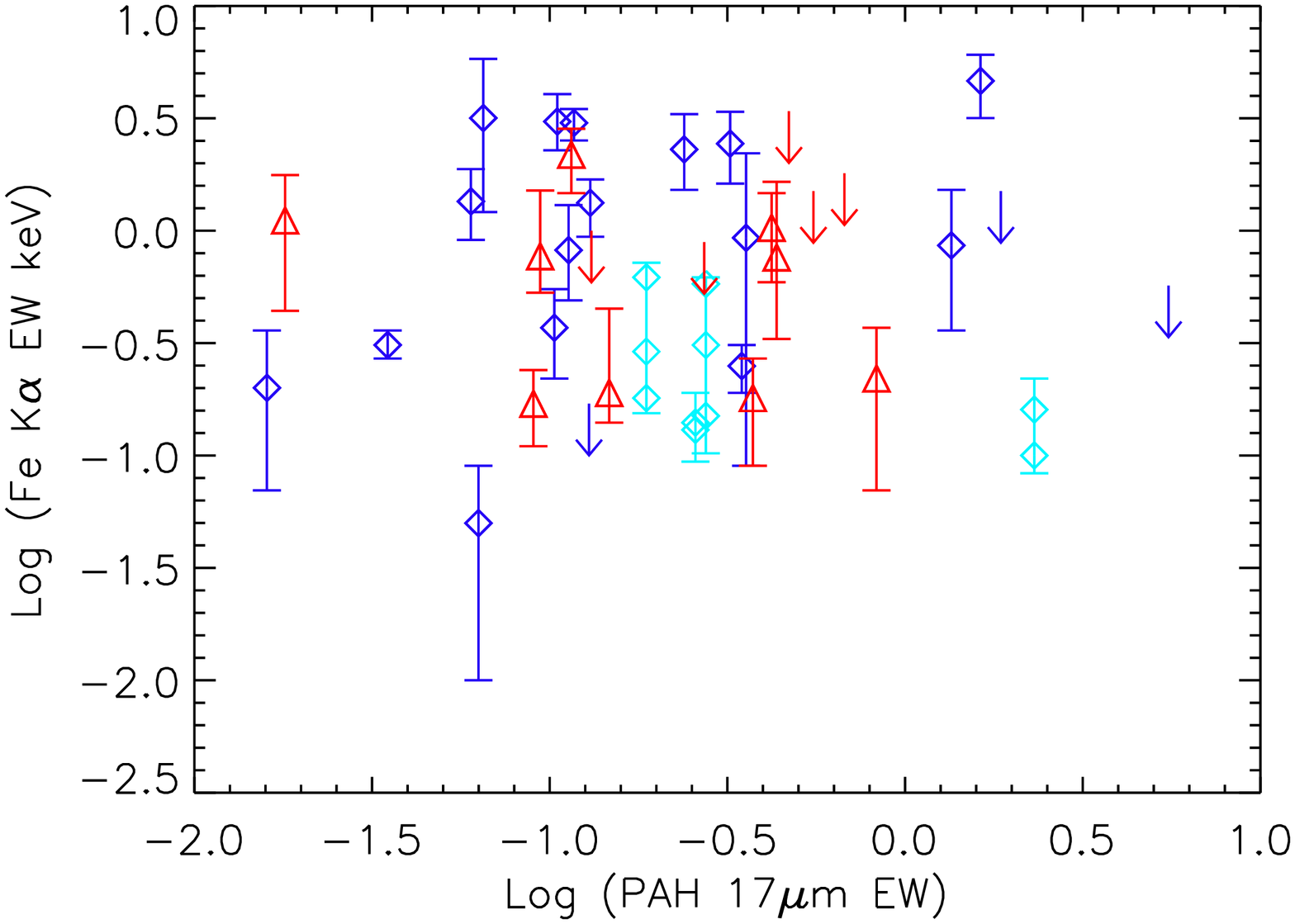}}
\caption{\label{pah_17_c_thick}Obscuration diagnostics vs. the EW of the PAH 17 $\mu$m feature, which parametrizes star formation rate. The lower limits on F$_{2-10keV}$/F$_{[OIV]}$ are displayed for illustrative purposes and not included in the survival analysis. No statistically significant trends are apparent: $\rho$=0.135, 0.062, -0.055 and -0.192, respectively. Color coding same as Figure \ref{lx_l_iso}.}
\end{figure}

\begin{figure}[ht]
\centering
\subfigure[]{\includegraphics[scale=0.30,angle=90]{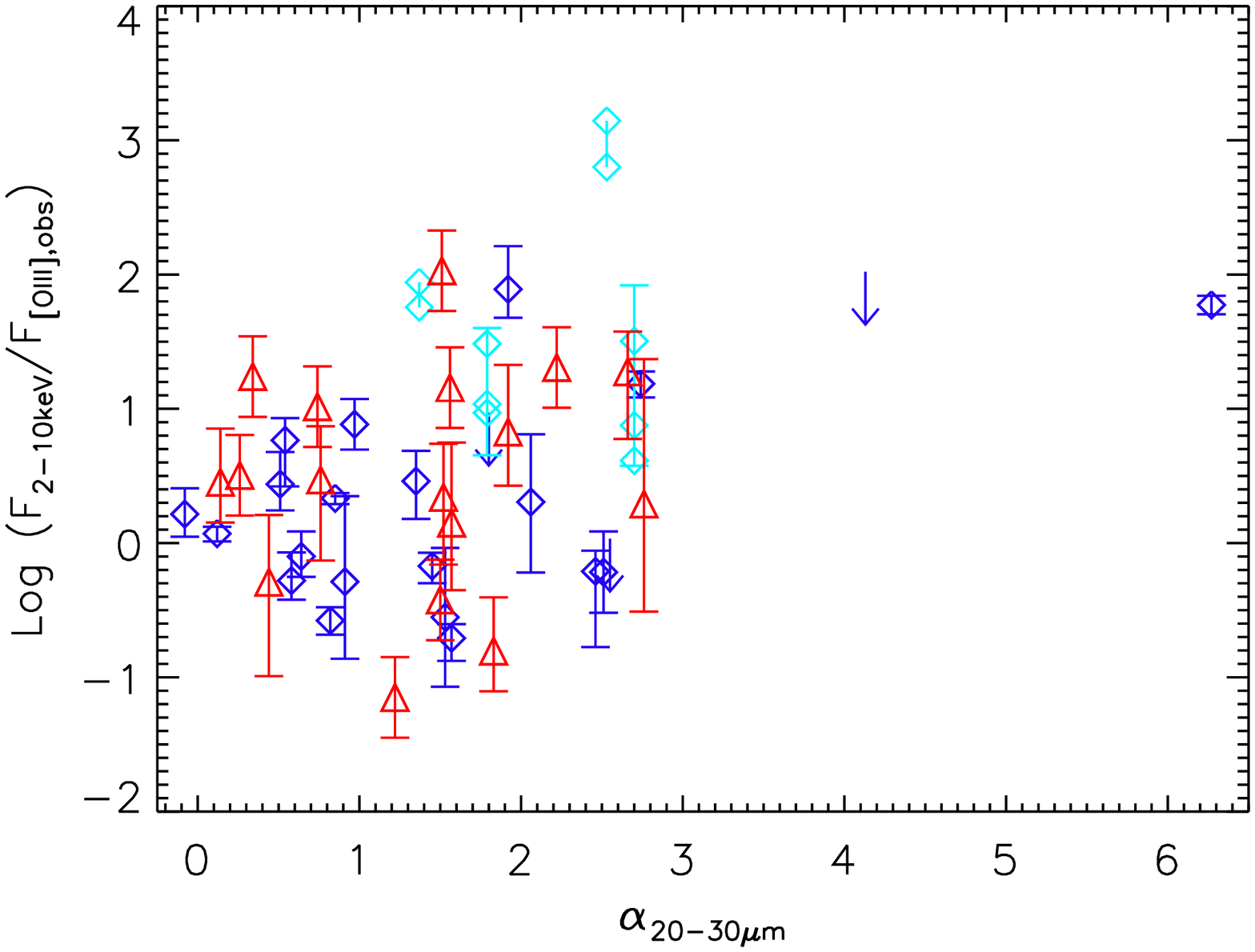}}
\subfigure[]{\includegraphics[scale=0.30,angle=90]{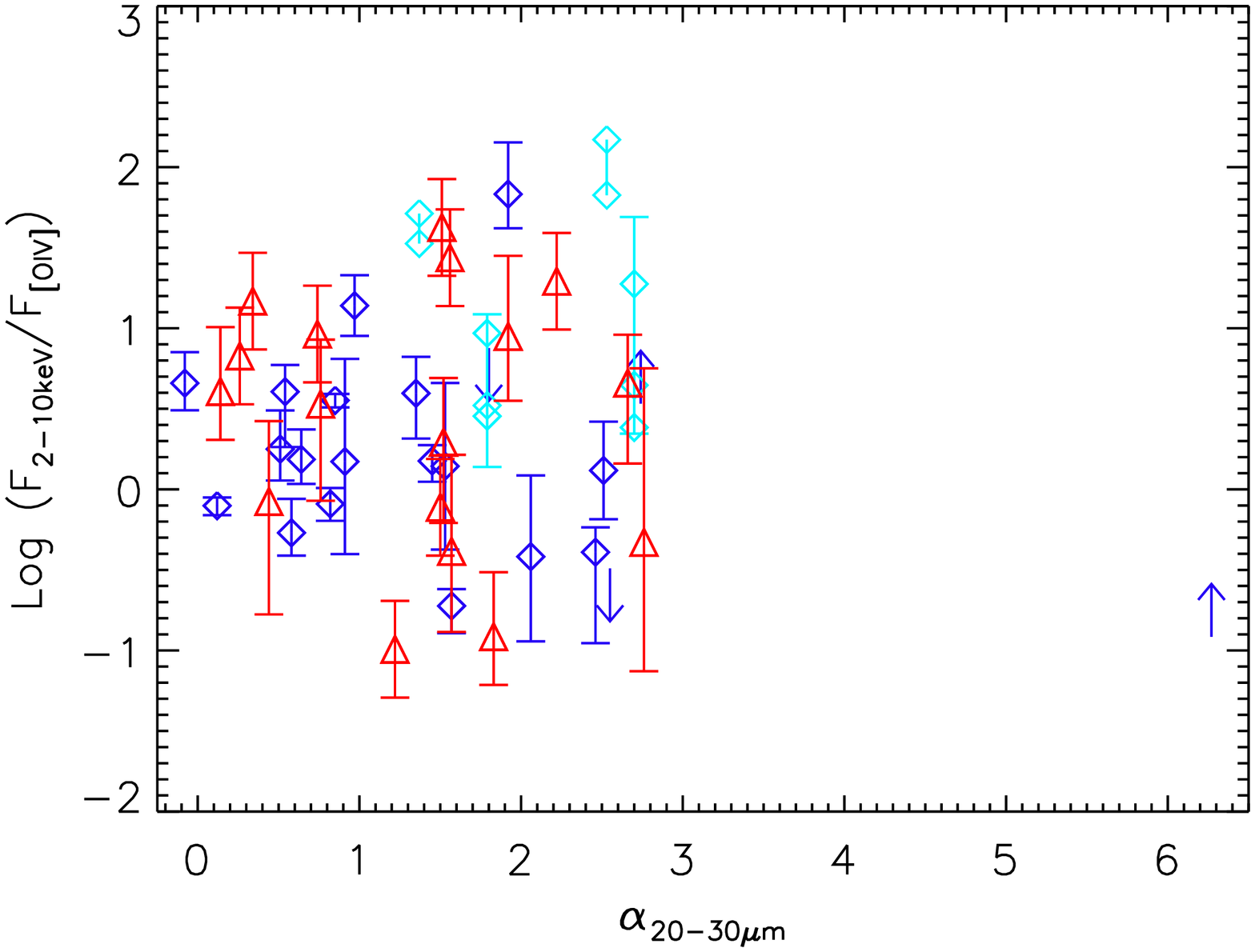}}
\subfigure[]{\includegraphics[scale=0.30,angle=90]{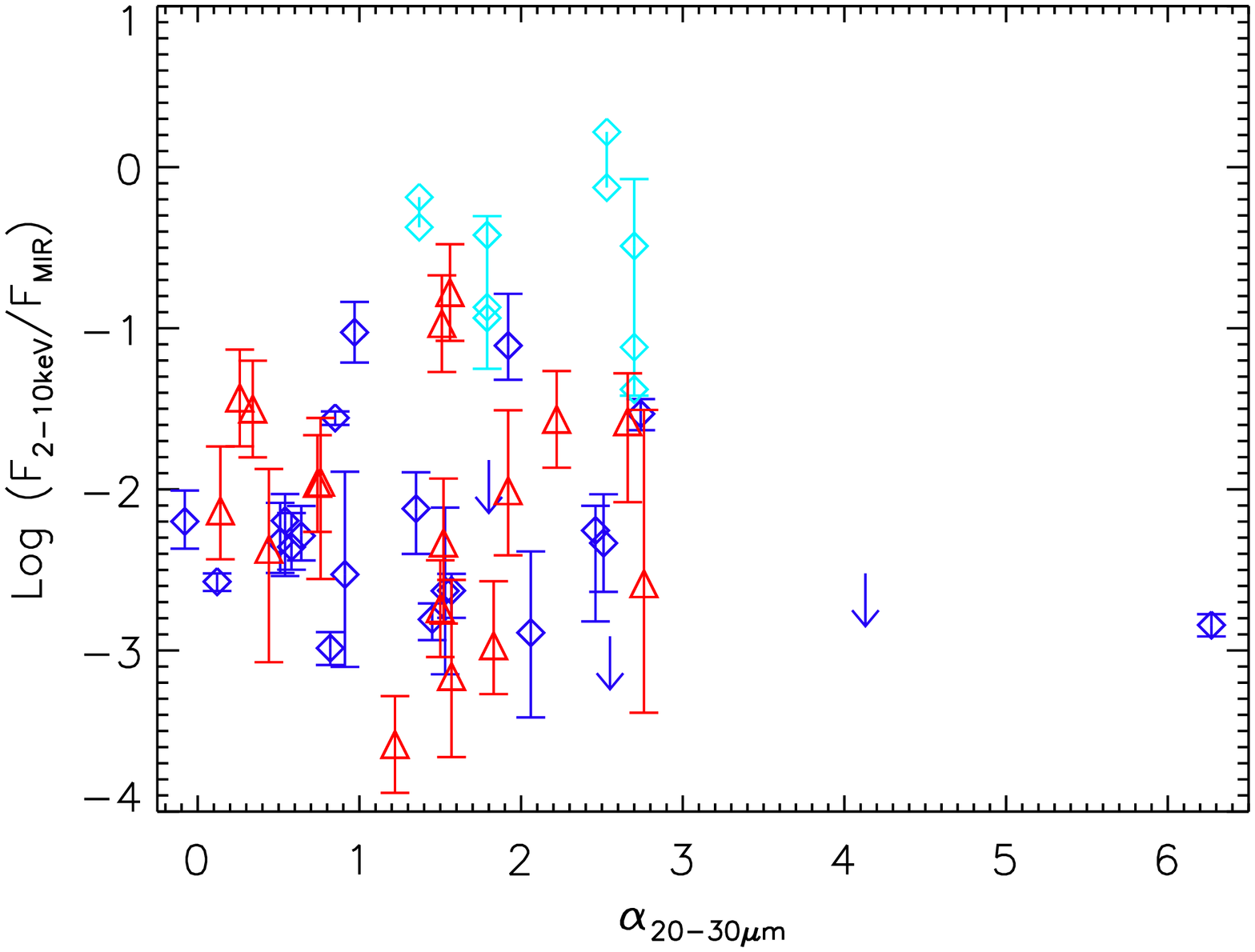}}
\subfigure[]{\includegraphics[scale=0.30,angle=90]{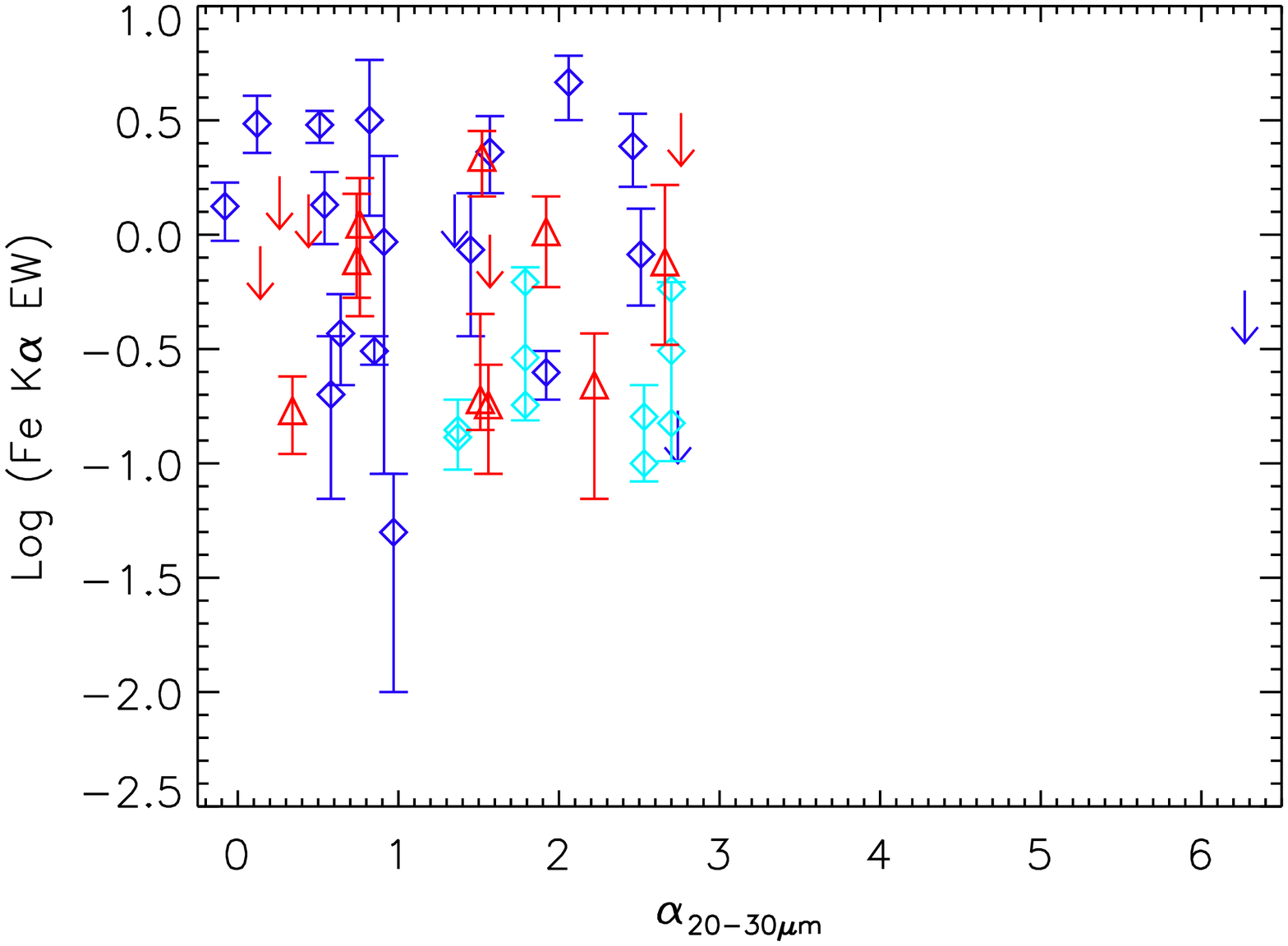}}
\caption{\label{alpha_c_thick}Obscuration diagnostics vs. alpha$_{20-30\mu m}$, which parametrizes star formation rate. The lower limits on F$_{2-10keV}$/F$_{[OIV]}$ are displayed for illustrative purposes and not included in the survival analysis. No statistically significant trends are apparent: $\rho$=0.272, 0.014, 0.060 and -0.235. Color coding same as Figure \ref{lx_l_iso}.}
\end{figure}


\clearpage

\begin{deluxetable}{lclllr}
\small
\tablewidth{0pt}
\tablecaption{\label{log}Sample and Observation Log}
\tablehead{
\colhead{Galaxy} & \colhead{Distance} & \colhead{Observatory} & \colhead{Observation Start Date} & 
\colhead{ObsID} & \colhead{Exposure Time\tablenotemark{1}} \\
\colhead{ } & \colhead{ } & \colhead{ } & \colhead{} & \colhead{} & \colhead{MOS1/MOS2/PN\tablenotemark{2}}\\
\colhead{ }   & \colhead{Mpc\tablenotemark{3}} & \colhead{} & \colhead{UT} & \colhead{} &\colhead{ks} }

\startdata

NGC 0424    & 51.2  & XMM     & 2001 Dec 12 & 00029242301 & 7.6/7.6/5.0 \\
            &       & Chandra & 2002 Feb 4  & 03146       & 9.1       \\
NGC 1068    & 16.9  & XMM     & 2000 Jul 29 & 0111200101  & 38.7/35.6/35.3 \\
            & 16.9  & XMM     & 2000 Jul 30 & 0111200201  & 37.8/35.0/32.2 \\

NGC 1144   & 120.8 & XMM     & 2006 Jan 28 & 0312190401  & 11.6/11.6/10.0 \\

NGC 1320    & 38.3  & XMM     & 2006 Aug 6  & 0405240201  & 16.8/16.8/13.7 \\

NGC 1386    & 12.7  & XMM     & 2002 Dec 29 & 0140950201  & 17.1/17.1/15.1 \\
            &       & Chandra & 2003 Nov 19 & 04076       & 19.6 \\

NGC 1667    & 64.1  & XMM     & 2004 Sep 20 & 0200660401  & 10.0/10.1/8.1  \\

F05189-2524 & 187.7 & XMM     & 2001 Mar 17 & 0085640101  & 10.7/10.6/7.6  \\
            &       & Chandra & 2001 Oct 30 & 02034       & 18.7 \\
            &       & Chandra & 2002 Jan 30 & 03432       & 14.9 \\

F08572+3915 & 256.0 & Chandra & 2006 Jan 26 & 06862      & 14.9 \\

NGC 3982    & 16.9  & XMM     & 2004 Jun 15 & 0204651201 & 11.5/11.5/9.7 \\
            &       & Chandra & 2004 Jan 3  & 04845      & 9.2 \\

NGC 4388    & 34.0  & Chandra & 2001 Jun 8  & 01619      & 20.0 \\
            &       & XMM     & 2002 Dec 12 & 0110930701 & 11.7/11.7/7.8 \\
            &       & XMM     & 2002 Jul 7  & 0110930301 & 9.0/9.2/2.8   \\

NGC 4501    & 34.0  & XMM     & 2001 Dec 4  & 0112550801 & 13.4/13.4/2.9 \\
            &       & Chandra & 2002 Dec 9  & 02922      & 17.9 \\

TOLOLO 1238-364  & 46.9 & Chandra & 2004 Mar 7 & 04844    & 8.7 \\

NGC 4968    & 42.6  & XMM     & 2001 Jan 5  & 0002940101 & 7.3/7.3/4.9 \\

            &       & XMM     & 2004 Jul 5  & 0200660201 & 4.5/4.7/5.2 \\

M-3-34-64   & 72.7  & XMM     & 2005 Jan 24 & 0206580101 & 44.6/44.6/42.9 \\

NGC 5135    & 59.8  & Chandra & 2001 Sep 4  & 02187      & 29.3 \\

NGC 5194    & 8.5 & Chandra  & 2000 Jun 20 & 00354      & 14.9 \\
            &     & Chandra  & 2001 Jun 23 & 01622      & 26.8 \\
            &     & Chandra  & 2003 Aug 7  & 03932      & 47.9 \\

NGC 5347    & 34.0  & Chandra & 2004 Jun 5 & 04867       & 36.9 \\

Mrk 463     & 219.4 & XMM     & 2001 Dec 12 & 0094401201 & 26.0/26.0/23.4 \\
             &      & Chandra & 2004 Jun 11 & 04913      & 49.3           \\

NGC 5506    & 25.5  & XMM     & 2001 Feb 2  & 0013140101 & 17.8/17.8/14.3 \\
            &       & XMM     & 2002 Jan 9  & 0013140201 & 13.2/13.2/10.6 \\
            &       & XMM     & 2004 Jul 11 & 0201830201 & 21.3/21.3/21.1 \\
            &       & XMM     & 2004 Jul 14 & 0201830301 & 20.2/20.2/19.7 \\
            &       & XMM     & 2004 Jul 22 & 0201830401 & 19.6/19.6/19.9 \\
            &       & XMM     & 2004 Aug 7  & 0201830501 & 20.2/20.2/20.0 \\
            &       & XMM     & 2008 Jul 27 & 0554170201 & 85.2/88.0/90.4 \\
            &       & XMM     & 2009 Jan 2  & 0554170101 & 75.1/76.0/87.0 \\

NGC 5953    & 29.7  & Chandra & 2002 Dec 12 & 04023      & 4.7 \\

Arp 220     & 77.1  & XMM     & 2002 Aug 11 & 0101640801 & 13.6/13.6/11.8 \\
            &       & XMM     & 2003 Jan 15 & 0101640901 & 14.6/14.6/9.3  \\
            &       & XMM     & 2005 Jan 14 & 0205510201 & 8.7/8.2/0.7    \\
            &       & XMM     & 2005 Feb 19 & 0205510401 & 8.1/8.3/4.3    \\
            &       & Chandra & 2000 Jun 6  & 00869      & 56.5           \\

NGC 6890    & 34.0  & XMM     & 2005 Sep 29 & 0301151001 & 9.3/9.2/2.4 \\

IC 5063     & 46.9  & Chandra & 2007 Jun 15 & 07878      & 34.1 \\

NGC 7130    & 68.4  & Chandra & 2001 Oct 23 & 02188      & 38.6 \\

NGC 7172    & 38.3  & XMM     & 2002 Nov 18 & 0147920601 & 13.6/13.6/12.0 \\
            &       & XMM     & 2004 Nov 11 & 0202860101 & 50.8/50.9/36.0 \\
            &       & XMM     & 2007 Apr 4  & 0414580101 & 48.9/48.8/31.7 \\

NGC 7582    & 21.2  & Chandra & 2000 Oct 14 & 00436      & 10.5 \\
            &       & Chandra & 2000 Oct 15 & 02319      & 5.9  \\
            &       & XMM     & 2001 May 25 & 0112310201 & 22.6/22.6/19.6 \\
            &       & XMM     & 2005 Apr 29 & 0204610101 & 80.2/79.7/71.8 \\

NGC 7590    & 21.2  & XMM     & 2007 Apr 30 & 0405380701 & 9.8/9.3/2.5 \\

NGC 7674    & 125.3 & XMM     & 2004 Jun 2  & 0200660101 & 8.4/9.2/8.3 \\

\enddata

\tablenotetext{1}{Net exposure time after filtering.}
\tablenotetext{2}{For {\it XMM-Newton} observations.}
\tablenotetext{3}{Distances based on optical spectroscopic redshift using $H_{0}$ = 70 km s$^{-1}$ Mpc$^{-1}$, $\Omega_M = 0.27$ and $\Omega_\Lambda$=0.73}
\end{deluxetable}

\clearpage
\begin{deluxetable}{lccccccc}
\footnotesize
\tablewidth{0pt}
\tablecaption{\label{apec}APEC model parameters (solar abundance)}
\tablehead{
\colhead{Galaxy} & \colhead{N$_{H,1}$} & \colhead{kT} & \colhead{$\Gamma$} & \colhead{$N_{H,2}$} &
\colhead{$\chi^2$} & \colhead{$\chi^2$ 2pow} & \colhead{$\chi^2$ 1pow }\\
\colhead{ }      & \colhead{10$^{22}$ cm$^{-2}$} & \colhead{keV} & \colhead{} & 
\colhead{10$^{22}$cm$^{-2}$} &\colhead{DOF} & \colhead{DOF} & \colhead{DOF} }

\startdata

NGC 0424\tablenotemark{2} & 0.05$^{+0.04}_{-0.03}$ & 0.82$^{+0.18}_{-0.17}$ & 2.85$^{+0.32}_{-0.28}$ & 16.8$^{+5.8}_{-3.5}$ &
 269.5 (171) & 273.8 (173) & 846.4 (178) \\

NGC 1068 & 0.31$^{+0.03}_{-0.03}$ & 0.61$^{+0.01}_{-0.01}$ & 2.02$^{+0.59}_{-0.45}$ & 9.33$^{+1.77}_{-2.58}$  &
450.4 (247) & 1013 (249) & 6634 (269) \\

NGC 1144\tablenotemark{1}  & 0.06 & 0.37$^{+0.29}_{-0.06}$ & 1.91$^{+0.37}_{-0.24}$ & 47.0$^{+3.5}_{-3.2}$  &
174.7 (149) & 216.8 (151) & 1347 (156) \\

NGC 1320 & 0.07$^{+0.03}_{-0.02}$ & 0.78$^{+0.07}_{-0.07}$ & 3.30$^{+0.22}_{-0.19}$ &  43.5$^{+81.5}_{-12.3}$ &
 269.1 (170) & 311.2 (172) & 639.9 (177) \\

NGC 1386\tablenotemark{3} & 0.04$^{+0.03}_{-0.02}$ & 0.66$^{+0.04}_{-0.03}$ & 2.97$^{+0.27}_{-0.22}$ & 35.8$^{+19.7}_{-13.3}$ &
 412.7 (340) & 591.1 (342) & 876.9 (347) \\

NGC 1667\tablenotemark{1} & 0.05 & 0.33$^{+0.07}_{-0.04}$ & 2.18$^{+0.34}_{-0.37}$ & ... &
49.8 (38) & ... & 82.3 (39) \\

F05189-2524\tablenotemark{1,3} &  0.02 & $<$0.104 & 2.08$^{+0.13}_{-0.13}$ & 6.75$^{+0.40}_{-0.41}$  & 
530.3 (376) & 607.6 (378) & 2212 (379) \\

NGC 3982\tablenotemark{4,7} & 0.53$^{+0.11}_{-0.16}$ & $<$0.12 & 0.57$^{+1.14}_{-0.90}$ & ... & 21.7 (16) & ... & 45.7 (18) \\

NGC 4388 (Chandra) & 1.47$^{+0.49}_{-0.53}$ & $<$0.18 & 0.92$^{+0.27}_{-0.45}$ & 29.2$^{+3.1}_{-4.3}$  & 
110.6 (92) & 121.7 (94) & 324.1 (99) \\
NGC 4388 (XMM)\tablenotemark{1,6}  & 0.03 & 0.30$^{+0.01}_{-0.01}$ & 1.35$^{+0.14}_{-0.10}$ & 26.2$^{+1.2}_{-0.9}$ &
580.2 (498) & 844.3 (500) & 3936 (506) \\

NGC 4501\tablenotemark{1,4,7} & 0.03 & 0.42$^{+0.16}_{-0.09}$ & 0.30$^{+0.45}_{-0.50}$  & ... & 27.5 (38) & ... & 58.9 (40) \\

TOLOLO 1238-364\tablenotemark{7} & 0.06 & 0.73$^{+0.11}_{-0.13}$ & 2.47$^{+0.31}_{-0.35}$  & ... & 51.0 (77) & ... & 72.6 (78) \\

NGC 4968\tablenotemark{7} & 0.84$^{+0.10}_{-0.08}$ & $<$0.13 & 1.50$^{+0.41}_{-0.31}$ & ... & 343.0 (267) & ... & 337.8 (270) \\

M-3-34-64 & 0.07$^{+0.01}_{-0.01}$ & 0.79$^{+0.02}_{-0.02}$ & 2.68$^{+0.10}_{-0.09}$ & 46.7$^{+1.6}_{-1.6}$ &
 847.5 (493) & 1660 (495) & 8590 (500)\\

NGC 5135\tablenotemark{1,7} & 0.05 & 0.77$^{+0.24}_{-0.22}$ & 2.78$^{+0.14}_{-0.12}$ & 104$^{+81}_{-70}$ &
194.8 (132) & 200.8 (134) & 317.8 (138) \\

NGC 5194\tablenotemark{1} & 0.02 &  0.65$^{+0.05}_{-0.04}$ & 2.20$^{+0.16}_{-0.17}$ & 90.1$^{+62.9}_{-43.1}$ & 
 274.0 (231) & 445.2 (232) & 944.9 (237) \\

NGC 5347 & 0.02 & $<$0.24 & 1.19$^{+0.24}_{-0.26}$ & 63.6$^{+37.4}_{-25.7}$  &
31.8 (22) & 36.9 (24) & 78.4 (26) \\

Mrk 463\tablenotemark{3} & $<$0.06 & 0.73$^{+0.03}_{-0.04}$ & 2.02$^{+0.27}_{-0.12}$ & 26.5$^{+4.9}_{-4.6}$  & 
334.2 (263) &  600.3 (268) & 1505 (270)  \\ 

NGC 5506\tablenotemark{8} & 0.11$^{+0.01}_{-0.01}$ & 0.77$^{+0.04}_{-0.05}$ & 1.71$^{+0.01}_{-0.01}$ & 2.68$^{+0.03}_{-0.03}$ &
 2720 (2385) & 2781 (2387) & 15637 (2389)\\

NGC 5506\tablenotemark{9} & 0.13$^{+0.01}_{-0.01}$ & 0.85$^{+0.10}_{-0.03}$ & 1.77$^{+0.01}_{-0.0}$ & 2.80$^{+0.01}_{-0.02}$ &
4171 (3143) & 4299 (3145) & 31991 (3147) \\

Arp 220 (XMM)\tablenotemark{1,5}  & 0.04 & 0.82$^{+0.05}_{-0.05}$ &  1.27$^{+0.15}_{-0.15}$ & ... &
146.0 (145) & ... & 248.4 (147) \\
Arp 220 (Chandra) & 0.47$^{+0.07}_{-0.06}$ & '' & '' &  ... & '' & ... & '' \\

NGC 6890\tablenotemark{7} & $<$0.10 & 0.78$^{+0.24}_{-0.19}$ & 3.28$^{+0.88}_{-0.74}$ & 27.4$^{+18.4}_{-11.3}$  &
164.0 (148) & 171.3 (150) & 197.4 (152) \\

IC 5063\tablenotemark{10} &  0.64$^{+0.26}_{-0.43}$ & 0.43$^{+0.17}_{-0.22}$ & 1.39$^{+0.41}_{-0.41}$ & 19.6$^{+2.3}_{-2.4}$ &
131.0 (116) & 135.6 (119) & 452.5 (120) \\

NGC 7130\tablenotemark{7}  & $<$0.08 & 0.76$^{+0.04}_{-0.04}$ & 2.41$^{+0.27}_{-0.26}$ & 64.1$^{+58.9}_{-23.3}$ & 
 220.7 (199) & 381.8 (201) & 563.9 (206) \\

NGC 7172\tablenotemark{1,6} & 0.02 & 0.26$^{+0.03}_{-0.02}$ & 1.55$^{+0.03}_{-0.01}$ & 7.74$^{+0.09}_{-0.08}$ & 
 2330 (1748) & 2530 (1750) & 5379 (1751) \\

NGC 7582 (XMM)\tablenotemark{1,5,12} & 0.01 & 0.71$^{+0.01}_{-0.01}$ & 1.95$^{+0.03}_{-0.02}$ & 26.0$^{+1.4}_{-1.5}$ & 
1586 (886) & 4044 (903) & 15004 (910) \\

NGC 7582 (Chandra)\tablenotemark{6} & 1.24$^{+0.07}_{-0.10}$ & $<$0.11 & 1.80$^{+0.42}_{-0.03}$ & 19.8$^{+2.3}_{-0.20}$ & 
 104.9 (81) & 117.5 (83) & 305.2 (85) \\

NGC 7674\tablenotemark{1}  & 0.04 & 0.70$^{+0.13}_{-0.09}$ & 2.92$^{+0.16}_{-0.15}$ & 34.7$^{+10.3}_{-7.3}$ &
 112.9 (72) & 129.6 (74) & 342.4 (75) \\

\enddata
\tablenotetext{1}{Best-fit $N_H$ was same as Galactic value and therefore frozen at this value.}
\tablenotetext{2}{Best-fit parameters between \textit{Chandra} and \textit{XMM-Newton} observations are consistent.}
\tablenotetext{3}{Best-fit parameters between \textit{Chandra} and \textit{XMM-Newton} observations are consistent except for the constant multiplicative factor, which is much lower for the \textit{Chandra} observations, indicating extended emission in the \textit{XMM} field of view.}
\tablenotetext{4}{Best-fit parameters between \textit{Chandra} and \textit{XMM-Newton} observations differ due to presence of extended
emission in \textit{XMM} field of view. Parameters for the \textit{Chandra} observation, which isolates the point source, are listed.}
\tablenotetext{5}{Best-fit parameters between \textit{Chandra} and \textit{XMM-Newton} observations differ due to variability.}
\tablenotetext{6}{Second power law component normalizations fit indepedently between the two \textit{XMM-Newton} observations.}
\tablenotetext{7}{Used c-stat.}
\tablenotetext{8}{\textit{XMM-Newton} observations from Feb 2, 2001; Jul 11, 2004; Jul 14, 2004 and Jul 22, 2004.}
\tablenotetext{9}{\textit{XMM-Newton} observations from Jan 9, 2002; Aug 7, 2004; Jul 27, 2008 and Jan 2, 2009.}
\tablenotetext{10}{Used pileup model.}
\end{deluxetable}

\clearpage
\begin{deluxetable*}{lcccc}
\footnotesize
\tablewidth{0pt}
\tablecaption{\label{pow}Power law model parameters}
\tablehead{
\colhead{Galaxy} & \colhead{N$_{H,1}$} & \colhead{$\Gamma$} & \colhead{$N_{H,2}$} &
  \colhead{$\chi^2$}  \\
\colhead{ }      & \colhead{10$^{22}$ cm$^{-2}$} &  \colhead{} & 
\colhead{10$^{22}$cm$^{-2}$} & \colhead{DOF} }

\startdata

NGC 0424\tablenotemark{1} &  0.07$^{+0.03}_{-0.03}$ & 2.97$^{+0.27}_{-0.26}$ & 16.9$^{+6.0}_{-3.0}$ & 273.8 (173)\\

NGC 4388 (Chandra) & 0.22$^{+0.24}_{-0.15}$ & 0.38$^{+0.39}_{-0.36}$ & 23.3$^{+3.5}_{-3.1}$ & 121.7 (94) \\

NGC 4968\tablenotemark{2,3} &  0.08 & 1.94$^{+0.14}_{-0.13}$  & ... & 337.8 (270) \\

NGC 5135\tablenotemark{2} & 0.05 & 2.75$^{+0.11}_{-0.10}$ & 118$^{+82}_{-60}$  & 200.8 (134) \\

NGC 5347\tablenotemark{2} & 0.02 & 1.41$^{+0.24}_{-0.22}$ & 56.2$^{+31.9}_{-22.9}$  & 36.9 (24) \\

NGC 5953\tablenotemark{2,3,4} & 0.03 & 2.10$^{+0.63}_{-0.65}$  & ... & 39.9 (21) \\

NGC 6890\tablenotemark{3} & 0.21$^{+0.11}_{-0.09}$ & 3.86$^{+0.75}_{-0.64}$ & 18.9$^{+16.5}_{-11.0}$ & 171.3 (150) \\

IC 5063\tablenotemark{2,5} & 0.06 & 1.48$^{+0.26}_{-0.25}$ & 20.5$^{+1.4}_{-1.4}$ & 135.6 (119) \\

NGC 7582 (Chandra)\tablenotemark{5} & $<$0.23 & 1.63$^{+0.50}_{-0.40}$ & 18.8$^{+2.9}_{-2.1}$  & 117.5 (83) \\

NGC 7674\tablenotemark{2} & 0.04 & 2.86$^{+0.12}_{-0.11}$ & 36.9$^{+12.4}_{-7.7}$  & 129.6 (74) \\

\enddata
\tablenotetext{1}{Best-fit parameters between \textit{Chandra} and \textit{XMM-Newton} observations are consistent.}
\tablenotetext{2}{Best-fit $N_H$ was same as Galactic value and therefore frozen at this value.}
\tablenotetext{3}{Used c-stat.}
\tablenotetext{4}{Only detected in soft band}
\tablenotetext{5}{Used pileup model.}
\end{deluxetable*}

\begin{deluxetable*}{lrrl}
\scriptsize
\tablewidth{0pt}
\tablecaption{\label{flux} 2 - 10 keV X-ray Flux and Luminosity for 12$\mu$m Sample}
\tablehead{
\colhead{Galaxy} & \colhead{F$_{2-10keV}$}  & \colhead{Log L$_{2-10keV}$} & \colhead{Comments} \\
                 & \colhead{10$^{-13}$ erg/s/cm$^{2}$} & \colhead{erg/s}}

\startdata

NGC 0424  & 11.5$^{+6.4}_{-3.7}$   & 41.56$^{+0.19}_{-0.17}$ \\

NGC 1068  & 54.2$^{+110}_{-32.0}$  & 41.27$^{+0.48}_{-0.39}$ \\

NGC 1144  & 33.4$^{+36.6}_{-12.9}$ & 42.77$^{+0.32}_{-0.21}$ \\

NGC 1320  & 3.84$^{+2.83}_{-1.39}$ & 40.83$^{+0.24}_{-0.20}$ \\

NGC 1386  & 1.55$^{+0.42}_{-0.50}$ & 39.48$^{+0.10}_{-0.17}$ & Chandra observation \\

NGC 1667  & 0.43$^{+0.11}_{-0.11}$ & 40.33$^{+0.10}_{-0.13}$ & \\

F05189-2524   & 23.5$^{+5.5}_{-4.9}$ & 43.00$^{+0.09}_{-0.10}$ & Chandra observations \\

F08572+3915   & $<$1.26            & $<$42.02  \\

NGC 3982  & 0.56$^{+1.28}_{-0.39}$ & 39.28$^{+0.52}_{-0.52}$ & Chandra observation \\

NGC 4388  & 74.6$^{+88.5}_{-38.5}$ & 42.01$^{+0.34}_{-0.32}$ & Chandra observation \\
          & 86.9$^{+28.6}_{-17.7}$ & 42.08$^{+0.12}_{-0.10}$ & XMM Jul 2002 observation \\
          & 244$^{+76}_{-47}$      & 42.53$^{+0.12}_{-0.09}$ & XMM Dec 2002 observation \\

NGC 4501  & 1.07$^{+0.73}_{-0.51}$ & 40.17$^{+0.23}_{-0.28}$ & Chandra observation\\

TOLOLO 1238-364  & 1.21$^{+0.31}_{-0.26}$ & 40.50$^{+0.10}_{-0.11}$ \\

NGC 4968  & 2.08$^{+0.26}_{-0.26}$ & 40.65$^{+0.05}_{-0.06}$ \\

M-3-34-64 & 32.5$^{+3.1}_{-3.1}$   & 42.31$^{+0.04}_{-0.04}$ \\

NGC 5135  & 2.31$^{+0.98}_{-1.68}$ & 40.99$^{+0.15}_{-0.56}$ \\

NGC 5194  & 1.04$^{+2.28}_{-0.73}$ & 38.95$^{+0.50}_{-0.53}$ \\

NGC 5347  & 2.58$^{+1.20}_{-1.41}$ & 40.55$^{+0.17}_{-0.34}$ \\

Mrk 463   & 2.95$^{+1.84}_{-0.82}$ & 42.23$^{+0.21}_{-0.14}$ & Chandra observation \\

NGC 5506  & 725$^{+68}_{-79}$      & 42.75$^{+0.04}_{-0.05}$ & 2001 \& Jul 2004 observations \\
          & 1113$^{+59}_{-59}$     & 42.94$^{+0.02}_{-0.02}$ & 2002, Aug 2004, 2008 \& 2009 observations \\

NGC 5953  & $<$0.51                & $<$39.73 \\

Arp 220   & 1.07$^{+0.18}_{-0.16}$ & 40.88$^{+0.07}_{-0.07}$ \\

NGC 6890  & 1.20$^{+4.01}_{-0.88}$ &  40.22$^{+0.64}_{00.57}$ \\

IC 5063   & 134$^{+73}_{-47}$      & 42.55$^{+0.19}_{-0.19}$ \\

NGC 7130  & 2.07$^{+2.09}_{-1.04}$ & 41.06$^{+0.30}_{-0.30}$ \\

NGC 7172  & 517$^{+43}_{-40}$      & 42.96$^{+0.03}_{-0.03}$ & 2007 observation \\
          & 234$^{+19}_{-18}$      & 42.61$^{+0.03}_{-0.03}$ & 2002 \& 2004 observations \\

NGC 7582  & 21.1$^{+1.7}_{-1.8}$   & 41.05$^{+0.03}_{-0.04}$ & 2005 XMM observation \\
          & 38.6$^{+3.0}_{-3.1}$   & 41.32$^{+0.03}_{-0.04}$ & 2001 XMM observation \\
          & 164$^{+263}_{-87}$     & 41.95$^{+0.42}_{-0.33}$ & Chandra observations \\

NGC 7590  & $<$2.72                & $<$40.17 \\

NGC 7674  & 5.71$^{+3.05}_{-1.69}$ & 42.03$^{+0.19}_{-0.15}$ \\

\enddata
\end{deluxetable*}

\begin{deluxetable*}{lcccccccc}
\small
\tablewidth{0pt}
\tablecaption{\label{kalpha}Fe K$\alpha$ Flux and EW}
\tablehead{
\colhead{Galaxy} & \multicolumn{4}{c}{Global Fit}  &  \multicolumn{4}{c}{Local Fit} \\
& \colhead{Energy} & \colhead{$\sigma$} & \colhead{EW} & \colhead{Flux\tablenotemark{1}} &  \colhead{Energy} &
 \colhead{$\sigma$} & \colhead{EW} & \colhead{Flux\tablenotemark{1}}  \\
\cline{2-4}  \cline{6-8}
& \multicolumn{3}{c}{keV} &  & \multicolumn{3}{c}{keV} &  }

\startdata

NGC 0424\tablenotemark{2}  & 6.45$^{+0.07}_{-0.07}$ & 0.42$^{+0.17}_{-0.11}$ & 4.22$^{+0.88}_{-0.97}$ & 3.95$^{+0.83}_{-0.91}$ &
6.36$^{+0.08}_{-0.05}$ & 0.21$^{+0.11}_{-0.17}$ & 1.33$^{+0.36}_{-0.39}$ & 2.37$^{+0.64}_{-0.69}$  \\

NGC 1068\tablenotemark{2} & 6.40$^{+0.00}_{-0.01}$ & $<$0.03 & 0.65$^{+0.05}_{-0.05}$ & 5.52$^{+0.41}_{-0.40}$ & 
 6.4$^{+0.00}_{-0.01}$ & $<$0.03 & 0.60$^{+0.05}_{-0.05}$ & 5.95$^{+0.47}_{-0.47}$ \\

NGC 1144\tablenotemark{3} & 6.24$^{+0.02}_{-0.02}$ & $<$0.07 & 0.26$^{+0.06}_{-0.06}$ &  1.99$^{+0.49}_{-0.44}$ &
6.24$^{+0.02}_{-0.02}$ & $<$0.07 & 0.25$^{+0.06}_{-0.06}$ & 1.89$^{+0.48}_{-0.43}$ \\

NGC 1320\tablenotemark{2} & 6.37$^{+0.02}_{-0.02}$ & 0.06$^{+0.03}_{-0.03}$ & 3.50$^{+0.49}_{-0.47}$ & 1.55$^{+0.22}_{-0.21}$ &
6.37$^{+0.02}_{-0.01}$ & 0.05$^{+0.02}_{-0.04}$ & 3.02$^{+0.46}_{-0.50}$ & 1.70$^{+0.26}_{-0.28}$ \\

NGC 1386 (Chandra)\tablenotemark{2} & 6.39$^{+0.02}_{-0.03}$ & $<$0.05 & 3.43$^{+1.39}_{-1.39}$ & 0.51$^{+0.21}_{-0.21}$ & 
6.39$^{+0.03}_{-0.03}$ & $<$0.05 & 2.30$^{+1.00}_{-0.78}$ & 0.72$^{+0.32}_{-0.25}$ \\

NGC 1667\tablenotemark{5,6} & 6.31 & 0.01 & - & 0.16$^{+0.10}_{-0.10}$ & 6.31 & 0.01 & 0.86$^{+0.66}_{-0.50}$ & 0.13$^{+0.10}_{-0.08}$  \\

F05189-25212 (Chandra)\tablenotemark{6} & 6.14 & 0.01 & 0.09$^{+0.09}_{-0.08}$ & 0.28$^{+0.27}_{-0.24}$ & 
6.14 & 0.01 & $<$0.17 & $<$0.59 \\

NGC 3982 (Chandra)\tablenotemark{6} & 6.37 & 0.01 & - & 0.31$^{+0.41}_{-0.22}$ & 
6.37 & 0.01 & - & 0.11$^{+0.13}_{-0.11}$  \\

NGC 4388 (Chandra)\tablenotemark{2} & 6.34$^{+0.02}_{-0.03}$ & $<$0.09 & 0.31$^{+0.08}_{-0.08}$ & 4.03$^{+1.06}_{-1.03}$ &
6.34$^{+0.02}_{-0.02}$ & $<$0.08 & 0.29$^{+0.11}_{-0.08}$ & 3.72$^{1.47}_{-0.99}$ \\
NGC 4388 (XMM Jul 2002)\tablenotemark{2} & 6.37$^{+0.01}_{-0.01}$ & $<$0.09 & 0.46$^{+0.08}_{-0.08}$ & 7.05$^{+1.19}_{-1.16}$ & 
6.37$^{+0.02}_{-0.02}$ & 0.08$^{+0.03}_{-0.03}$ & 0.62$^{+0.10}_{-0.10}$ & 8.69$^{+1.41}_{-1.35}$ \\
NGC 4388 (XMM Dec 2002)\tablenotemark{2}   & '' & 0.06$^{+0.02}_{-0.02}$ & 0.20$^{+0.03}_{-0.03}$ & 9.20$^{+1.24}_{-1.21}$ &
6.37$^{+0.01}_{-0.01}$ & 0.05$^{+0.02}_{-0.02}$ & 0.18$^{+0.03}_{-0.02}$ & 8.39$^{+1.26}_{-1.14}$  \\

NGC 4501 (Chandra)\tablenotemark{6} & 6.35 & 0.01 & $<$2.28 & $<$0.29 & 
6.35 & 0.01 & $<$1.50 & $<$0.31 \\

TOLOLO 1238-364\tablenotemark{3}  & 6.30$^{+0.29}_{-0.23}$ & 0.39$^{+0.34}_{-0.28}$ & - & 0.70$^{+0.38}_{-0.30}$ &
6.38$^{+0.09}_{-0.10}$ & $<$0.25 & 3.17$^{+2.64}_{-1.96}$ & 0.56$^{+0.47}_{-0.35}$ \\

NGC 4968\tablenotemark{2}  &  6.38$^{+0.08}_{-0.03}$ & 0.13$^{+0.16}_{-0.05}$ & - & 0.95$^{+0.23}_{-0.20}$ &
6.37$^{+0.03}_{-0.02}$ & 0.07$^{+0.04}_{-0.04}$ & 3.06$^{+0.99}_{-0.78}$ & 0.91$^{+0.29}_{-0.23}$ \\

M-3-34-64\tablenotemark{4} & 6.30$^{+0.02}_{-0.01}$ & $<$0.08 & 0.17$^{+0.04}_{-0.03}$ & 1.27$^{+0.32}_{-0.21}$ & 
6.30$^{+0.02}_{-0.01}$ & 0.10$^{+0.04}_{-0.03}$ & 0.31$^{+0.05}_{-0.04}$ & 1.78$^{+0.31}_{-0.23}$ \\

NGC 5135\tablenotemark{2}  &  6.34$^{+0.04}_{-0.04}$ & $<$0.12 & 1.18$^{+0.56}_{-0.45}$ & 0.46$^{+0.22}_{-0.18}$ &
6.35$^{+0.04}_{-0.05}$ & 0.09$^{+0.08}_{-0.06}$ & 2.44$^{+0.94}_{-0.82}$ & 0.61$^{+0.23}_{-0.20}$\\

NGC 5194\tablenotemark{2} & 6.39$^{+0.02}_{-0.01}$ & $<$0.04 & 3.05$^{+0.82}_{-0.65}$ & 0.40$^{+0.11}_{-0.08}$ &
6.39$^{+0.02}_{-0.01}$ & $<$0.05 & 4.64$^{+1.42}_{-1.47}$ & 0.57$^{+0.17}_{-0.18}$ \\

NGC 5347\tablenotemark{2,6} & 6.35 & 0.01 & 1.04$^{+0.49}_{-0.45}$ & 0.37$^{+0.17}_{-0.16}$ & 
6.35 & 0.01 & 1.35$^{+0.53}_{-0.44}$ & 0.45$^{+0.18}_{-0.15}$ \\

Mrk 463 (Chandra)\tablenotemark{3,6}  & 6.10 & 0.01 & $<$0.38 & $<$0.27 & 
6.10 & 0.01 & 0.20$^{+0.16}_{-0.13}$ & 0.15$^{+0.12}_{-0.10}$ \\

NGC 5506 (2001 Feb 2)\tablenotemark{2}  & 6.38$^{+0.01}_{-0.01}$ & 0.09$^{+0.02}_{-0.01}$ & 0.11$^{+0.01}_{-0.01}$ & 8.02$^{+0.69}_{-0.68}$ &
6.38$^{+0.02}_{-0.01}$ & 0.08$^{+0.03}_{-0.02}$ & 0.12$^{+0.02}_{-0.02}$ & 7.35$^{+1.15}_{-1.06}$ \\ 

NGC 5506 (2004 Jul 11)\tablenotemark{2} & '' & '' & '' & '' & 
6.40$^{+0.03}_{-0.03}$ & 0.20$^{+0.06}_{-0.05}$ & 0.17$^{+0.03}_{-0.03}$ & 10.9$^{+2.2}_{-1.9}$ \\

NGC 5506 (2004 Jul 14)\tablenotemark{2} & '' & '' & '' & '' & 
6.38$^{+0.02}_{-0.02}$ & 0.10$^{+0.03}_{-0.03}$ & 0.13$^{+0.02}_{-0.02}$ & 8.30$^{+1.37}_{-1.15}$ \\

NGC 5506 (2004 Jul 22)\tablenotemark{2} & '' & '' & '' & '' & 
6.38$^{+0.03}_{-0.02}$ & 0.11$^{+0.05}_{-0.04}$ & 0.14$^{+0.03}_{-0.04}$ & 7.90$^{+1.85}_{-2.06}$ \\

NGC 5506 (2002 Jan 9)\tablenotemark{2}  & 6.45$^{+0.01}_{-0.01}$ & 0.13$^{+0.02}_{-0.01}$ & 0.10$^{+0.01}_{-0.01}$ & 10.3$^{+0.8}_{-0.7}$ &
6.42$^{+0.03}_{-0.04}$ & 0.11$^{+0.08}_{-0.07}$ & 0.09$^{+0.03}_{-0.03}$ & 8.75$^{+2.74}_{-3.22}$ \\

NGC 5506 (2004 Aug 7)\tablenotemark{2}  & '' & '' & '' & '' &
6.38$^{+0.04}_{-0.03}$ & 0.17$^{+0.07}_{-0.04}$ & 0.12$^{+0.03}_{-0.02}$ & 11.1$^{+3.2}_{-1.9}$ \\

NGC 5506 (2008 Jul 27)\tablenotemark{2} & '' & '' & '' & '' &
6.46$^{+0.02}_{-0.02}$ & 0.15$^{+0.04}_{-0.02}$ & 0.13$^{+0.02}_{-0.01}$ & 11.7$^{+1.8}_{-1.4}$ \\

NGC 5506 (2009 Jan 2)\tablenotemark{2}  & '' & '' & '' & ''& 
6.51$^{+0.03}_{-0.02}$ & 0.28$^{+0.05}_{-0.04}$ & 0.18$^{+0.02}_{-0.02}$ & 18.0$^{+2.4}_{-2.1}$ \\

Arp 220\tablenotemark{6}  &  6.29 & 0.01 & $<$0.66 & $<$0.09 &
6.29 & 0.01 & $<$0.57 & $<$0.09 \\

NGC 6890\tablenotemark{5,6} &  6.35 & 0.01 & 1.21$^{+1.46}_{-1.01}$ & 0.15$^{+0.18}_{-0.13}$ &
6.35 & 0.01 & 0.93$^{+1.28}_{-0.84}$ & 0.17$^{+0.23}_{-0.15}$ \\

IC 5063\tablenotemark{3,6} & 6.33 & 0.01 & 0.05$^{+0.04}_{-0.04}$ & 1.14$^{+0.96}_{-0.94}$ & 
6.33 & 0.01 & 0.05$^{+0.04}_{-0.04}$ & 0.58$^{+0.42}_{-0.42}$ \\

NGC 7130\tablenotemark{3} & 6.30$^{+0.04}_{-0.04}$ & $<$0.09 & 0.70$^{+0.39}_{-0.31}$ &  0.20$^{+0.11}_{-0.09}$ & 
6.30$^{+0.04}_{-0.04}$ & $<$0.10 & 0.82$^{+0.48}_{-0.33}$ & 0.26$^{+0.15}_{-0.10}$ \\

NGC 7172 (2007)\tablenotemark{2}  & 6.33$^{+0.02}_{-0.01}$ & $<$0.06 & 0.05$^{+0.01}_{-0.01}$ & 3.10$^{+0.56}_{-0.48}$ &
6.33$^{+0.02}_{-0.02}$ & 0.11$^{+0.04}_{-0.03}$ & 0.10$^{+0.02}_{-0.01}$ & 5.48$^{+1.00}_{-0.68}$ \\
NGC 7172 (2004)\tablenotemark{2}  & 6.38$^{+0.01}_{-0.01}$ & 0.09$^{+0.02}_{-0.02}$ & 0.12$^{+0.01}_{-0.01}$ & 4.11$^{+0.49}_{-0.49}$ &
6.37$^{+0.02}_{-0.01}$ & 0.09$^{+0.02}_{-0.02}$ &  0.12$^{+0.01}_{-0.01}$ & 3.30$^{+0.38}_{-0.40}$ \\
NGC 7172 (2002)\tablenotemark{2}  & 6.37$^{+0.03}_{-0.04}$ & 0.14$^{+0.06}_{-0.04}$ & 0.14$^{+0.03}_{-0.03}$ & 4.88$^{+1.17}_{-0.91}$ & 
6.35$^{+0.04}_{-0.06}$ & 0.21$^{+0.11}_{-0.06}$ & 0.20$^{+0.06}_{-0.04}$ & 5.51$^{+1.61}_{-1.17}$  \\

NGC 7582 (XMM 2005)\tablenotemark{2} & 6.37$^{+0.01}_{-0.01}$ & $<$0.04 & 0.41$^{+0.03}_{-0.03}$ & 1.97$^{+0.16}_{-0.16}$ & 
6.38$^{+0.0}_{-0.01}$ & 0.05$^{+0.01}_{-0.01}$ & 0.58$^{+0.04}_{-0.04}$ & 2.40$^{+0.17}_{-0.17}$ \\
NGC 7582 (XMM 2001)\tablenotemark{2} & 6.37$^{+0.02}_{-0.01}$ & 0.11$^{+0.03}_{-0.02}$  & 0.62$^{+0.08}_{-0.07}$ & 3.93$^{+0.52}_{-0.45}$ &
6.37$^{+0.01}_{-0.01}$ & $<$0.07 & 0.31$^{+0.05}_{-0.05}$ & 2.50$^{+0.38}_{-0.38}$ \\
NGC 7582 (Chandra)\tablenotemark{2,6}  & 6.37 & 0.01 & 0.18$^{+0.10}_{-0.07}$ & 4.37$^{+2.31}_{-1.62}$ &
6.37 & 0.01 & 0.15$^{+0.07}_{-0.07}$ & 1.62$^{+0.76}_{-0.75}$ \\

NGC 7674\tablenotemark{2,6} & 6.22 & 0.01 & 0.58$^{+0.25}_{-0.27}$ & 0.49$^{+0.21}_{-0.23}$ & 
6.22 & 0.01 & 0.37$^{+0.18}_{-0.15}$ & 0.48$^{+0.23}_{-0.20}$  \\

\enddata
\tablenotetext{1}{Flux in units of 10$^{-13}$erg s$^{-1}$cm$^{-2}$. Line energies are reported in observed frame. Upper limits on parameters
refer to the 90\% confidence level whereas upper limits on the EW and flux signify 3$\sigma$ error bars. ``-'' denotes unconstrained parameter.}
\tablenotetext{2}{Fe K$\alpha$ line detected at greater than the 3$\sigma$ level.}
\tablenotetext{3}{Fe K$\alpha$ line detected at greater than the 2.5$\sigma$ level.}
\tablenotetext{4}{Fe K$\alpha$ line detected at greater than the 2$\sigma$ level.}
\tablenotetext{5}{Fe K$\alpha$ line detected at greater than the 1.5$\sigma$ level.}
\tablenotetext{6}{XSpec model ZGAUSS used, with E frozen at 6.4 keV (rest-frame) and $\sigma$ frozen at 0.01 keV.}
\end{deluxetable*}

\clearpage

\begin{deluxetable*}{lrrr}
\scriptsize
\tablewidth{0pt}
\tablecaption{\label{obs_table} Obscuration Diagnostic Ratios}
\tablehead{
\colhead{Galaxy} & \colhead{Log($\frac{F_{2-10keV}}{F_{[OIII],obs}}$)}  & \colhead{Log($\frac{F_{2-10keV}}{F_{[OIV]}}$)} & 
\colhead{Log($\frac{F_{2-10keV}}{F_{MIR}}$)} }

\startdata

NGC 0424      & 0.22    & 0.66    & -2.20 \\

NGC 1068      & -0.40   & -0.51   & -     \\

NGC 1144      & 1.89    & 1.83    & -1.11 \\
 
NGC 1320      & 0.44    & 0.25    & -2.32 \\

NGC 1386      & -0.71   & -0.72   & -2.63 \\

NGC 1667      & -0.17   & 0.18    & -2.81 \\

F05189-2524   & 1.19    & $<$0.53 & -1.53 \\

F08572+3915   & $>$2.02 & -       &$>$-2.52 \\

NGC 3982      & -0.55   & 0.14    & -2.63 \\

NGC 4388      & 0.97    & 0.45    & -0.94 \\
              & 1.04    & 0.52    & -0.87 \\
              & 1.48    & 0.97    & -0.42 \\

NGC 4501         & 0.46  & 0.60  & -2.12  \\

TOLOLO 1238-364  & -0.58 & -0.09 & -2.99  \\

NGC 4968         & 0.07  & -0.10 & -2.57  \\

M-3-34-64        & 0.33  & 0.55  & -1.56  \\

NGC 5135         & -0.21 & -0.39 & -2.26  \\ 

NGC 5194         & 0.31  & -0.42 & -2.89  \\

NGC 5347         & 0.77  & 0.61  & -2.19  \\

Mrk 463          & -0.28 & -0.27 & -2.36  \\

NGC 5506         & 1.76 & 1.53 & -0.37 \\
                 & 1.94 & 1.71 & -0.19 \\

NGC 5953 & $>$0.03 & $>$-0.49 & $>$-2.91 \\

Arp 220  & 1.77    & $<$-0.92 & -2.84    \\

NGC 6890 & -0.29   & 0.17     & -2.53    \\

IC 5063  & 0.88    & 1.14     & -1.03    \\

NGC 7130 & -0.22   & 0.12     & -2.33    \\

NGC 7172 & 3.15    & 2.17     & 0.22     \\
         & 2.80    & 1.83     & -0.13    \\

NGC 7582 & 0.61    & 0.38     & -1.38   \\
         & 0.88    & 0.65     & -1.12   \\
         & 1.50    & 1.27     & -0.49   \\

NGC 7590 & $>$0.97 & $>$0.88 & $>$-1.82  \\

NGC 7674 & -0.10   & 0.19    & -2.29     \\

\enddata
\end{deluxetable*}

\clearpage

\begin{table}[h]
\small
\caption{\label{host_prob} Correlation of AGN Properties and Star Formation Activity with
Obscuration Diagnostics}
\begin{tabular}{lcc}
\hline
\hline
L$_{[OIV]}$ vs. & $\rho$ & P$^{1}$ \\
\hline
F$_{2-10keV}$/F$_{[OIII],obs}$ & 0.273  & 0.053 \\
F$_{2-10keV}$/F$_{[OIV]}$      & 0.185  & 0.206 \\
F$_{2-10keV}$/F$_{MIR}$        & 0.349  & 0.015 \\
Fe K$\alpha$ EW                & -0.302 & 0.048 \\
\hline
L$_{[OIV]}$/M$_{BH}$ vs. \\
\hline
F$_{2-10keV}$/F$_{[OIII],obs}$ & 0.110  & 0.439 \\
F$_{2-10keV}$/F$_{[OIV]}$      & 0.056  & 0.703 \\
F$_{2-10keV}$/F$_{MIR}$        & 0.280  & 0.050 \\
Fe K$\alpha$ EW                & -0.219 & 0.151 \\
\hline
M$_{BH}$ vs. \\
\hline
F$_{2-10keV}$/F$_{[OIII],obs}$ & 0.148  & 0.295 \\
F$_{2-10keV}$/F$_{[OIV]}$      & 0.115  & 0.432 \\
F$_{2-10keV}$/F$_{MIR}$        & -0.012 & 0.932 \\
Fe K$\alpha$ EW                & -0.234 & 0.124 \\
\hline
F$_{[OIV]}$/F$_{[NeII]}$ vs.\\
\hline
F$_{2-10keV}$/F$_{[OIII],obs}$ & 0.036  & 0.804  \\
F$_{2-10keV}$/F$_{[OIV]}$      & 0.016  & 0.913  \\
F$_{2-10keV}$/F$_{MIR}$        & 0.269  & 0.067  \\
Fe K$\alpha$ EW                & -0.130 & 0.404 \\
\hline
PAW EW 17 $\mu$m vs.  \\
\hline
F$_{2-10keV}$/F$_{[OIII],obs}$ & 0.135  & 0.346 \\
F$_{2-10keV}$/F$_{[OIV]}$      & 0.062  & 0.675 \\
F$_{2-10keV}$/F$_{MIR}$        & -0.055 & 0.700 \\
Fe K$\alpha$ EW                & -0.192 & 0.213 \\
\hline
$\alpha_{20-30\mu m}$  vs.  \\
\hline
F$_{2-10keV}$/F$_{[OIII],obs}$ & 0.272  & 0.057 \\
F$_{2-10keV}$/F$_{[OIV]}$      & 0.014  & 0.922 \\
F$_{2-10keV}$/F$_{MIR}$        & 0.060  & 0.675 \\
Fe K$\alpha$ EW                & -0.235 & 0.127 \\
\hline
\hline
\multicolumn{3}{l}{$^1$Probability that the null hypothesis,}\\
\multicolumn{3}{l}{that the two quantities are uncorrelated,}\\
\multicolumn{3}{l}{is correct. Quantities are statistically}\\
\multicolumn{3}{l}{significantly correlated if P$<$0.05.}\\
 \end{tabular}
\end{table}

\clearpage

\appendix
\section{Notes on Individual Sources}

\textit{NGC 424} - The default \textit{Chandra} aperture extraction spectrum and \textit{XMM-Newton} spectra were consistent and fit simultaneously using a double absorbed power law with a Gaussian component to accommodate the Fe K$\alpha$ emission; including a thermal component did not statistically significantly improve the fit. We note that fitting the 3 - 8 keV continuum with a power law plus a Gaussian led to a tighter constraint on the Fe K$\alpha$ emission and a more realistic EW value. Matt et al. (2003) analyzed the \textit{Chandra} and \textit{XMM-Newton} spectra independently using a slightly more complicated model, including Gaussian components at 0.55 and 0.90 keV to account for emission features, possibly from the OVIII recombination line and the OVIII recombination continuum or Ne IX recombination line, respectively. They also added a component for cold reflection, the PEXRAV model in XSpec. However, their 2-10 keV flux (1.6$\times10^{-12}$ erg s$^{-1}$ cm$^{-2}$) and Fe K$\alpha$ EW values ($\sim$0.88 keV) are consistent with the values we obtained (1.15$^{+0.64}_{-0.37} \times 10^{-12}$ erg s$^{-1}$ cm$^{-2}$ and 1.33$^{+0.36}_{-0.39}$ keV, respectively) using a simpler model.
\\

\textit{NGC 1068} - The PN and MOS1 \textit{XMM-Newton} spectra for both observations showed evidence of pileup according to \textit{epatplot} and were therefore not included in the spectral fit; archival \textit{Chandra} data also exists for NGC 1068, but was not used in this study due to the effects of pileup. Also, as several strong emission features were present below 1 keV (which are not important for the purposes of this study), we fitted the spectrum from 1 keV to 8 keV. The MOS2 spectra were best fit by an absorbed double power law with a thermal model and Gaussian components at 2.0 keV, 2.43 keV, 6.4 keV (neutral Fe K$\alpha$), 6.66 keV (likely ionized Fe K$\alpha$) and 6.95 keV. The neutral Fe K$\alpha$ line was detected at a statistically significant level. Pounds \& Vaughan fitted the 3.5 - 15 keV \textit{XMM-Newton} spectra with two continuum components, a cold reflection model (PEXRAV) and a series of Gaussian emission features from 6 - 8 keV (where nine of these features were detected at a significant level). Based on this fit, they find a 3-15 keV flux of 63 $\times 10^{-13}$ erg s$^{-1}$ cm$^{-2}$, which is consistent with our 2 - 10 keV flux of 54.2$^{+110}_{-32.0} \times 10^{-13}$ erg s$^{-1}$ cm$^{-2}$, and a neutral Fe K$\alpha$ EW of 0.60$\pm$0.10 keV, which agrees with our neutral Fe K$\alpha$ EW value (0.60$\pm$0.05 keV). However, as noted in the main text, \citet{M04} obtained an EW value of 1.2 keV, using a PEXRAV model, power law and a series of Gaussians for emission line features.
\\

\textit{NGC 1144} - The \textit{XMM-Newton} spectra were best fit by a double absorbed power law with a thermal component and a Gaussian component for the Fe K$\alpha$ emission. When fitting the 3 - 8 keV continuum to obtain a local fit for the Fe K$\alpha$ component, a double power law was needed to accommodate the spectrum shape; the power law indices of the two components were tied together with the normalizations and an absorption component attenuating the second power law allowed to vary. Winter et al. (2008) fit NGC 1144 with the partial covering model in XSpec (which is akin to a double absorbed power law model with the photon indices tied together, which we have done) and a blackbody component for the soft emission. They derived comparable 2 - 10 keV flux (3$\times 10^{-12}$ erg s$^{-1}$ cm$^{-2}$) and Fe K$\alpha$ EW (0.22 keV) values as us (33.4$^{+36.6}_{-12.9} \times 10^{-13}$ erg s$^{-1}$ cm$^{-2}$ and 0.25$^{+0.06}_{-0.06}$ keV, respectively).
\\

\textit{NGC 1320} - The \textit{XMM-Newton} spectrum to be best fit by an absorbed double power law component with a thermal component and a Gaussian component to accommodate the Fe K$\alpha$ emission. Greenhill et al. (2008) used a cold reflection model (PEXRAV) as well as two thermal components (using MEKAL whereas we used APEC) for the soft emission; a Gaussian had also been included to model the Fe K$\alpha$ emission. We derive consistent Fe K$\alpha$ EWs (3.02$^{+0.46}_{-0.50}$ keV vs. 2.20$^{+0.44}_{-0.43}$ keV), yet their 2 - 10 keV flux is over an order of magnitude higher than ours (3.84$^{+2.83}_{-1.39} \times10^{-13}$ erg s$^{-1}$ cm$^{-2}$ vs. $\sim$43$\times 10^{-13}$ erg s$^{-1}$ cm$^{-2}$).
\\

\textit{NGC 1386} - As noted in the main text, the spectral parameters between the \textit{XMM-Newton} and \textit{Chandra} observations were consistent, except for a lower multiplicative factor for the \textit{Chandra} observation, indicating extended emission contaminated the \textit{XMM-Newton} observation. Though the \textit{Chandra} data were grouped by a minimum of 5 counts per bin, the \textit{XMM-Newton} spectra were grouped by a minimum of 15 counts, so we used $\chi^2$ statistics in this analysis. To derive the 2 - 10 keV flux, we fitted spectra from both observatories simultaneously to better constrain the \textit{Chandra} spectrum, using a double absorbed power law with a thermal component and a Gaussian feature at 6.4 keV to accommodate the Fe K$\alpha$ emission, yet we report the \textit{Chandra} flux only. We fit the \textit{Chandra} spectrum independently, both globally and locally, to derive the Fe K$\alpha$ parameters, binning the data by a minimum of three counts and using C-stat. Levenson et al. (2006) fit the \textit{Chandra} 4 - 8 keV nuclear spectrum with a reflection model (PEXRAV) and a Gaussian component at the Fe K$\alpha$ energy. We obtain consistent 2 - 10 keV flux values (1.55$^{+0.74}_{-0.33} \times 10^{-13}$ erg s$^{-1}$ cm$^{-2}$ vs. 2.1$\pm0.1 \times 10^{-13}$ erg s$^{-1}$ cm$^{-2}$) and Fe K$\alpha$ EWs (2.30$^{+1.00}_{-0.78}$ vs. 2.3$\pm$1.5 keV).
\\

\textit{NGC 1667} - The \textit{XMM-Newton} spectra were best fit with a single absorbed power law plus a thermal component. To constrain the Fe K$\alpha$ EW in the local continuum fit (i.e. 3 - 8 keV), the spectra were binned by a minimum of 2 counts versus the 15 counts used for the global fit; C-stat was utilized in this local fit. Bianchi et al. (2005) fit the spectra with a reflection model (PEXRAV) with a soft excess, including a line at $\sim$0.9 keV. Our Fe K$\alpha$ EWs are consistent (0.86$^{+0.66}_{-0.50}$ vs. $<$0.60 keV), however our 2 - 10 keV flux values disagree by about a factor of two (0.43$^{+0.15}_{-0.11} \times 10^{-13}$ erg cm$^{-2}$ s$^{-1}$ vs 10$^{-13}$ erg cm$^{-2}$ s$^{-1}$).
\\

\textit{F05189-2524} - Since the spectral parameters were consistent between the \textit{XMM-Newton} and \textit{Chandra} observations, other than a lower constant multiplicative factor for the \textit{Chandra} spectra, we fit these spectra simultaneously to constrain the 2 - 10 keV flux. However, we only report the flux from the \textit{Chandra} observation as we have demonstrated that extended emission contaminates the \textit{XMM-Newton} field of view. The spectra were best fit by a double absorbed power law plus a thermal component, though the temperature was not constrained. The \textit{Chandra} spectra were fit independently to model the Fe K$\alpha$ emission. Though this feature was not detected, we derived an upper limit on the EW of 0.17 keV, consistent with the results of Ptak et al. (2003) who fit the 2002 \textit{Chandra} observation with a single power law plus thermal component (MEKAL). We also obtain similar 2 - 10 keV fluxes (23.5$^{+5.5}_{-4.9} \times 10^{-13}$ erg cm$^{-2}$ s$^{-1}$ vs. 37$\times 10^{-13}$ erg cm$^{-2}$ s$^{-1}$).
\\

\textit{NGC 3982} - As noted in the main text, the parameters, flux and Fe K$\alpha$ EW  values listed are from the model fits to \textit{Chandra} spectrum only as extended emission is present in the \textit{XMM-Newton} field of view. The spectrum was best-fit by an absorbed power law model with a thermal component, using the C-statistic on data grouped by 3 counts. We used ZGAUSS to test for the presence of an Fe K$\alpha$ line, but the EW was unconstrained in both the global and local fit, where in the latter, it was necessary to group by 1 count per bin to fit the continuum. Guainazzi et al. (2005) fit the \textit{Chandra} spectrum in a similar fashion (single absorbed power law with a thermal component) and obtained an upper limit on the 2 - 10 keV X-ray flux of 0.5$\times 10^{-13}$ erg s$^{-1}$ cm$^{-2}$, which is consistent with our value of 0.56$^{+1.15}_{-0.35}$ erg s$^{-1}$ cm$^{-2}$.  They also report a 1$\sigma$ detection on the Fe K$\alpha$ EW of 8$\pm$5 keV, which may indicate this parameter is unconstrained.
\\

\textit{NGC 4388} - This source varied between the \textit{XMM-Newton} observations from July to December 2002, increasing in flux and decreasing in Fe K$\alpha$ EW. Both observations were best fit by a double absorbed power law (allowing the normalization of the second power law component to vary between the two observations), a thermal component (necessary to fit the soft emission), and a Gaussian component to accommodate the Fe K$\alpha$ line. The spectral shape of the local continuum (3 - 8 keV,  to constrain the Fe K$\alpha$ EW) for the December observation and \textit{Chandra} observation required a base model of a double power law with an absorption component attenuating the second power law; a single power law base model was sufficient to fit the local continuum for the July observation. Beckmann et al. (2004) fit the \textit{XMM-Newton} spectra with a single absorbed power law and a Raymond-Smith thermal plasma model; they also detect Fe K$\alpha$ and a possible Fe K$\beta$ line at $\sim$6.89 keV. Our derived Fe K$\alpha$ EWs are consistent (0.62$^{+0.10}_{-0.10}$ keV vs. 0.57 keV and 0.18$^{+0.03}_{-0.02}$ keV vs. 0.22 keV for the July and December observations, respectively); they do not report a 2 - 10 keV flux or luminosity. The \textit{Chandra} flux for NGC 4388, from the June 2001 observation, is consistent with the \textit{XMM-Newton} July 2002 flux, though the Fe K$\alpha$ EW increased, which could be due to the presence of extended emission the \textit{XMM-Newton} field of view. Iwasawa et al. (2003) found the nucleus from the \textit{Chandra} observation to be moderately affected by pileup, but we did not see evidence of this when we applied the jdpileup model to the spectrum in Sherpa. We obtain consistent Fe K$\alpha$ EW values as Iwasawa et al. (0.29$^{+0.11}_{-0.10}$ vs. 0.44$\pm$0.09 keV), though a somewhat higher 2 - 10 keV flux (74.6$^{+88.5}_{-38.5} \times 10^{-13}$ erg s$^{-1}$ cm$^{-2}$) than their reported 2 - 7 keV flux (27$\times10^{-13}$ erg s$^{-1}$ cm$^{-2}$), though these values are likely consistent given the error bars on their flux and the more limited energy range over which they integrated.
\\

\textit{NGC 4501} - We report the parameters from the \textit{Chandra} spectral fit as the \textit{XMM-Newton} observation is contaminated by extended emission. The spectrum was best fit by an absorbed power law with a thermal component and we utilized the C-statistic as the data were grouped by 3 counts per bin. Brightman \& Nandra (2008) also find the \textit{XMM-Newton} field of view to be contaminated by extended emission. They fit the \textit{Chandra} spectrum with a reflection component, ontop of an absorbed thermal power law model. Similar to our work, they do not detect the Fe K$\alpha$ emission line in the global spectral fit.
\\

\textit{TOLOLO 1238-364} - The \textit{Chandra} spectrum was best fit by an absorbed power law with a thermal component and a Gaussian to accommodate the Fe K$\alpha$ emission. The data were binned by a minimum of 2 counts and we therefore employed the C-statistic. \citet{tololo1238} fit this spectrum with an absorbed power law plus thermal brehmsstrahlung model after binning by a minimum of 20 counts which washes out the Fe K$\alpha$ feature. They detected the line at low signal to noise after re-binning by constant width, but obtain an unconstrained EW whereas we detected this feature.
\\

\textit{NGC 4968} - The two \textit{XMM-Newton} observations for this source were best fit by an absorbed single power law model with a Gaussian component at the Fe K$\alpha$ energy, using the C-statistic on data binned by 2 - 3 counts; we saw no evidence for variability between the two observations. \citet{n4968} fitted these spectra were fit independently with a reflection model (PEXRAV) with fixed photon index ($\Gamma$ = 1.7), a power law component for the soft excess and gaussian for the Fe K$\alpha$ line. We obtained consistent 2 - 10 keV flux (2.08$^{+0.26}_{-0.26}$ $\times 10^{-13}$ erg cm$^{-2}$ s$^{-1}$ vs. 2.7$\pm$0.08, 2.3$\pm$0.08 $\times 10^{-13}$ erg cm$^{-2}$ s$^{-1}$) and Fe K$\alpha$ EW values (3.06$^{+0.99}_{-0.78}$ keV vs. 1.9$\pm$0.9, 3.2$\pm$1.1 keV) using the simpler power law model.
\\

\textit{M-3-34-64} - The \textit{XMM-Newton} spectra were fit by a double absorbed powerlaw with a thermal component and a Gaussian at the Fe K$\alpha$ line energy. Miniutti et al. (2007) fit this source with a reflection model, with the soft emission accommodated by  a power law model with two thermal components and a photoionized gas model and Gaussian components at 6.4 and 6.8 keV. Our observed 2 - 10 keV flux values approximately agree (32.5$^{+3.1}_{-3.1} \times 10^{-13}$ erg s$^{-1}$ cm$^{-2}$ vs. 21$\pm2 \times 10^{-13}$ erg s$^{-1}$ cm$^{-2}$), as well as our Fe K$\alpha$ EWs derived from the global fits (0.17$^{+0.04}_{-0.03}$ keV vs. 0.11$\pm0.02$ keV), though our local continuum fit results in a higher EW (0.31$^{+0.05}_{-0.04}$ keV).
\\

\textit{NGC 5135} - \textit{Chandra} observations of NGC 5135 reveal two X-ray point sources near the nucleus of the galaxy. The northern source was identified by Levenson et al. (2004) as the active nucleus, so we restrict our analysis to this source, using an extraction region of 1.2.'' They find the AGN spectrum to be best fit by a model consisting of two thermal components, a Gaussian component at $\sim$2 keV and at the Fe K$\alpha$ energy, and an absorbed power law. We grouped the data by a minimum of 3 counts per bin, employed the C-statistic and find that a double absorbed power law with a Gaussian component to accommodate the Fe K$\alpha$ emission reasonably fits the spectrum. Despite the different models used between Levenson et al. (2004) and us, we obtain consistent 2 - 10 keV fluxes (2.31$^{+0.98}_{-1.68} \times 10^{-13}$ erg cm$^{-2}$ s$^{-1}$ vs. 2.10$^{+0.19}_{-0.68} \times 10^{-13}$ erg cm$^{-2}$ s$^{-1}$) and Fe K$\alpha$ EWs (2.44$^{+0.94}_{-0.82}$ keV vs. 2.4$^{+1.8}_{-0.5}$ keV).
\\

\textit{NGC 5194} - The nuclear region of NGC 5194 contains several X-ray emitting features: the AGN and diffuse emission to the North and South (see Terashima et al. 2001). \textit{Chandra} is necessary to isolate the Seyfert nucleus and we therefore present the results of the \textit{Chandra} analysis only, and do not include the archival data from \textit{XMM-Newton}. Similar to Terashima et al. (2001), we extracted a source region centered on the optical center of the galaxy with a 1.5'' radius from the \textit{Chandra} data. The data were binned by a minimum of 3 counts and we utilized the C-statistic. The spectra were best fit by a double absorbed power law with a thermal component and a Gaussian component to accommodate the Fe K$\alpha$ emission. Terashima et al. fit the 2001 observation with an absorbed power law model and a reflection model, both of which yield consistent fluxes ($\sim$1.2$\times10^{-13}$ erg s$^{-1}$ cm$^{-2}$) and high EW values (3.5$^{+2.7}_{-1.6}$ keV and 4.8$^{+4.3}_{-2.5}$ keV, respectively) which agree with the values we obtain (1.04$^{+2.28}_{-0.73}$ erg s$^{-1}$ cm$^{-2}$ and 4.64$^{+1.42}_{-1.47}$ keV).
\\

\textit{NGC 5347} - The \textit{Chandra} spectrum is best fit by a double absorbed power law. To fit the local continuum to test for the presence of the Fe K$\alpha$ line, we rebinned the spectrum by a minimum of 3 counts, utilized the C-statistic and detected the line at the 3$\sigma$ level. Levenson et al. (2006) applied a reflection model (PEXRAV) to the higher energy range of the spectrum (4 - 8 keV) and fit the lower energy portion with power laws, a thermal component and line emission. Our 2 - 10 keV flux and Fe K$\alpha$ EW values agree, 2.58$^{+1.20}_{-1.41} \times 10^{-13}$ erg s$^{-1}$ cm$^{-2}$ vs. 2.2$\pm$ 0.4 erg s$^{-1}$ cm$^{-2}$ and 1.35$^{+0.53}_{-0.44}$ keV vs. 1.3$\pm$0.5 keV , respectively.
\\

\textit{Mrk 463} - Extended emission was evident in the \textit{XMM-Newton} field of view as indicated by the lower observed flux from the \textit{Chandra} spectrum. Indeed, the \textit{XMM-Newton} is likely contaminated by the double nucleus (see Bianchi et al. 2008), which the \textit{Chandra} observation is able to resolve. However, other than the constant multiplicative factor, the spectral parameters were consistent among the \textit{Chandra} and three \textit{XMM-Newton} spectra. The observations were consequently fit simultaneously to better constrain the \textit{Chandra} parameters, though only the flux for the \textit{Chandra} spectrum (for the Eastern source) is reported. To test for the presence of the Fe K$\alpha$ line, the \textit{Chandra} spectrum was rebinned by a minimum of 3 counts and the C-statistic was employed in the local (3 - 8 keV) fit; the line was detected at greater than the 99\% confidence level according to our simulations. Bianchi et al. (2008) also employed a double absorbed power law model to fit the spectrum and we obtain consistent 2 - 10 keV flux values and Fe K$\alpha$ EWs, 2.95$^{+1.84}_{-0.82} \times10^{-13}$ erg s$^{-1}$ cm$^{-2}$ vs. 4.1$\pm$1.8 $\times10^{-13}$ erg s$^{-1}$ cm$^{-2}$ and 0.20$^{+0.16}_{-0.13}$ keV vs. 0.21$^{+0.15}_{-0.12}$ keV, respectively.
\\

\textit{NGC 5506} - The \textit{XMM-Newton} MOS1 spectra showed evidence of mild pile-up above 6 keV according to the SAS tool \textit{epatplot} for all eight observations and were therefore not fit. For four of these (the 2001, 2002, 2008 and 2009 observations), the MOS2 spectra was also slightly piled and we excluded them from fitting. Archival \textit{Chandra} data for this source does exist, but were not included in this study because they were severely affected by pile-up. We find the spectra to be best fit by a double absorbed power law with a thermal component and several Gaussian components to fit Fe K emission, at energies $\sim$6.4 keV (neutral Fe K$\alpha$), $\sim$6.7 keV  and $\sim$6.95-7.0 keV; however we note that for the local continuum fits, sometimes only two components were needed. The source varies by a factor of $\sim$1.5 in flux on the time scale of approximately several months, though the Fe K$\alpha$ EW remains relatively constant. For the purposes of our analysis, we use the average flux and Fe K$\alpha$ EW among the 2001 and July 2004 observations ($<$EW$>$ = 0.14$^{+0.05}_{-0.06}$ keV) and among the 2002, August 2004, 2008 and 2009 observations ($<$EW$>$ = 0.13$^{+0.05}_{-0.04}$ keV; i.e. Figures \ref{ew_ratios} - \ref{alpha_c_thick}).  Guainazzi et al. (2010) studied the Fe K$\alpha$ emission of these observations in depth, using a suite of physically motivated models (i.e. a combination of relativistically/non-relativistically broadened Fe K$\alpha$ emission with relativistic/non-relativistic Compton reflection); the Fe K$\alpha$ EWs are consistent among these various fits. We derive EW values that agree with Guainazzi et al. (2010) using simpler modeling of the local (4 - 8 keV) spectrum with a power law and Gaussian components.
\\

\textit{Arp 220} - The \textit{XMM-Newton} and \textit{Chandra} spectra were best fit by a single absorbed power law model with a thermal component; only the absorption varied between the \textit{XMM-Newton} and \textit{Chandra} observations, though this had a negligible impact on the observed flux. We discarded the \textit{XMM-Newton} spectra from 2005 from our fitting due to low signal-to-noise, though we note the best-fit parameters were consistent with the other \textit{XMM-Newton} spectra. To test for the presence of Fe K$\alpha$ emission in the nuclear region, we utilized the \textit{Chandra} data only and rebinned by a minimum of 3 counts, employing the C-statistic, but the line was not detected. We note, however, that an emission line for ionized Fe K$\alpha$ at E = 6.51 keV was detected, consistent with \citet{arpFe}. The \textit{Chandra} data were first analyzed by Clements et al. (2002) who report a double nucleus and a halo of extended emission. We obtain consistent fluxes between their 3'' extraction area (which encompasses the double nuclei) and our 4.5'' extraction region: 1.07$^{+0.18}_{-0.16} \times 10^{-13}$ erg s$^{-1}$ cm$^{-2}$ vs 1.0$ \times 10^{-13}$ erg s$^{-1}$ cm$^{-2}$
\\

\textit{NGC 6890} - The \textit{XMM-Newton} spectra were grouped by a minimum of 3 counts per bin and were fit wit the C-statistic. The data were best fit by a double absorbed power law model and the Fe K$\alpha$ line was marginally detected at the $\sim$93\% confidence level. Shu et al. (2008) fit this source with a single power law and did not detect the Fe K$\alpha$ line, though this likely due to their choice of binning the data by a minimum of 20 counts which would eradicate the weak Fe K$\alpha$ feature. We obtain consistent 2 - 10 keV flux values given the error range on our derived flux: 1.20$^{+4.01}_{-0.88} \times 10^{-13}$ erg s$^{-1}$ cm$^{-2}$ vs. 0.69$\times 10^{-13}$ erg s$^{-1}$ cm$^{-2}$.
\\

\textit{IC 5063} - The \textit{Chandra} spectrum was moderately affected by pileup, $\sim$14\% according to the jdpileup model in \textit{Sherpa}. We therefore included a pileup model in the XSpec spectral fits, allowing only the grade migration parameter ($\alpha$) to be free, and obtained an $\alpha$ value of 0.37; excluded the pileup component when calculating the observed 2 - 10 keV flux. The jdpileup model indicated no pileup in the local 4 - 8 keV spectral fit, so it was modeled without a pileup component in XSpec. The broadband spectrum was best fit by a double absorbed power law and our simulations indicate that the Fe K$\alpha$ emission feature is significant at greater than the 2.5$\sigma$ level. This \textit{Chandra} spectrum has not been previously analyzed.
\\

\textit{NGC 7130} - The \textit{Chandra} spectrum of NGC 7130 was best fit by a double absorbed power law with a thermal component for the soft emission and a Gaussian feature at the Fe K$\alpha$ energy; we used the C-statistic with the data binned by 2 counts. This source was studied in detail by Levenson et al. (2005) where they fit the AGN spectrum with a double absorbed power law with a gaussian component as well as two thermal components. Though they added an extra thermal component, our derived 2 - 10 keV flux  and Fe K$\alpha$ EW values are consistent (2.07$^{+2.09}_{-1.04} \times 10^{-13}$ erg cm$^{-2}$ s$^{-1}$  vs. 1.6$^{+0.3}_{-0.4} \times 10^{-13}$ erg cm$^{-2}$ s$^{-1}$ and 0.82$^{+0.48}_{-0.33}$ keV vs 1.8$^{+0.7}_{-0.8}$, respectively).
\\

\textit{NGC 7172} - The \textit{XMM-Newton} observations were best fit by a double absorbed power law with a thermal component and Gaussian feature at the Fe K$\alpha$ energy. The normalization of the second power law was consistent between the the 2002 and 2004 observations, yet had to be fit independently for the 2007 observation. We only fit the PN and MOS2 spectra for the 2007 observation as the MOS1 spectrum showed evidence for milder pileup at higher energies from the task \textit{epatplot}. Our results indicate that the source increased by about a factor of 2 in flux between 2004 and 2007. The Fe K$\alpha$ emission features were fit independently among the three observations, though we use the average Fe K$\alpha$ EW for the 2002 and 2004 observations in the plots ($<$EW$>$=0.16$^{+0.06}_{-0.04}$ keV; i.e. Figures \ref{ew_ratios} - \ref{alpha_c_thick}) as the 2 - 10 keV flux values are consistent between the two observations and the EW values are not widely discrepant in both the local and global fits, indicating that the variations in the independent Gaussian fits are likely not significant. Noguchi et al. (2009) fit the 2007 observation with a double absorbed power law, thermal component (using MEKAL whereas we used APEC) and two Gaussian components, one at the Fe K$\alpha$ energy and the other at 1.7 keV; we obtain consistent 2 - 10 keV fluxes (517$^{+43}_{-40} \times 10^{-13}$ erg cm$^{-2}$ s$^{-1}$ vs. 423$ \times 10^{-13}$ erg cm$^{-2}$ s$^{-1}$) and Fe K$\alpha$ EWs (0.10$^{+0.02}_{-0.01}$ keV vs. 0.07$\pm$0.01 keV). Shu et al. fit the spectra from the 2002 \textit{XMM-Newton} observation by a double absorbed power law with a Gaussian component at the Fe K$\alpha$ energy, similar to our analysis, though they do not include a thermal component. Our 2 - 10 keV flux values are consistent (234$^{+19}_{-18} \times 10^{-13}$ erg cm$^{-2}$ s$^{-1}$ vs. 220$\times 10^{-13}$ erg cm$^{-2}$ s$^{-1}$), but our Fe K$\alpha$ EWs are not (0.14$^{+0.03}_{-0.03}$ keV from the global continuum fit vs. 0.04$\pm$0.03 keV). This discrepancy could be due to constraints placed on their modeling of the Fe K$\alpha$ line: they froze the energy at 6.4 keV, whereas we allowed this parameter to be free and it is not clear whether they placed a similar constraint on $\sigma$. Regardless of this discrepancy, both EW values are consistent with a Compton-thin source. Analysis of the EPIC data for the 2004 observation has not been previously published, making our analysis the first for this dataset.
\\

\textit{NGC 7582} - This source varied between the 2000 \textit{Chandra} observations and later \textit{XMM-Newton} observations, as well as between the 2001 and 2005 \textit{XMM-Newton} observations; the \textit{Chandra} and \textit{XMM-Newton} observations were fit independently. The \textit{Chandra} spectra were moderately affected by pileup ($\sim$30 - 49\% according to the jdpileup model in Sherpa) and were therefore fit with a pileup model component in XSpec, with only $\alpha$, the grade migration parameter, allowed to vary; the pileup component was discarded before calculating the flux and the Fe K$\alpha$ EW. The broad-band \textit{Chandra} spectra were best fit by a double absorbed power law whereas the \textit{XMM-Newton} spectra required a thermal component and Gaussian components at the Fe K$\alpha$ energy and at ionized  Fe K$\alpha$ energies ($\sim$6.72 keV for the 2005 observation and $\sim$6.97 keV for the 2001 observation)\footnote{However, in the local (3 - 8 keV) fits, this higher energy Gaussian is more consistent with Fe K$\beta$ emission, with a best-fit centroid energy of 7.08 keV for the 2001 observation, and shifts to a best-fit centroid energy of 6.91 keV for the 2005 observation.}, $\sim$1.84 keV and $\sim$2.47 keV. The Fe K$\alpha$ feature was detected at a statistically significant level for all observations according to our simulations. NGC 7582 dimmed between the \textit{Chandra} observation and each subsequent \textit{XMM-Newton} observation and the Fe K$\alpha$ EW increased. This decrease in flux with increase in Fe K$\alpha$ EW could reflect a variation in the obscuring medium, where the obscuration enhanced over time. Indeed, such an interpretation is favored by Piconelli et al. (2007), who postulate the existence of multiple absorption components in this system: a higher column-density absorber (possibly mildly Compton-thick) attributed to the putative torus and a lower-column density absorber acting as a screen to both the reflected and transmitted radiation. Though Piconelli et al. (2007) fit the \textit{XMM-Newton} observations with a more complex model (PEXRAV) we obtain consistent fluxes (21.1$^{+1.7}_{-1.8} \times10^{-13}$ erg s$^{-1}$ cm$^{-2}$ vs. 23.5$\times10^{-13}$ erg s$^{-1}$ cm$^{-2}$ for the 2005 observation and 38.6$^{+3.0}_{-3.1} \times10^{-13}$ erg s$^{-1}$ cm$^{-2}$ vs. 40.2$\times10^{-13}$ erg s$^{-1}$ cm$^{-2}$ for the 2001 observation), though lower Fe K$\alpha$ EW values (2005: 0.58$^{+0.04}_{-0.04}$ keV vs. 0.77$^{+0.05}_{-0.04}$ keV; 2001: 0.31$^{+0.05}_{-0.05}$ keV vs 0.62$^{+0.07}_{-0.08}$). Dong et al. (2004) fit the \textit{Chandra} spectra independently with a double absorbed power law, yet obtain an observed flux about half of ours (164$^{+263}_{-87} \times10^{-13}$ erg s$^{-1}$ cm$^{-2}$ vs. $\sim$75 $\times10^{-13}$ erg s$^{-1}$ cm$^{-2}$), though this discrepancy could result from us modeling pileup whereas they did not; we obtain consistent Fe K$\alpha$ EW values (0.15$^{+0.07}_{-0.07}$ keV vs. an averaged $\sim$0.19 keV).
\\

\textit{NGC 7674} - The global \textit{XMM-Newton} spectra were best fit with a double absorbed power law. To fit the local continuum between 3 - 8 keV, the data were binned to a minimum of 3 counts and the C-statistic was utilized; the Fe K$\alpha$ feature was detected a statistically significant level. Analysis of the broad-band \textit{XMM-Newton} spectra for this source has not been previously published.

\section{Simulated NuSTAR Detections: 10-40 keV}

Using ARF, RMF and background files provided by the NuSTAR team, which include separate ARF and background files for a  45'' and 101'' point spread function (useful for weak and strong sources, respectively) we generated a simulated NuSTAR spectrum in the 10-40 keV energy range as described in the main text. If the net count rate was $>10^{-2}$ s$^{-1}$, we utilized the simulated spectrum corresponding to the larger PSF, otherwise we used the spectrum generated with the smaller PSF. In Tables \ref{nustar_12} and \ref{nustar_o3}, we list the simulated source count rate and corresponding exposure time for the target to be detected at greater than the 5$\sigma$ level above the background, which was either $\sim8 \times10^{-4}$ s$^{-1}$ or $\sim4 \times10^{-3}$ s$^{-1}$, depending on the PSF. For the sources where this derived exposure time is under 5 ks, we instead list the exposure time necessary for at least 100 counts to be detected; if this exposure time is also under 5 ks, we adopt a minimum exposure time of 5 ks. For the three 2-10 keV non-detections from the 12$\mu$m sample, we list the exposure times using a spectrum simulated from the PLCABS model in XSpec, with N$_H = 10^{24}$ cm$^{-2}$, the number of scatterings set to 5, $\Gamma$=1.8 and the normalization adjusted such that the 2-10 keV flux equals the 3$\sigma$ upper limit. Again, we note that such a flux estimate is quite optimistic and the corresponding derived exposure times necessary for detection should be considered lower limits and are listed as such in Table \ref{nustar_12}.

\begin{table}[h]
\small
\caption{\label{nustar_12} NuSTAR Simulation Summary for 12$\mu$m Sample}
\begin{tabular}{lll}
\hline
\hline
Galaxy & Simulated Source Count Rate  & Exposure Time$^1$ \\
       & counts/sec                   & ks \\
\hline

NGC 0424                     & 3.48$\times10^{-3}$ & 8.8 \\

NGC 1068$^3$                 & 6.53$\times10^{-2}$ & 5.0 \\

NGC 1144$^3$                 & 1.77$\times10^{-1}$ & 5.0 \\

NGC 1320                     & 1.33$\times10^{-3}$ & 30  \\

F05189-2524$^3$              & 3.58$\times10^{-2}$ & 5.0 \\

F08572+3915$^2$              & 1.10$\times10^{-2}$ & $>$9.1  \\

NGC 3982                     & 1.19$\times10^{-3}$ & 35  \\

NGC 4388$^3$                 & 9.60$\times10^{-1}$ & 5.0 \\

NGC 4501                     & 5.04$\times10^{-3}$ & 5.7 \\

M-3-34-64$^3$                & 1.08$\times10^{-2}$ & 5.0 \\

NGC 5135$^2$                 & 6.88$\times10^{-3}$ & 15  \\

NGC 5194                     & 1.82$\times10^{-3}$ & 20  \\

NGC 5347$^2$                 & 1.77$\times10^{-2}$ & 5.6 \\

Mrk 463                      & 3.90$\times10^{-3}$ & 7.7 \\

NGC 5506$^3$                 & 1.51                & 5.0 \\

NGC 5953                     & 3.72$\times10^{-3}$ & $>$8.2  \\

Arp 220                      & 9.00$\times10^{-4}$ & 53  \\

IC 5063$^3$                  & 6.22$\times10^{-1}$ & 5.0 \\

NGC 7130                     & 3.14$\times10^{-3}$ & 10  \\

NGC 7172$^3$                 & 7.67$\times10^{-1}$ & 5.0 \\

NGC 7582$^3$                 & 4.68$\times10^{-2}$ & 5.0 \\

NGC 7590$^3$                 & 3.52$\times10^{-2}$ & $>$5.0  \\

NGC 7674$^2$                 & 6.17$\times10^{-3}$ & 16  \\

\hline
\hline

\multicolumn{3}{l}{Exposure time needed for source to be detected above the background at}\\
\multicolumn{3}{l}{greater than the 5$\sigma$ level.}\\
\multicolumn{3}{l}{Exposure time listed is for a detection of 100 counts as a 5$\sigma$ detection}\\
\multicolumn{3}{l}{above the background is $<$5 ks.}\\
\multicolumn{3}{l}{Minimum exposure time of 5.0 ks is adopted as the exposure times for both a 5$\sigma$}\\
\multicolumn{3}{l}{detection above the background and for a detection of at least 100 counts are $<$5 ks.}\\
\end{tabular}
\end{table}

\begin{table}[h]
\small
\caption{\label{nustar_o3} NuSTAR Simulation Summary for the [OIII] Sample}
\begin{tabular}{lll}
\hline
\hline
Galaxy & Simulated Source Count Rate  & Exposure Time$^{1}$ \\
       & counts/sec                   & ks \\
\hline

NGC 0291                    &  1.40$\times10^{-3}$ & 28  \\

Mrk 0609$^2$                &  1.72$\times10^{-2}$ & 5.8 \\

IC 0486$^3$                 &  1.27$\times10^{-1}$ & 5.0 \\

2MASX J08244333+2959238$^2$ &  7.36$\times10^{-3}$ & 14  \\

CGCG 064-017$^2$            &  1.12$\times10^{-2}$ & 8.9 \\

2MASX J11110693+0228477     &  1.68$\times10^{-3}$ & 22  \\

CGCG 242-028$^2$            &  1.22$\times10^{-2}$ & 8.2 \\

SBS 1133+572                &  1.14$\times10^{-3}$ & 37  \\

Mrk 1457$^2$                &  6.84$\times10^{-3}$ & 15  \\

2MASX J11570483+5249036$^2$ &  5.87$\times10^{-3}$ & 17  \\

2MASX J12183945+4706275     &  3.23$\times10^{-3}$ & 9.7 \\

2MASX J12384342+0927362$^3$ &  2.31$\times10^{-2}$ & 5.0 \\

CGCG 218-007$^2$            &  1.34$\times10^{-2}$ & 7.5 \\

\hline
\hline
\multicolumn{3}{l}{$^1$Exposure time needed for source to be detected above the background at greater than}\\
\multicolumn{3}{l}{the 5$\sigma$ level.}\\
\multicolumn{3}{l}{$^2$Exposure time listed is for a detection of 100 counts as a 5$\sigma$ detection above}\\
\multicolumn{3}{l}{the background is $<$5 ks.}\\
\multicolumn{3}{l}{$^3$Minimum exposure time of 5.0 ks is adopted as the exposure times for both a 5$\sigma$}\\
\multicolumn{3}{l}{detection above the background and for a detection of at least 100 counts are $<$5 ks.}\\
\end{tabular}
\end{table}

\end{document}